\documentclass[12pt,preprint]{aastex}

\slugcomment{Submitted to ...}

\shorttitle{The Brightest Young Star Clusters in NGC\,5253}
\shortauthors{Calzetti et al.}

\begin{document}

\title{The Brightest Young Star Clusters in NGC\,5253.\altaffilmark{1}}

\author{D. Calzetti\altaffilmark{2}, K.E. Johnson\altaffilmark{3}, A. Adamo\altaffilmark{4}, J.S. Gallagher III\altaffilmark{5}, 
J.E. Andrews\altaffilmark{6},  L.J. Smith\altaffilmark{7},  
G.C. Clayton\altaffilmark{8}, J.C. Lee\altaffilmark{9, 10},  E. Sabbi\altaffilmark{9}, L. Ubeda\altaffilmark{9},  H. Kim\altaffilmark{11,12},  J.E. Ryon\altaffilmark{5}, D. Thilker\altaffilmark{13}, S.N. Bright\altaffilmark{9}, E. Zackrisson\altaffilmark{14},  R.C. Kennicutt \altaffilmark{15}, S.E. de Mink\altaffilmark{16},  B.C. Whitmore\altaffilmark{9}, A. Aloisi\altaffilmark{9}, R. Chandar\altaffilmark{17}, M. Cignoni\altaffilmark{9}, D. Cook\altaffilmark{18}, D.A. Dale\altaffilmark{18}, B.G. Elmegreen\altaffilmark{19}, D.M. Elmegreen\altaffilmark{20}, A.S. Evans\altaffilmark{3,21}, M. Fumagalli\altaffilmark{22}, D.A. Gouliermis\altaffilmark{23,24},  K. Grasha\altaffilmark{2}, E.K. Grebel\altaffilmark{25}, M.R. Krumholz\altaffilmark{26},  R. Walterbos\altaffilmark{27}, A. Wofford\altaffilmark{28}, T.M. Brown\altaffilmark{9},  C. Christian \altaffilmark{9}, C. Dobbs\altaffilmark{29},  A. Herrero\altaffilmark{30,31}, L. Kahre\altaffilmark{27}, M. Messa\altaffilmark{4}, P. Nair\altaffilmark{32}, A. Nota\altaffilmark{7}, G. \"Ostlin\altaffilmark{4}, A. Pellerin\altaffilmark{33}, E. Sacchi\altaffilmark{34,35}, D. Schaerer\altaffilmark{36}, M. Tosi\altaffilmark{35}}
  
\altaffiltext{1}{Based on observations obtained with the NASA/ESA Hubble Space Telescope, at the Space Telescope Science Institute, which is operated by the Association of Universities for Research in Astronomy, Inc., under NASA contract NAS 5-26555. } 
\altaffiltext{2}{Dept. of Astronomy, University of Massachusetts -- Amherst, Amherst, MA 01003;\\ calzetti@astro.umass.edu}
\altaffiltext{3}{Dept. of Astronomy, University of Virginia, Charlottesville, VA}
\altaffiltext{4}{Dept. of Astronomy, The Oskar Klein Centre, Stockholm University, Stockholm, Sweden }
\altaffiltext{5}{Dept. of Astronomy, University of Wisconsin--Madison, Madison, WI}
\altaffiltext{6}{Dept. of Astronomy, University of Arizona, Tucson, AZ}
\altaffiltext{7}{European Space Agency/Space Telescope Science Institute, Baltimore, MD}
\altaffiltext{8}{Dept. of Physics and Astronomy, Louisiana State University, Baton Rouge, LA}
\altaffiltext{9}{Space Telescope Science Institute, Baltimore, MD}
\altaffiltext{10}{Visiting Astronomer, Spitzer Science Center, Caltech. Pasadena, CA}
\altaffiltext{11}{Department of Astronomy, The University of Texas at Austin, Austin, TX}
\altaffiltext{12}{Korea Astronomy and Space Science Institute, Daejeon, Republic of Korea}
\altaffiltext{13}{Dept. of Physics and Astronomy, The Johns Hopkins University, Baltimore, MD}
\altaffiltext{14}{Dept. of Physics and Astronomy, Uppsala University, Uppsala, Sweden}
\altaffiltext{15}{Institute of Astronomy, University of Cambridge, Cambridge, United Kingdom}
\altaffiltext{16}{Anton Pannekoek Institute for Astronomy, University of Amsterdam, Amsterdam, The Netherlands}
\altaffiltext{17}{Dept. of Physics and Astronomy, University of Toledo, Toledo, OH}
\altaffiltext{18}{Dept. of Physics and Astronomy, University of Wyoming, Laramie, WY}
\altaffiltext{19}{IBM Research Division, T.J. Watson Research Center, Yorktown Hts., NY}
\altaffiltext{20}{Dept. of Physics and Astronomy, Vassar College, Poughkeepsie, NY}
\altaffiltext{21}{National Radio Astronomy Observatory, Charlottesville, VA}
\altaffiltext{22}{Institute for Computational Cosmology and Centre for Extragalactic Astronomy, Department of Physics, Durham University, Durham, United Kingdom}
\altaffiltext{23}{Centre for Astronomy, Institute for Theoretical Astrophysics, University of Heidelberg, Heidelberg, Germany}
\altaffiltext{24}{Max Planck Institute for Astronomy, Heidelberg, Germany}
\altaffiltext{25}{Astronomisches Rechen-Institut, Zentrum f\"ur Astronomie der Universit\"at Heidelberg, Heidelberg, Germany}
\altaffiltext{26}{Dept. of Astronomy \& Astrophysics, University of California -- Santa Cruz, Santa Cruz, CA}
\altaffiltext{27}{Dept. of Astronomy, New Mexico State University, Las Cruces, NM}
\altaffiltext{28}{UPMC--CNRS, UMR7095, Institut d'Astrophysique de Paris, Paris, France}
\altaffiltext{29}{School of Physics and Astronomy, University of Exeter, Exeter, United Kingdom}
\altaffiltext{30}{Instituto de Astrofisica de Canarias, La Laguna, Tenerife, Spain}
\altaffiltext{31}{Departamento de Astrofisica, Universidad de La Laguna, Tenerife, Spain}
\altaffiltext{32}{Dept. of Physics and Astronomy, University of Alabama, Tuscaloosa, AL}
\altaffiltext{33}{Dept. of Physics and Astronomy, State University of New York at Geneseo, Geneseo, NY}
\altaffiltext{34}{Dipartimento di Fisica e Astronomia, Universit\`a degli Studi di Bologna, Bologna, Italy}
\altaffiltext{35}{INAF -- Osservatorio Astronomico di Bologna, Bologna, Italy}
\altaffiltext{36}{Observatoire de Gen\`eve, University of Geneva, Geneva,Switzerland}

\begin{abstract}
The nearby dwarf starburst galaxy NGC~5253 hosts a number of young, massive star clusters, the two youngest of which are centrally concentrated and surrounded by thermal radio emission (the `radio nebula'). To investigate the role of these clusters  in the starburst energetics, we combine new and archival Hubble Space Telescope images of NGC~5253 with  wavelength coverage from 1500~\AA\ to 1.9~$\mu$m in 13 filters. These include H$\alpha$, P$\beta$, and P$\alpha$, and the imaging from the Hubble Treasury Program LEGUS (Legacy Extragalactic UV Survey).  The extraordinarily well-sampled spectral energy distributions enable modeling with unprecedented accuracy the ages, masses, and extinctions of the 9 optically brightest clusters (M$_V<-$8.8) and  the two young radio nebula clusters. The clusters have ages  $\sim$1--15~Myr and masses $\sim1\times10^4 - 2.5\times10^5$~M$_{\odot}$. The clusters'  spatial location and ages indicate that star formation has become more concentrated towards the radio nebula over the last $\sim 15$~Myr.  The most massive  cluster is in the radio nebula; with a mass $\sim2.5\times10^5$~M$_{\odot}$ and an age $\sim$1~Myr, it is  2--4 times less massive and younger  than previously estimated. It is within a dust cloud with $A_V\sim 50$~mag, and shows a clear nearIR excess, likely from hot dust.  The second radio nebula cluster  is also $\sim$1~Myr old, confirming the extreme youth of the starburst region. These two clusters account for about half of the ionizing photon rate in the radio nebula, and will eventually supply  about 2/3 of the mechanical energy in present-day shocks. Additional sources are required to supply the remaining ionizing radiation, and  may include very massive stars.  
\end{abstract}


\section{Introduction}

Local dwarf starburst galaxies are  close  counterparts to the high--redshift  star forming systems  that built today's galaxies via interactions and mergers. The 
investigation of nearby dwarfs that are undergoing starburst events may, thus, shed light on the way galaxies assemble their stellar populations across cosmic times, and on the 
role young massive star clusters have in the energy and mechanical output of  star formation.  

The extreme youth of the starburst in the center of the dwarf galaxy NGC\,5253 has been established by many investigators 
\citep[e.g.,][]{vdBergh1980,Moorwood1982,Rieke1988,Caldwell1989,Beck1996,Calzetti1997,Pellerin2007},   and continues to be supported by recent 
data. The majority of the star clusters located within the central $\sim$250--300~pc has ages in the range 
from $\sim$10$^6$~yrs to a few 10$^7$~yrs  \citep{Calzetti1997,Tremonti2001,Harris2004,Chandar2005,Cresci2005,deGrijs2013}. 
A few older clusters, up to $\sim$10$^{10}$~yr in age, are located farther away from the galaxy's center \citep{Harbeck2012,deGrijs2013}. 
The youth of the central starburst is further supported by the absence of detectable non--thermal radio emission \citep{Beck1996} and 
the presence of strong signatures from Wolf--Rayet stars \citep{Campbell1986,Kobulnicky1997,Schaerer1997,LopezSanchez2007,MonrealIbero2010,Westmoquette2013}, 
which set a limit of $\lesssim$3--4~Myr to the most recent episode of star formation. The age range of the diffuse UV stellar population \citep{Tremonti2001,Chandar2005} 
and the recent star formation history of NGC\,5253 \citep{McQuinn2010,Harbeck2012} indicate that the star formation has been elevated, relative to the 
mean Hubble time value, for the past $\sim$5$\times$10$^8$~yr. 

The question of how to sustain continuously elevated star formation, possibly in the form of subsequent bursts, in NGC\,5253 has been 
tackled by many authors. An encounter with the relatively nearby grand--design spiral M\,83 about 1~Gyr ago has been suggested as one of the 
potential initial triggers \citep[e.g.][]{vdBergh1980,Caldwell1989}. M\,83 is located at a distance of 4.5 Mpc \citep{Thim2003} and is 1$^o$54$^{\prime}$ 
to the NW of NGC\,5253; thus, M\,83 is separated from NGC\,5253 \citep[at a distance of 3.15~Mpc][]{Freedman2001,Davidge2007} by about 1.35~Mpc. Although 
the distance is significant, and although \citet{Karachentsev2007} place NGC\,5253 in the neighboring Cen~A subgroup, \citet{LopezSanchez2012} argue 
that NGC\,5253 is located at the boundary between the two subgroups of Cen~A and M\,83\footnote{There are 
still uncertainties in the actual distances of both M\,83 and NGC\,5253. \citet{Karachentsev2007} place M\,83 at a distance of 5.2~Mpc, and \citet{Sakai2004} place
NGC\,5253 at a distance of 3.6~Mpc, the latter much closer to the distance of Cen~A, 3.8~Mpc. This has led \citet{Karachentsev2007} to associate NGC\,5253 to 
the Cen~A subgroup. In the latter case, NGC\,5253 may have interacted with Cen~A, instead of M\,83, in the past.}. A past interaction with the latter galaxy could 
explain the tidal extension in HI to the SE of M\,83 and the extension to the North of the HI distribution in NGC\,5253. These tails could be providing 
the fuel for the past and current bursts of star formation in NGC\,5253, in the form of in--falling metal--poor HI clouds \citep{LopezSanchez2012}.  The in--falling clouds convert to 
higher density molecular gas once they enter the central galaxy region and mix with the local interstellar medium \citep[ISM;][]{Turner1997,Meier2002,Turner2015}. The 
potential entrance `channel' for the gas  is defined by the only prominent dust lane,  which bisects the galaxy roughly along the minor axis  and emits in CO \citep{WalshRoy1989,Meier2002,Turner2015}.

Thus, the current starburst in NGC\,5253 is possibly the latest episode of a series of such feeding events, which are still 
ongoing. The dust--corrected UV and H$\alpha$ luminosities both provide a consistent value of the star formation rate, SFR=0.1--0.13~M$_{\odot}$~yr$^{-1}$ 
\citep{Calzetti2004,Calzetti2015}, also in agreement with the SFR derived from the total infrared emission, SFR(TIR)= 0.1~M$_{\odot}$~yr$^{-1}$ 
 \citep[using L$_{TIR}$=3.7$\times$10$^{42}$~erg~s$^{-1}$, which we calculate from the Spitzer imaging data of][]{Dale2009}. 
Radio measurements at 0.3, 0.7, 1.3, and 2 cm of the free--free emission \citep{Turner2000,Meier2002,Turner2004} yield a SFR$\sim$0.3--0.36~M$_{\odot}$~yr$^{-1}$ which is roughly 
a factor of three higher than what is obtained from the TIR and from the dust--attenuation--corrected UV and H$\alpha$.
The relatively small H$\alpha$ and UV half--light radii, $\sim$100~pc and $\sim$160~pc, 
respectively \citep{Calzetti2004}, imply a high star formation rate density, $\Sigma_{SFR}\sim$3.5~M$_{\odot}$~yr$^{-1}$~kpc$^{-2}$, confirming 
the star bursting nature of the galaxy \citep{KennicuttEvans2012}. The specific SFR of NGC\,5253 is sSRF$\sim$0.6--1.4$\times$10$^{-9}$~yr$^{-1}$, 
for a stellar mass  M$_*\simeq$2.2$\times$10$^8$~M$_{\odot}$ \citep{Calzetti2015}; the galaxy lies above the Main Sequence of star formation, i.e., the SFR versus stellar 
mass relation, for local galaxies \citep{Cook2014},  as expected for a starburst. 

Most of the current activity is coincident with a centrally concentrated, dusty radio source about 15--20~pc in extent, which we term the `radio nebula'. This has enough free-free emission to require 
one or more $\lesssim$3~Myr old star clusters with total mass  M$\sim$10$^6$~M$_{\odot}$ \citep{Turner2000,Turner2004}, for  a 0.1--120~M$_{\odot}$ 
Kroupa stellar Initial Mass Function \citep[IMF][]{Kroupa2001,Leitherer1999}. The ratio of the stellar mass to gas mass in the region suggests a star formation efficiency 
around 60\%, or about 10 times higher than that of Milky Way clouds \citep{Turner2015}. At least two distinct young star clusters are identifiable in the region, one of which is heavily attenuated 
by dust, and has been associated with the peak of emission at 1.3, and 2 cm by \citet{AlonsoHerrero2004}; this source has angular size 
 0.05$^{\prime\prime}\times$0.1$^{\prime\prime}$ ($\sim$0.8$\times$1.6~pc$^2$) and is associated with $\sim$20\%--30\% of the ionizing photons in the radio nebula 
\citep{Turner2004}.  The other cluster is also affected by the dust contained in the radio nebula, but to a much smaller degree; it is relatively bright in the UV, 
and it corresponds to the peak of observed H$\alpha$ emission in the galaxy \citep{Calzetti1997}. \citet{AlonsoHerrero2004} associate this UV--bright cluster with the 
secondary peak of emission at 1.3~cm \citep{Turner2004}. The radio nebula is driving most of the ionization in the galaxy, and the past and on--going starburst has been 
stirring the surrounding ISM, both chemically and energetically. 

NGC\,5253 is one of the few known cases containing regions of well--detected nitrogen enhancement, likely due to localized pollution from Wolf--Rayet stars in 
the area of the radio nebula \citep{WalshRoy1989,Kobulnicky1997,Schaerer1997,MonrealIbero2010,MonrealIbero2012,Westmoquette2013}. However, no other chemical `anomalies' have been 
convincingly detected. Tentative reports of He enhancement \citep{Campbell1986,LopezSanchez2007}, also a potential sign of pollution from Wolf--Rayet or 
other very massive stars, have been recently cast into doubt \citep{MonrealIbero2013}.
The galactocentric profile of the oxygen abundance is fairly flat \citep{Westmoquette2013}, with a mean value of 12$+$log(O/H)=8.25 \citep{MonrealIbero2012}, 
or about 35\% solar\footnote{We adopt 12$+$log(O/H)$_{\sun}$=8.69, for the solar oxygen abundance value \citep{Asplund2009}}, and with some scatter depending on 
assumptions for the electron temperature zone model \citep{Westmoquette2013}. This value of the oxygen abundance is similar to the one  reported by
\citet{Bresolin2011}, 12$+$log(O/H)=8.20$\pm$0.03. 

The ionized gas emission shows evidence of feedback from previous activity in the galaxy: filaments, shells, and arches  
characterize the H$\alpha$ distribution \citep{Marlowe1995,Martin1998,Calzetti1999}, closely followed by the X--ray emission tracing the hot gas 
\citep{Strickland1999,Summers2004}. The H$\alpha$ is mostly photo--ionized but also includes a non--negligible fraction, up to 
15\% in luminosity, of shock--ionization \citep{Calzetti1999,Calzetti2004,Hong2013};  the $\sim$kpc--size shells expand at a velocity of 
$\sim$35~km~s$^{-1}$ \citep{Marlowe1995} and have ages around 10--15~Myr \citep{Martin1998}. Thus, the clusters and stars located in the starburst 
 have a major impact on a number of observable characteristics of this galaxy, which would otherwise appear to be a rather 
unremarkable early--type dwarf. 

Despite ample evidence for mechanical feedback, we will assume in this work that only a small fraction of ionizing photons escapes the galaxy.
 This is true for local starburst galaxies in general, where escaping fractions are less than 3\% \citep[e.g.,][]{Grimes2009,Leitet2013}. Recently, \citet{Zastrow2013} 
 have suggested that these fractions may be lower limits due to the presence, in several starburst galaxies, of optically thin ionization cones, which may act as channels for the escape 
of ionizing photons. These will remain mostly undetected due to the random orientation of the cones relative to the line of sight. In NGC\,5253, the putative ionization cone is 
coincident with the dust lane \citep{Zastrow2011} and with  optically--thick CO(3-2) emission \citep{Turner2015}. Thus, while the properties of this feature are consistent with photoionization 
by escaping radiation, the ionization cone may be dusty and optically thick. Furthermore, the high ionization levels that mark this feature in NGC\,5253 are found in only one direction, 
so the solid angle of the escape zone is likely to be small \citep{Zastrow2013}. In what follows, we assume negligible escape of ionizing photons from NGC\,5253, although the issue remains open.

The new high--spatial resolution UV observations presented here provide an essential wavelength for probing the massive star population and the impact of  dust extinction 
in the radio nebula. Our goal is to quantify the properties of the star clusters in the radio nebula, in order to better understand their energetics and role within the NGC\,5253 starburst. 
To this end, we  study the stellar population content of the two star clusters using SED--modeling techniques on 
UV--optical--nearIR HST photometry. The photometric stellar continuum bands are supplemented with measurements of the emission lines in the light 
of H$\alpha$, P$\beta$, and P$\alpha$, also from HST imaging, which help to further constrain the ages and masses of the star clusters. 
The robustness of the SED modeling is first tested against other bright  stellar clusters within 
the starburst region of NGC\,5253, which are less affected by dust attenuation than the clusters within the radio nebula, and can, thus, 
provide a handle on potential degeneracies in the results for the latter. 

This paper is organized as follows: Section~2 describes the observations and the archival HST data used in this investigation; Section~3 presents the cluster 
selection and photometry; Section~4 presents the synthetic photometry and the fitting approach to the observed one; Section~5 describes the results of the 
SED fitting, and provides the ages, masses, and extinctions of the clusters, which are further discussed in Section~6.  A summary and the conclusions are 
provided in Section~7.  

\section{Observations and Archival Data}

\subsection{New Observations}

NGC\,5253 was observed with the HST Wide Field Camera~3 (WFC3) in the UVIS channel, in the filters F275W and F336W, on 2013--08--28, as 
part of the HST Treasury program LEGUS (Legacy ExtraGalactic UV Survey, GO--13364). A description of the survey, the observations, and the 
image processing is given in \citet{Calzetti2015}.

Briefly, the WFC3/UVIS datasets were 
processed through the {\sc calwf3} pipeline version 3.1.2 once all the relevant calibration files (bias and dark frames) for the date of observation were available in MAST. 
The calibrated, flat-fielded individual exposures were corrected for charge transfer efficiency (CTE) losses by using a publicly available stand-alone 
program\footnote{Anderson, J., 2013, http://www.stsci.edu/hst/wfc3/tools/cte\_tools}. 
These corrections were small because we used the post-flash 
facility\footnote{http://www.stsci.edu/hst/wfc3/ins\_performance/CTE/ANDERSON\_UVIS\_POSTFLASH\_EFFICACY.pdf}  to increase the background to a level 
near 12~e$^-$. The processed individual dithered images were then  
aligned, cosmic--ray cleaned, sky--subtracted, and combined at the native pixel scale using the {\sc astrodrizzle} routine\footnote{see: http://drizzlepac.stsci.edu/}, 
to an accuracy of better than 0.1 pixels. 
The World Coordinate System of the WFC3 F336W image was propagated to the other image, to obtain aligned images across filters, and the images in both 
filters were aligned with North up and East left. The images are in units of  e$^-$\,s$^{-1}$, which are converted to physical units using the WFC3 photometric zeropoints,
 included as keywords in the headers of the data products and posted at: http://www.stsci.edu/hst/wfc3/phot\_zp\_lbn. The basic details of the LEGUS images 
 for NGC\,5253 used in this paper are given in Table~\ref{tab1}.

\subsection{Archival Images}

The HST Archive contains a rich collection of images for NGC\,5253. For this paper, we retrieved images spanning from the UV to the H--band through the 
Hubble Legacy Archive\footnote{http://hla.stsci.edu} (HLA), both broad and narrow--band, to cover stellar continuum as well as optical and nearIR emission lines. 
When images in similar bands were available, preference was given to those at the higher angular resolution (e.g., ACS/HRC images were preferred over 
ACS/WFC images). Because of the extended 
wavelength coverage, the images used here have been obtained with different HST instruments, including the ACS/SBC, ACS/HRC, WFC3/IR, and NICMOS. 
Level~2 products were retrieved for each instrument/filter combination, implying that the individual post-pipeline exposures have been aligned, cosmic--ray cleaned, and 
combined using either {\it MultiDrizzle} or {\sc astrodrizzle}. The retrieved images have also been geometrically corrected and aligned with North up and East 
left. All images are provided by the HLA in units of e$^-$\,s$^{-1}$. We convert all images, except those from NICMOS, to physical units using the photometric 
zero points appropriate for each instrument/filter combination. The photometric zero points of the NICMOS images are referred to the default calibration of the instrument 
in DN\,s$^{-1}$,  thus we divide the HLA images first by the NICMOS Camera~2 gain (5.4 e$^-$/DN) in order to apply the published zero points. 
The details for each image product are listed in Table~\ref{tab1}. 

Although the archival data display a range of depths (as indicated by the large range of exposure times in Table~\ref{tab1}), all sources we study are 
detected with S/N$>$100 
in the broad and medium band filters. The uncertainty in the photometry is driven by crowding and uncertainties in the 
aperture corrections, rather than S/N limitations. 

\subsection{Additional Processing}

Improved alignment of all the images, both new and archival, is accomplished using the IRAF\footnote{IRAF is distributed by the National Optical Astronomy Observatory, which is operated by the Association of Universities for Research in Astronomy (AURA) under a cooperative agreement with the National Science Foundation.} 
tasks {\it geomap} and {\it geotran} and a sample of stellar sources in 
the ACS/HRC images as reference. 
The HRC images are preferred over others, because they have the smallest native pixel, which, in this case, drives the angular resolution of the final images. Thus, we elect to preserve 
as much as possible the highest angular resolution, even if it results in oversampling some of the lower resolution images. For the same reason, the aligned images are 
all re--sampled to the pixel scale of the ACS/HRC, 0.025$^{\prime\prime}$~pix$^{-1}$ (Table~\ref{tab1}).

After alignment, all images dominated by stellar continuum, i.e., all filters except F658N, F129N, and F187N, are converted to physical units of erg~s$^{-1}$~cm$^{-2}$~\AA$^{-1}$ 
using the most up--to--date values of the PHOTFLAM keyword as posted on the relevant webpage for each instrument (see example for WFC3 in previous section).

The nebular continuum and line emission from the central radio nebula significantly contaminate the fluxes in the broad band filters, for the clusters both within the nebula 
and in the surrounding  region. For instance, the presence of line emission in the F814W filter 
increases the flux measured for our individual sources between a few percent and a factor $>$3, depending on the location of the source. This effect has been  
noted as a problem for measurements of young sources by others \citep{Johnson1999, Reines2010}. Contamination by  emission lines of broad band filters, in turn, affects the derivation of the 
line flux intensities themselves, since the broad band images are used for the subtraction of the stellar continuum from narrow band images.  Nebular continuum will not 
have the same effect, since it  is present in both broad and narrow band filters. 

We derive emission--line--free images for the most affected among our filters: F435W, F814W, and F110W. The F110W filter receives most of the  emission line contribution 
from P$\beta$, and we use iterative subtraction between the F110W and F128N filters to remove the line contamination. For the lines affecting the F435W and F814W filters we do not 
have direct imaging in the corresponding narrow--band filters. We thus use the 3,200--10,000~\AA\ spectrum of \citet{StorchiBergmann1995} of the central 
10$^{\prime\prime}\times$20$^{\prime\prime}$ region of NGC\,5253 convolved with the F435W and F814W transmission curves to estimate the emission line contamination in these 
filters. The spectrum by \citet{StorchiBergmann1995} covers a sizable fraction of the region of interest here, along the E--W direction, and is thus representative of the excitation conditions 
in the center of NGC\,5253. The H$\alpha$ image derived from the F658N filter (see below) is then rescaled to the intensity of the emission lines and subtracted from both the F435W and 
F814W images. This process converges within two iterations. The remaining broad and medium band images are not significantly contaminated by emission lines, 
as estimated from the same spectrum. 

Emission line images are derived directly from the narrow--band filters, after subtracting the underlying stellar and nebular continuum. All narrow--band filters are converted to monochromatic fluxes (erg~s$^{-1}$~cm$^{-2}$~\AA$^{-1}$), before performing continuum subtraction. The continuum images are derived as follows. 
For the F128N image, which contains the P$\beta$ line, the rescaled, nebular--line--subtracted F110W image is used. Although straightforward, this method can include hard--to--quantify uncertainties, 
if there are significant color changes in the stellar population across the field-of-view. For the F187N image, which contains the P$\alpha$ line, we employ the 
rescaled F190N narrow--band image (Table~\ref{tab1}), which is free of emission lines and of any complications induced by potential color changes across the field. For both the F110W and 
F190N images, the rescaling factors are determined from emission--free point sources. For the F658N filter, 
which contains H$\alpha+$[NII], we create a continuum image by interpolating the flux--calibrated F550M and line--emission--subtracted F814W images. The resulting image is then 
subtracted from the flux--calibrated F658N image, without rescaling.  

The line emission images are then converted to units of erg~s$^{-1}$~cm$^{-2}$ by: (a) multiplying each image by the filter bandpass\footnote{The filter bandpass is defined as the filter rectangular width, i.e., the equivalent width divided by the maximum throughput within the filter bandpass. See, e.g., http://www.stsci.edu/hst/wfc3/documents/handbooks/currentIHB/c07\_ir06.html. } (72\AA, 159\AA, and 188\AA, for F658N, F128N, and 
F187N, respectively); and (b) correcting for the filter transmission curve values at the location of the redshifted lines. We remove the [NII] emission from the F658N image using 
[NII](6584\AA)/H$\alpha$=0.084 from \citet{Moustakas2006}, which is close to the value obtained from the spectrum of \citet{StorchiBergmann1995}, and the atomic ratio 
[NII](6548\AA)/[NII](6584\AA)=0.3. 

\section{Cluster Selection and Photometry}

\begin{figure}[h]
\figurenum{1a}
\plotone{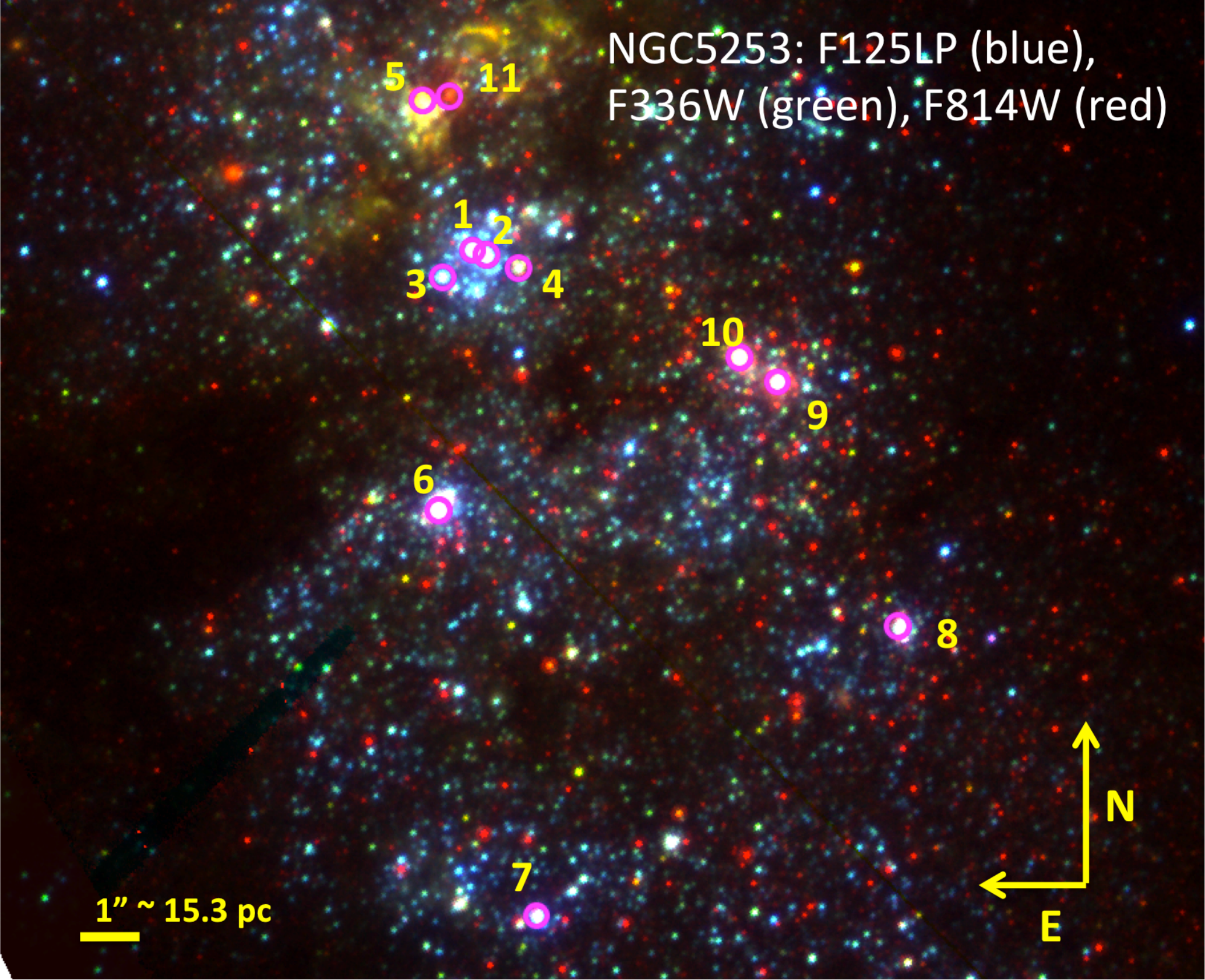}
\caption{A three--color composite of the central 20$^{\prime\prime}\times$16$^{\prime\prime}$ ($\sim$300~pc$\times$250~pc) of NGC\,5253, combining the ACS/SBC/F125LP (blue), 
WFC3/UVIS/F336W (green), and ACS/HRC/F814W (red) bands. The 11 star clusters in this study are identified with magenta circles and numbered. The circle radii are 7.5~pixels (0.19$^{\prime\prime}\sim$2.9~pc), i.e., 50\% larger than the baseline photometric apertures used in this work. Clusters~5 and ~11 are located within the central radio emission region \citep[see Figure~\ref{fig2} and][]{Turner2000}. A ruler of 1$^{\prime\prime}$ size ($\sim$15.3~pc) is provided at the bottom--left of the figure.} 
\label{fig1a} 
\end{figure}

Two cluster candidates are selected within the radio nebula (Figures~\ref{fig1a} and \ref{fig2}): one corresponding to the observed peak in H$\alpha$ and the other corresponding to the observed peak in 
P$\alpha$. Measurements at 7~mm indicate a size of  $\sim$1.$^{\prime\prime}$2 ($\sim$18~pc) for the radio nebula \citep{Turner2004}, as shown by the orange circle in the left--hand-size panel of  Figure~\ref{fig2}. The two peaks, H$\alpha$ and P$\alpha$, are separated by about 0.46$^{\prime\prime}$  mainly along the E--W direction (Figure~\ref{fig2}), corresponding to a 
spatial separation of $\sim$7~pc, with the P$\alpha$ peak emission located to the west of the H$\alpha$ one. 
For each peak, the other line is also present, but not as prominently. The H$\alpha$ peak, called `5' in Figure~\ref{fig1a}, has both P$\beta$ and P$\alpha$ emission 
spatially coincident with each other, and also with the continuum emission, within the accuracy that can be established from the image--to--image resolution differences (column~3 of Table~\ref{tab1}).  

The cluster candidate corresponding to the P$\alpha$ peak, called `11' in Figure~\ref{fig1a}, is slightly offset, by about 0.1$^{\prime\prime}$ ($\sim$1.5~pc), to the East of the centroid of the H$\alpha$ emission closest to it, while the P$\beta$ centroid falls in--between the peak locations of  the other two 
lines\footnote{Centroids of local emission peaks can be determined with an accuracy of about 1/5th--1/10th 
pixel, which, for the low--resolution WFC3/IR images, corresponds to a location accuracy of better than $\sim$0.02$^{\prime\prime}$.}.  This gradual transition as a function of increasing wavelength 
suggests that the offsets between the peaks of the hydrogen emission 
lines are likely due to variations in the dust optical depth, rather than the presence of separate sources of emission. Although the latter scenario cannot be completely ruled out, 
we will assume in this work that the slightly spatially--shifted lines all originate from the same source. A visual inspection of the continuum images shows that the spatial shift 
occurs between the J (F110W) and H (F160W) images, and no shift is obviously present at shorter wavelengths; the centroid of the source in the 
NICMOS F110W image coincides with the centroids in the shorter wavelength images, 
while the centroid in the NICMOS F160W image coincides with the centroid of the P$\alpha$ peak. 

Both clusters 5 and 11 are close to the peaks of free--free emission at cm wavelengths studied by \citet{Turner2000} (Figure~\ref{fig2}). 
Cluster 11 is within 0.18$^{\prime\prime}$, towards the S--W direction, of the peak at both 1.3~cm and 2~cm, while cluster~5 is $\sim$0.18$^{\prime\prime}$ to the South of 
the secondary peak at 1.3~cm. The coincidence between the sources would be increased if the relative astrometry 
 between the HST and the cm--wavelength observations were off by about 0.2$^{\prime\prime}$ along the N--S direction. This is consistent with the 0.1$^{\prime\prime}$--0.3$^{\prime\prime}$ uncertainty of the absolute astrometry for HST images \citep[e.g.,][]{Koekemoer2006}. We thus believe the optical and radio peaks to be actually coincident, in agreement with the assumption of \citet{AlonsoHerrero2004}; the observed offsets are likely due to small errors in  the absolute reference frames of the two sets of data.

\begin{figure}[h]
\figurenum{1b}
\plotone{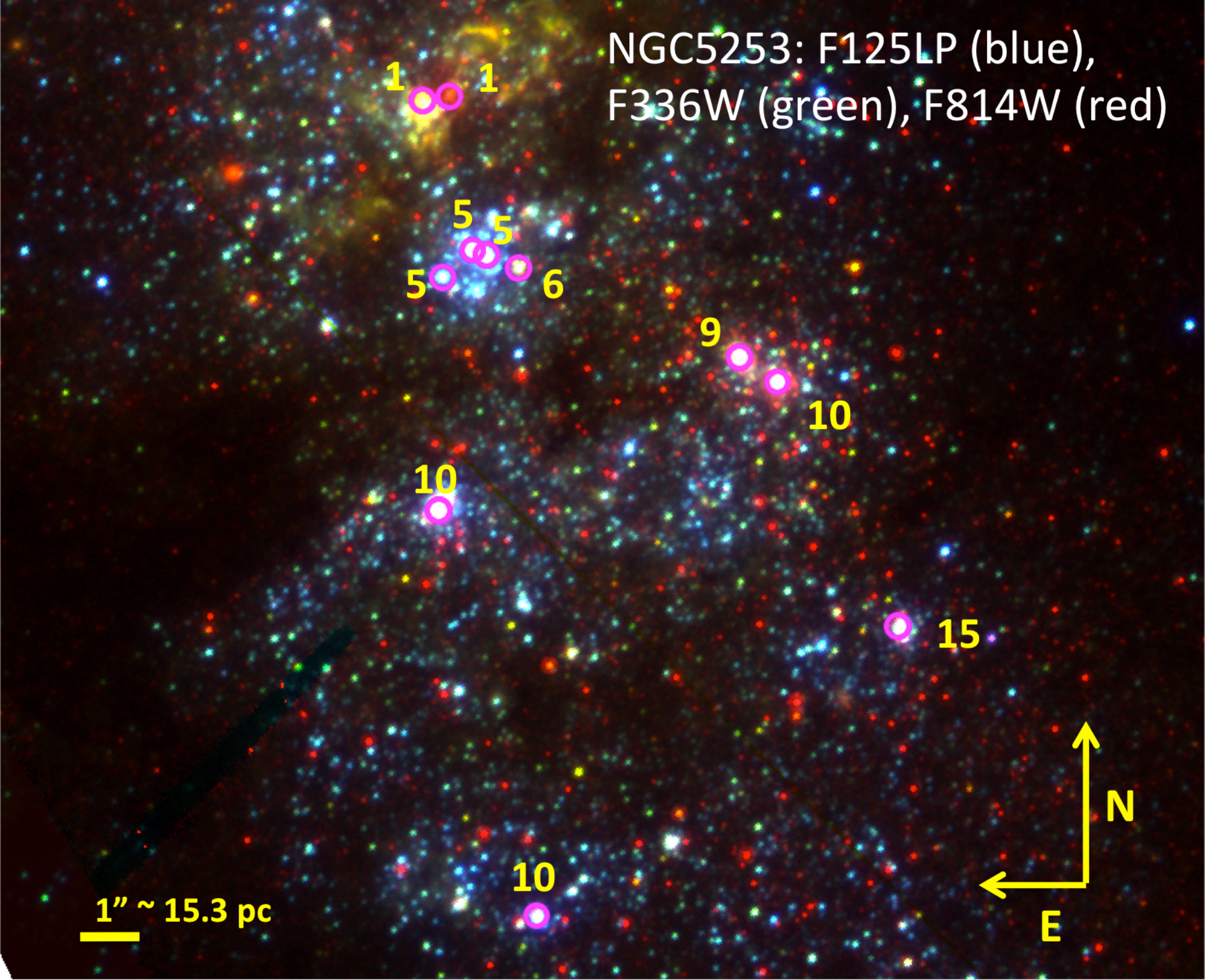}
\caption{The same as Figure~\ref{fig1a}, where the clusters' labels have been replaced by their best--fit ages.} 
\label{fig1b} 
\end{figure}

\begin{figure}[h]
\figurenum{2}
\plottwo{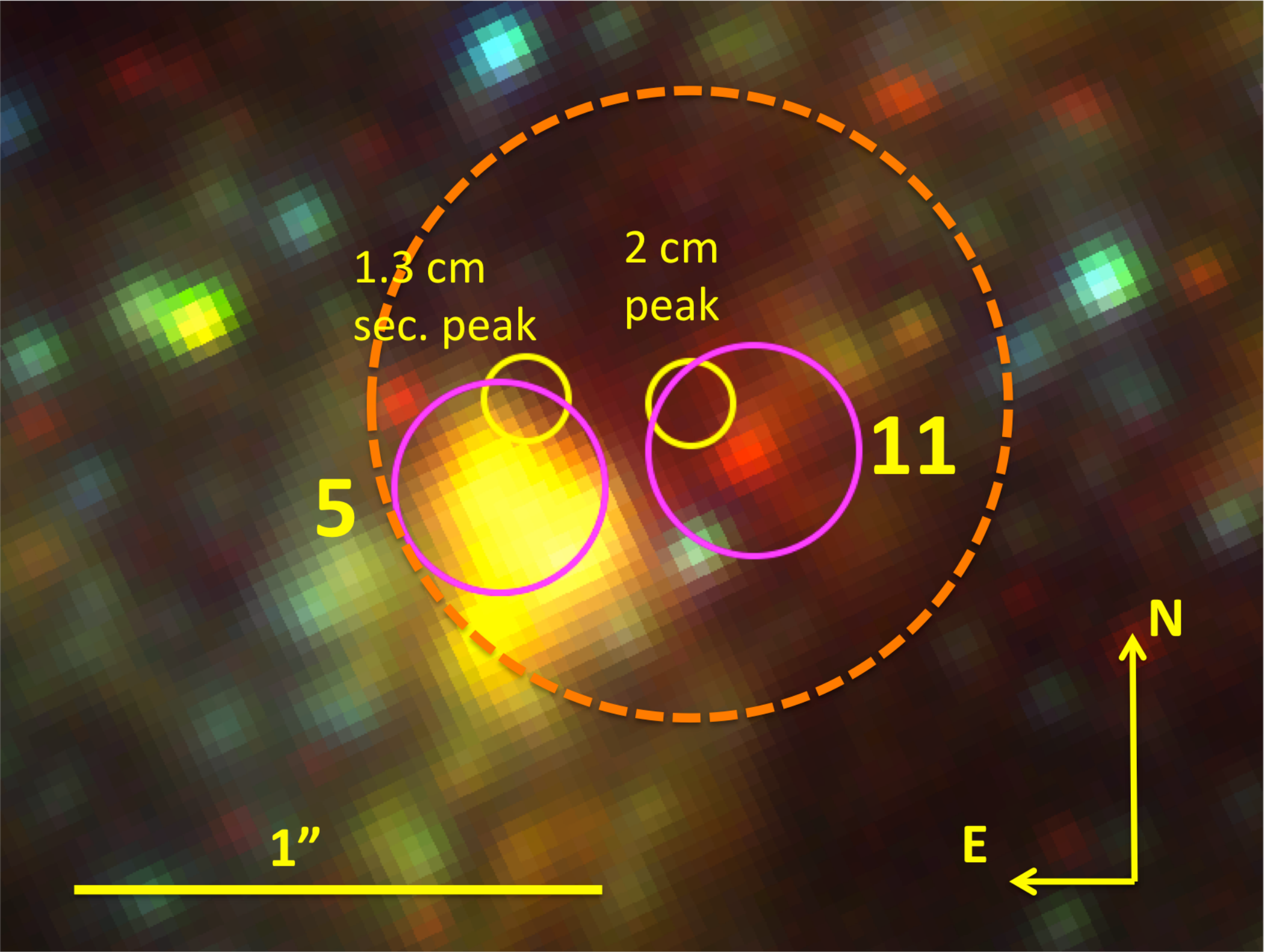}{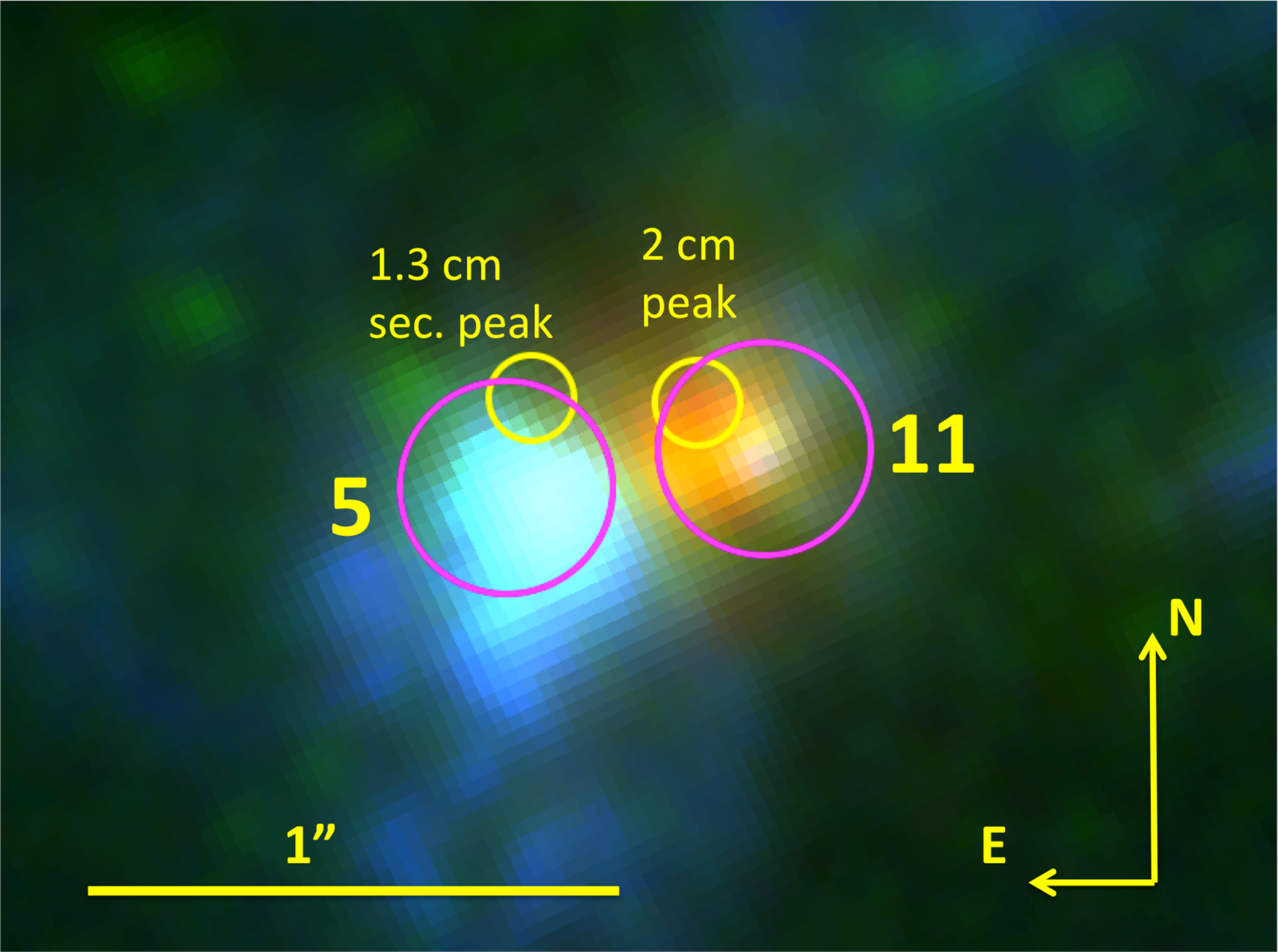}
\caption{A detail of the region surrounding clusters~5 and ~11, using {\bf (Left)} the same color composite as in Figure~\ref{fig1a} and 
{\bf (Right)} a three color--composite using the H$\alpha$ emission line (blue), the NICMOS/NIC2/F110W (green), and the P$\alpha$ emission line 
(red).  The region has size 2.4$^{\prime\prime}\times$1.8$^{\prime\prime}$ ($\sim$37~pc$\times$27~pc).  The two clusters are located at the center of the magenta circles, in correspondence of the 
H$\alpha$ (cluster~5) and the P$\alpha$ (cluster~11) emission peaks, respectively. The yellow circles identify the 
nominal positions, relative to the astrometry of the LEGUS WFC3 images, of the peak at 1.3 and 2~cm (right-hand--side circle) and the secondary peak at 1.3~cm (left--hand--side circle) identified by \citet{Turner2000}. The absolute astrometric uncertainty of the HST images, 0$^{\prime\prime}$.1--0$^{\prime\prime}$.3,  is comparable to the radius of the magenta circles (0$^{\prime\prime}$.2).  The diameters of the yellow circles, 0$^{\prime\prime}$.15, are comparable to the size of the mm  and cm beams \citep{Turner2000,Turner2004}. The orange circle in the left--hand--side panel marks the extent of the 7~mm emitting region, 1$^{\prime\prime}$.2 \citep{Turner2004}, which we term the ``radio nebula''. The color--composite in the right--hand--side panel 
highlights: (1) the differential intensity of the H$\alpha$ and P$\alpha$ emission in correspondence of the two clusters, with H$\alpha$ being stronger in cluster~5 and P$\alpha$ 
being stronger in cluster~11; and (2) the color gradient shown by the emission lines in cluster~11. The emission map in P$\beta$ is not shown, because of the significantly 
lower angular resolution of the WFC3/IR images relative to the ones shown here.} 
\label{fig2} 
\end{figure}

An additional nine star clusters, all visually identified as local peaks of emission in the V (F550M) band and all brighter than m$_V=$18.7~mag (M$_V<-$8.8~mag), are 
selected in order to perform tests 
on the SED fitting approach we adopt for this study. We ensure that the selected sources are clusters by requiring that each source's  FWHM is at least 50\% broader than the stellar PSF. 
Our compilation brings the total number of star clusters investigated here to 11, whose locations are identified in Figure~\ref{fig1a} and best--fit ages in Figure~\ref{fig1b} (see next section).  

Photometry is performed for all 11 clusters in multiple ways. Our default photometry uses an aperture of 5 pixels radius (0.125$^{\prime\prime}\sim$1.9~pc) with the background measured 
in an annulus with inner radius of 20 pixels and 3 pixels wide. We perform visual inspection of the sky annuli for each cluster to ensure that they are not affected by contamination from surrounding bright stars/clusters. We also run tests using sky annuli with inner radius in the range 15--20 pixels and width in the range 3--6 pixels, in order to quantify the effects of background contamination. 
The resulting photometry varies by less than 8\%, a much smaller uncertainty than those introduced by other effects (e.g. aperture corrections) as discussed below.  For the broad--band filters, we perform photometry on both the nebular--line subtracted and unsubtracted images, 
which we will compare to appropriate synthetic photometry from stellar population synthesis models (with and without nebular line emission, see next section). For the 
emission lines, we perform photometry on the stellar--continuum subtracted images. 

As some of the clusters show a complex structure (typically elongated), we also perform larger--aperture photometry, 
with 10-15 pixels radii (and up to 20 pixels for the emission lines), for the more spatially--isolated clusters. We use this larger aperture photometry as a check 
for our aperture corrections, especially for the WFC3/IR images, which have pixel size comparable to the radius of the default photometric aperture. We choose not to adopt 
the larger radius apertures as default for photometry, because a few of the 11 star clusters, including both clusters in the radio nebula, are located in crowded regions. 

The aperture corrections are determined from isolated star clusters found around the region where our 11 target clusters are located. We derive separate corrections for 
each instrument/filter combination. For the medium/broad--band filters (stellar continuum), 
the aperture corrections needed to bring the 5--pixels radius photometry to the infinite--aperture equivalent range from a factor 1.7 (WFC3/UVIS/F275W) to a factor 2.25 
(NICMOS/NIC2/F160W), with a larger value, 2.81, for the WFC3/IR/F110W instrument/filter combination.  For the emission lines, the aperture corrections for a 5--pixels 
radius are significantly larger, between a factor 3.7 and 7.8, which accounts for the more extended nature of the nebular emission. As expected, the aperture corrections 
decrease significantly, with values ranging from 5\% to 20\% for the stellar continuum filters, and from  40\% to 70\% for the emission lines, when a 15--pixel radius aperture is used 
for photometry. Comparisons between our default aperture and larger--aperture photometry indicates uncertainties of $\sim$15\% for all UV--optical medium/broad--band 
filters, 20\% for the NICMOS stellar continuum filters, and 35\% for the WFC3/IR/F110W filter; for the emission lines, we derive: $\sim$30\%--55\%--35\%  uncertainty for H$\alpha$, P$\beta$, 
and P$\alpha$, respectively. The larger uncertainty for the WFC3/IR photometry simply reflects the larger pixel scale of these images. The aperture corrections are the largest 
source of uncertainty for the stellar continuum filters;  the emission lines suffer from an additional (smaller) uncertainty due to the underlying stellar continuum subtraction.  Together with 
small registration offsets, this is especially a limitation for the P$\beta$ photometry, despite having one of the deepest among our exposures. The shallow depth of the exposure is an 
additional limitation for the P$\alpha$ image (Table~\ref{tab1}). The combination of all uncertainties, excluding the aperture correction ones, gives 
1$\sigma$ depths of: L(H$\alpha$)=2.3$\times$10$^{35}$~erg~s$^{-1}$,  L(P$\beta$)=3.5$\times$10$^{35}$~erg~s$^{-1}$, and L(P$\alpha$)=2.8$\times$10$^{35}$~erg~s$^{-1}$.

All photometry is corrected for foreground Milky Way extinction, using the extinction curve values listed in Table~\ref{tab2} and the color excess E(B$-$V)=0.049 from 
\citet[][as retrieved from the NASA/IPAC Extragalactic Database]{SchlaflyFinkbeiner2011}. The values of Table~\ref{tab2} can be directly applied to photometric 
measurements only for small values of the color excess, typically E(B$-$V)$\lesssim$0.1, since color variations across the filter bandpass are typically small; for larger 
values of the color excess, the extinction correction should be applied to the source's SED before convolution with the telescope/instrument/filter response curve. 

Table~\ref{tab3} lists the 11 star clusters, their coordinates, and the cross--IDs with other studies \citep{Calzetti1997,Harris2004,deGrijs2013}, where available\footnote{Our coordinates are slightly 
offset relative to those of \citet{deGrijs2013} by $\Delta\alpha=-0.057$~s and $\Delta\delta=-0.15^{\prime\prime}$.}, together 
with photometry,  H$\alpha$ equivalent widths (EW), and the color excess as inferred from the hydrogen emission 
line ratios. The listed photometry is for the measurements performed in the 5--pixels apertures, corrected for foreground Milky Way extinction and 
for aperture effects; in the case of continuum images, the photometry is from the original images, which include contribution from emission 
lines. Overall, the photometry of cluster~11 has larger uncertainties than that of the other 
clusters in our sample, due to its low flux densities, which are from a few times to over an order of magnitude fainter, depending on wavelength.

The EWs of H$\alpha$, calculated from the ratio of the emission line flux  to the stellar continuum flux density (interpolated from emission--line--subtracted 
images, see previous section), are given as a range: the smaller 
value corresponds to the ratio of line--to--continuum for  measurements within the 5--pixels radius aperture; the larger value corresponds to the ratio obtained after 
both line and continuum have been corrected for aperture effects. The color excess values are derived from the line ratios H$\alpha$/P$\beta$ and H$\alpha$/P$\alpha$ using the selective extinction values that can be derived from Table~\ref{tab2}, i.e., k(H$\alpha$)$-$k(P$\beta$)=1.70 and k(H$\alpha$)$-$k(P$\alpha$)=2.08, and the simple assumption of a foreground dust screen. 
For the intrinsic line ratios we adopt H$\alpha$/P$\beta$=17.57 and H$\alpha$/P$\alpha$=8.64, which are appropriate for HII regions with electron temperature T$_e\sim$11,500~K, 
measured for NGC\,5253 \citep{LopezSanchez2007}. We do not report the color excess derived from the ratio P$\beta$/P$\alpha$, since the selective extinction between the 
wavelengths of these two lines is small, and thus the resulting colors excess is subject to large uncertainties.

Two sets of values correspond to measurements performed at similar or close wavelengths, but with different instruments  (Table~\ref{tab1}): the WFC3/UVIS/F336W and the WFC3/IR/F110W 
measurements can be compared with the ACS/HRC/F330W and NICMOS/NIC2/F110W measurements, respectively. A close inspection of the photometry listed in Table~\ref{tab3} shows that 
the photometry in the two blue filters, WFC3/UVIS/F336W and  ACS/HRC/F330W, is usually comparable to better than 15\% ($\sim$0.07 in log scale), with the exception of cluster~11, where the difference 
is about 25\% ($\sim$0.1 in log scale).  We attribute the discrepancy to the difficulty of determining the background level around this highly obscured star cluster; however, even in this 
case the difference in photometry is still within the combined 1~$\sigma$ error of the two measurements. Conversely, the photometric values  in WFC3/IR/F110W and NICMOS/NIC2/F110W tend to be 
more discrepant with each other, with differences that range from 10\% to 40\% (0.04 to 0.15 in log scale). There is no obvious trend for one measurement to be systematically higher or 
lower than the other, although the NICMOS/NIC2/F110W measurement is more frequently the lower value. As the NICMOS /NIC2/F110W filter is at slightly shorter wavelength than the WFC3/IR/F110W, 
its photometry values should be higher, thus the observed discrepancy is likely a combination of measurement uncertainties and, possibly, some systematic calibration offset. Similarly to the other pair of filters, 
the discrepancies are within the combined 1~$\sigma$ error of the two measurements. 

\begin{figure}[h]
\figurenum{3}
\plottwo{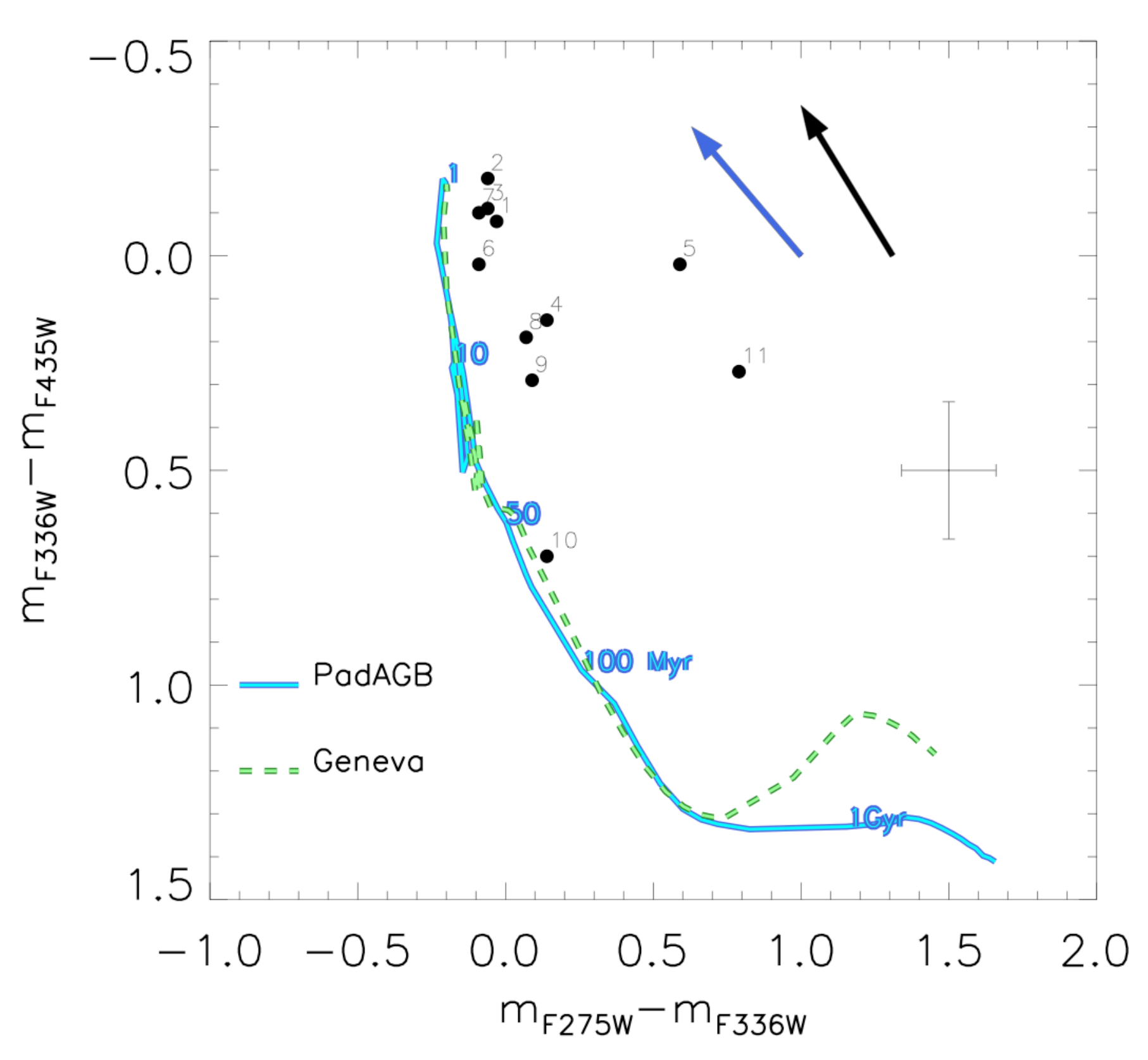}{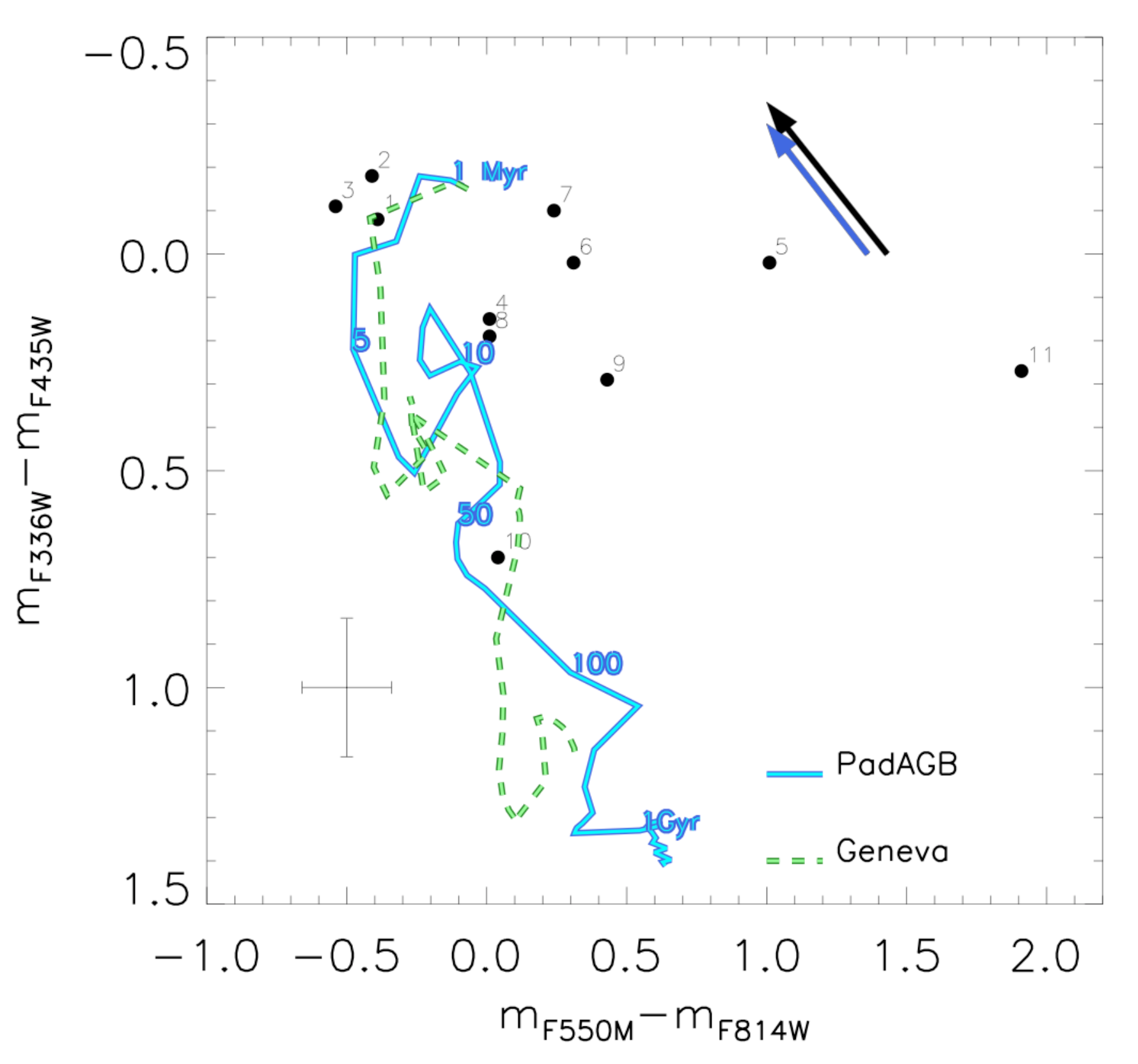}
\caption{Selected color--color plots for the 11 star clusters shown in Figure~\ref{fig1a}. The U--B (m$_{F336W}$--m$_{F435W}$) is shown as a function of both the 
NUV--U (m$_{F275W}$--m$_{F336W}$; left panel) and the V--I ( (m$_{F550M}$--m$_{F814W}$; right panel). All magnitudes are on the Vega photometric scale. Typical photometric uncertainties are shown as thin crosses in the two panels. The location of synthetic colors for a range of ages, from 1~Myr to 1~Gyr , is also shown for comparison, for 
both the  Padova stellar evolutionary tracks with AGB treatment and the Geneva tracks (section~4).  Vectors showing the effect on the observed colors of  a dust correction equivalent to a 
color excess E(B--V)=0.3 are reported for a starburst attenuation curve (black arrow) and for an LMC extinction curve (blue arrow).}
\label{fig3} 
\end{figure}

Color--color plots  of the 11 clusters in selected bands are shown in Figure~\ref{fig3}, together with the tracks of model stellar populations (section~4). These plots are only shown to guide intuition, and will not be used to derive the physical properties of the star clusters.

\section{Synthetic Photometry and Fitting Approach}

Spectral energy distributions (SEDs) from the UV to the nearIR  are generated using the Starburst99 \citep[][version as available in early 2014]{Leitherer1999} spectral synthesis models, using instantaneous star formation, 
with a \citet{Kroupa2001} IMF in the range 0.1--120~M$_{\odot}$ and metallicity Z=0.004 ($\sim$30\% solar), which is the closest value to the measured oxygen abundance of NGC\,5253 and for which models are available.  We produce models using both the Padova with AGB treatment\footnote{Our clusters are young enough, $\lesssim$15~Myr,  that use of the Padova tracks without AGB treatment yields identical results.} and the Geneva tracks \citep{Meynet1994, Girardi2000, Vazquez2005}. Since the clusters under consideration tend to be massive, M$\gtrsim$10$^4$~M$_{\odot}$, we expect minimal impact from stochastic sampling of the IMF \citep{Cervino2004}, and use the default deterministic models. The Starburst99 models include nebular continuum, but not nebular emission lines. 
These are added by Yggdrasil \citep{Zackrisson2011}, which uses Starburst99 stellar populations as an input for CLOUDY \citep{Ferland2013}. For Yggdrasil, we adopt a 50\% covering factor 
for the ionized gas, meaning that only 50\% of the nebular emission is spatially coincident with the star cluster. This attempts to reproduce the observed trend for nebular 
emission to be more extended than the stellar continuum (see previous section). 
Models with and without emission lines are generated for the age range 1~Myr--1~Gyr in steps of 1~Myr in the 1--15~Myr range, 
10~Myr in the 20--100~Myr range, and 100~Myr in the 200--1,000~Myr range. Instantaneous models are assumed here to reasonably represent the population of individual star clusters.

The SEDs produced by both Starburst99 and Yggdrasil are attenuated with: a starburst attenuation curve \citep{Calzetti2000}, and a Milky Way, an LMC and an SMC extinction curve \citep[as parametrized by][]{Fitzpatrick1999}. For the extinction curves, we adopt a foreground dust geometry \citep{Calzetti2001} of the form:
\begin{equation}
F(\lambda)_{out} = F(\lambda)_{model} 10^{[-0.4 E(B-V) k(\lambda)]},
\end{equation}
and both cases of equal and 
differential attenuation for the nebular gas and stellar continuum; for the differential attenuation, we assume that the stellar continuum is subject to half the attenuation of the 
nebular gas \citep{Calzetti1994,Kreckel2013}. For the case of the starburst attenuation curve, the dust geometry is `built--in' into the functional form of the curve, and the differential attenuation between gas and stars 
is part of the way the curve itself was derived. We thus end up with seven different models for the dust attenuation: one attenuation curve and six extinction curves (three times two 
different ways of attenuating gas and stars). We generate the models in the color excess range E(B$-$V)=0--3~mag, with step 0.01.

We will see in the next section that cluster~11 cannot be easily explained by foreground extinction/attenuation only. For this case, we generate models in which the dust and stars/gas 
are uniformly mixed together, according to the formula:
\begin{equation}
F(\lambda)_{out} = F(\lambda)_{model} \{1- e^{[-0.921 E(B-V) k(\lambda)]}\}/[0.921 E(B-V) k(\lambda)],
\end{equation}
with the color excess E(B$-$V) in the range 0--20~mag. Although the uniformly mixed geometry is likely to be an oversimplification of the complex environment surrounding 
cluster~11, it helps explain many of the properties of the star cluster. Throughout this paper, we will call `front--to--back optical depth' the quantity A$_V$=3.1 $E(B-V)\  k(V)$ from the 
mixed geometry.

The dust--attenuated SEDs are then convolved with the transmission curve of the filter plus the HST optics to produce synthetic luminosities, that are normalized to the default 
mass of Starburst99, 10$^6$~M$_{\odot}$. 

We use $\chi^2$--minimization between the models and the data, taking into account the measurement uncertainties, to obtain the distribution of solutions and the reduced 
$\chi^2$ value for each. We then plot the distribution of solutions within the  99\% significance level for the appropriate number of degrees of freedom, and select the best values 
and the uncertainty for the age, color excess, and mass of each star cluster based on the shape of the reduced $\chi^2$ probability distribution. We fit only the stellar continuum (medium/broad--band filter)  photometry up to and not including the H--band. Both the J--band and the  H--band can be heavily affected by the presence of small numbers of red supergiant stars \citep{Cervino2004,Gazak2013,deGrijs2013}.  In order to retain as much as possible of the wavelength baseline, we include the J--band in our fits, but exclude the H--band, and we only use it as a sanity check on our results.   We use the hydrogen emission 
line intensities and the H$\alpha$ EW as a check on our solutions, by deriving an approximate age from the H$\alpha$ EW and a range of color excesses from the 
emission line ratios. We do not include the emission lines in the fit directly, since these can be affected by feedback effects from the star clusters (e.g., supernova explosions, which begin within the first 3 Myr, can eject gas from the cluster's surroundings and lead to an underestimate of the emission line intensity), especially for the massive clusters we are studying. 

As presented in the next section, some of the star clusters have best fit ages around 1~Myr. This implies that pre--Main--Sequence stars could be present and contribute to the observed SEDs. Our models do not include pre--Main--Sequence stars, and this should be taken as a limitation to our approach.

We derive three solutions for the age, color excess, and mass of each star cluster from SED fitting. Two are based on the full wavelength coverage from $\sim$1,500~\AA\ to $\sim$11,000~\AA\ (7 data points =  
3 degrees of freedom, we average together the two measurements in U and the two measurements in J, to produce one single data point at each wavelength), using Starburst99 and Yggdrasil models for the nebular--line--subtracted and unsubtracted photometry, respectively. The solutions from the comparison of the 
unsubtracted photometry with the Yggdrasil models are our reference values. We use the sets of solutions from the subtracted photometry plus Starburst99 models as 
a comparison, in order  to 
evaluate how well CLOUDY reproduces the conditions of the nebular gas in each star cluster. This is particularly important for the central clusters in NGC\,5253, where the strong ionized gas emission 
 can affect the measurements  (e.g., by leaving residual emission in the stellar continuum bands). 
A third set of solutions is based on using only 5 bands (F275W, F336W, F435W, F550M, and F814W = 1 degree of freedom) for the best fits. This third set  enables 
us to compare the solutions obtained from the more restricted wavelength range (which is the common situation for galaxies in the LEGUS and other projects) against those, possibly more secure, 
obtained from the more extended wavelength coverage. 

\section{The Ages, Masses, and Extinctions of Bright Star Clusters}

\subsection{Clusters outside the Radio Nebula}

Clusters 1--4 and 6--10 are located outside the radio nebula, although still within the starburst region. All except for cluster~4 have been investigated before by 
\citet{Calzetti1997}, \citet{Tremonti2001}, \citet{Harris2004}, and \citet{deGrijs2013}. All are younger than 15--20~Myr, as determined by those authors, using either lower resolution HST data, from the WFPC2, or 
UV spectroscopy, or  a combination of ACS/HRC, WFPC2, and NICMOS data. Those earlier papers using broad and narrow-band photometry employ a more restricted wavelength range, 
and generally only one emission line (H$\alpha$). In our case, the availability of filters further in the UV (F125LP and F275W) provides 
better leverage for constraining ages of the star clusters from photometry, and the presence of multiple emission lines enables additional 
considerations on the physical conditions surrounding the clusters. 

\begin{figure}
\figurenum{4}
\plottwo{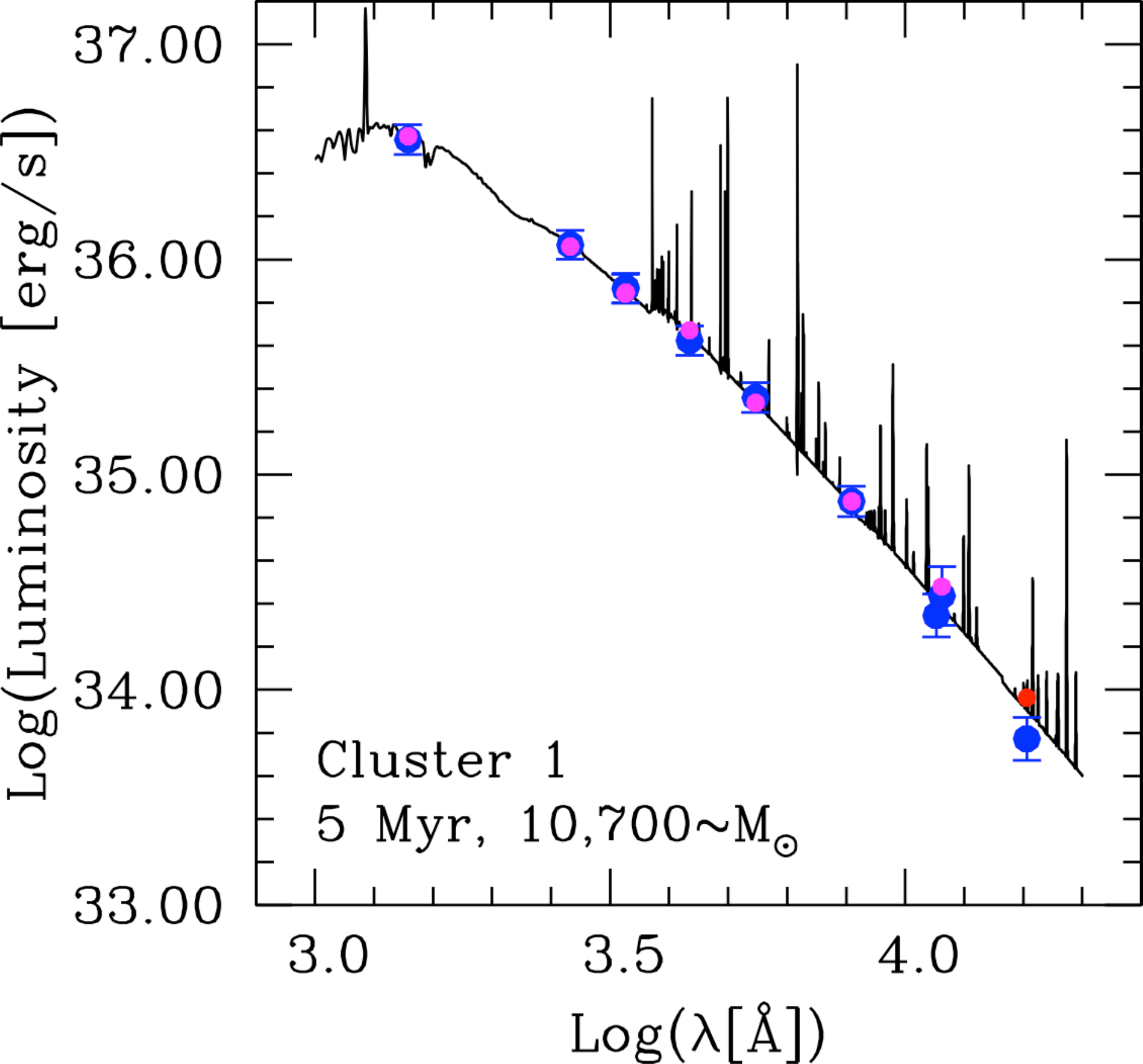}{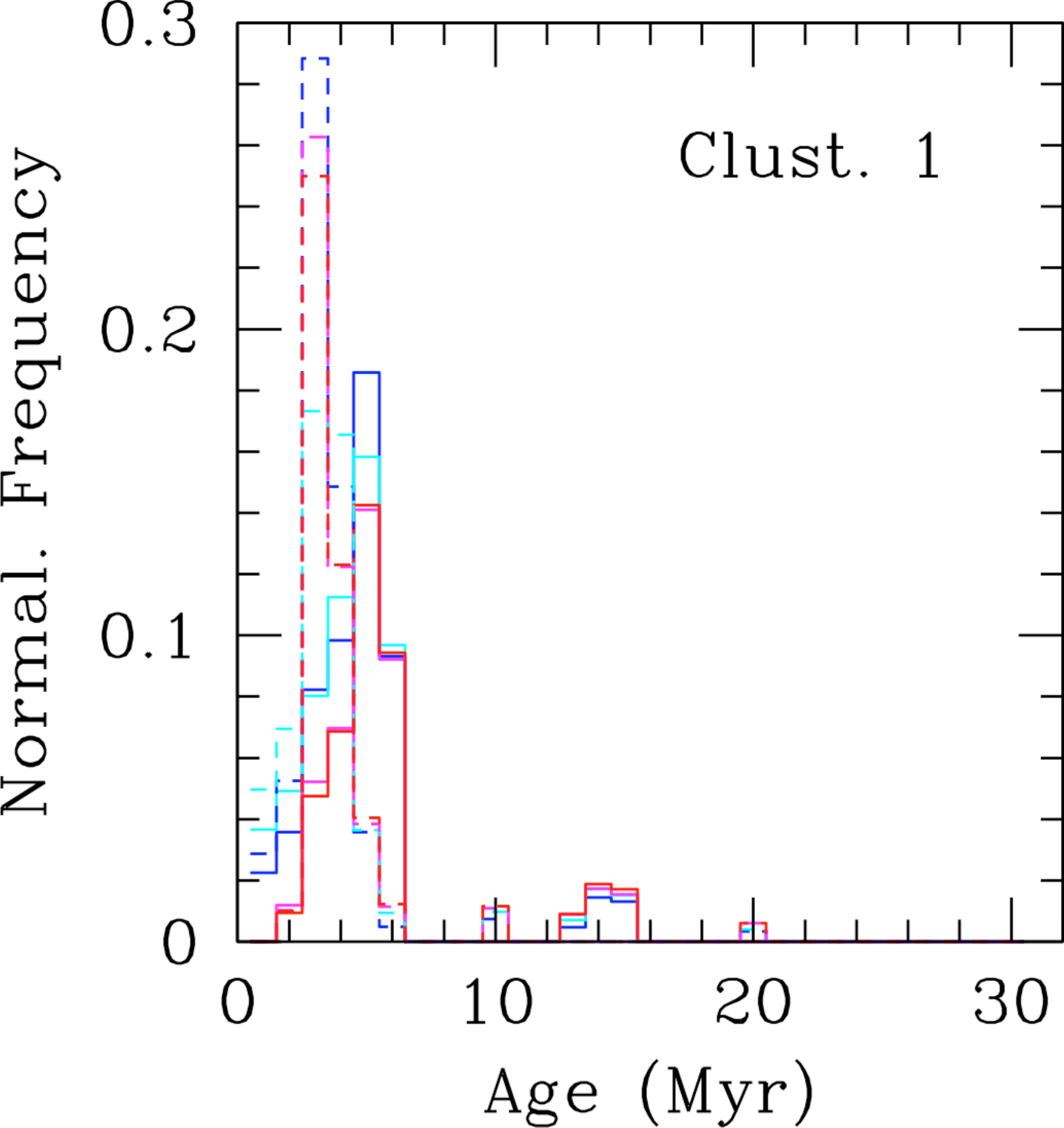}
\plottwo{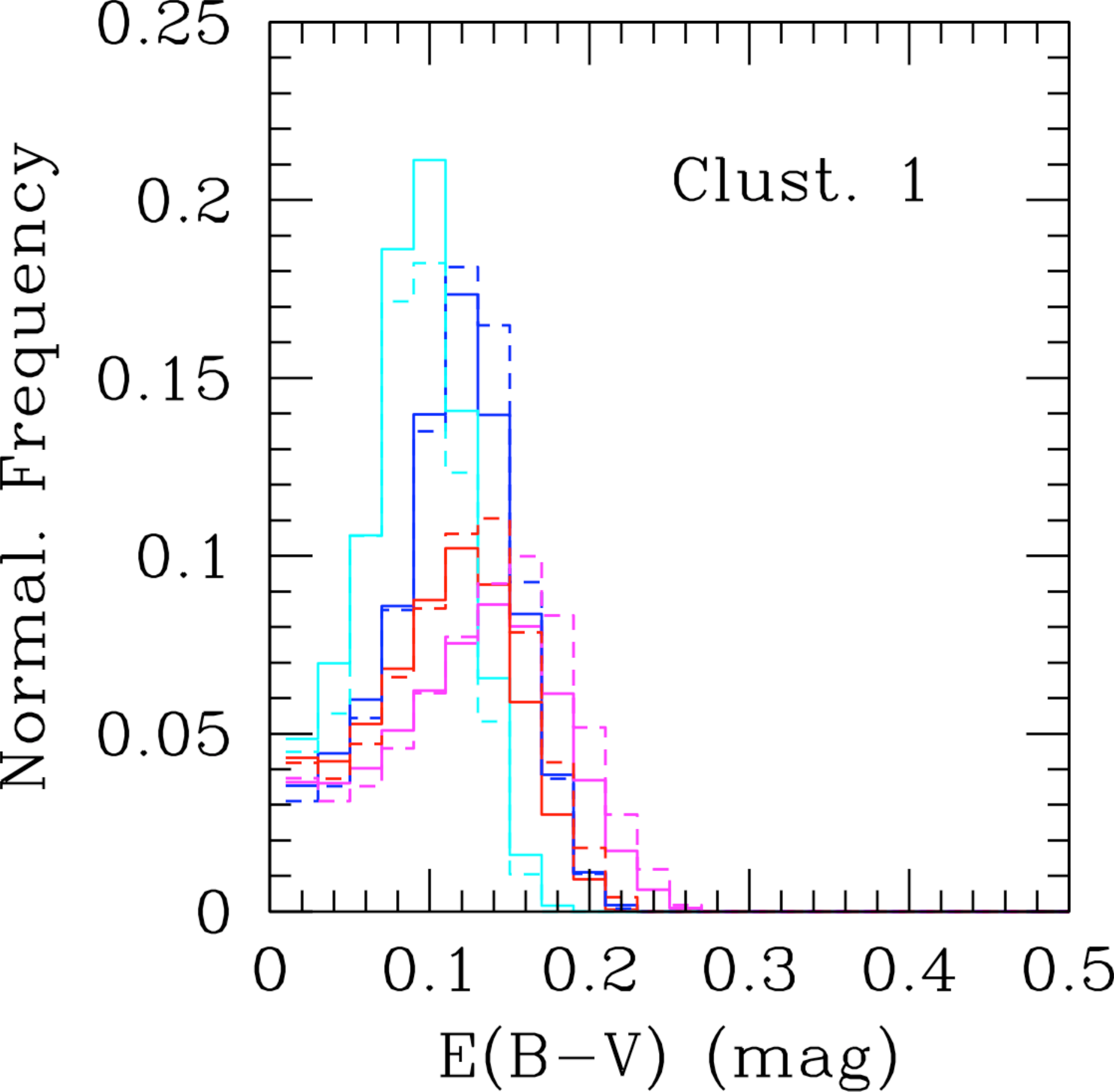}{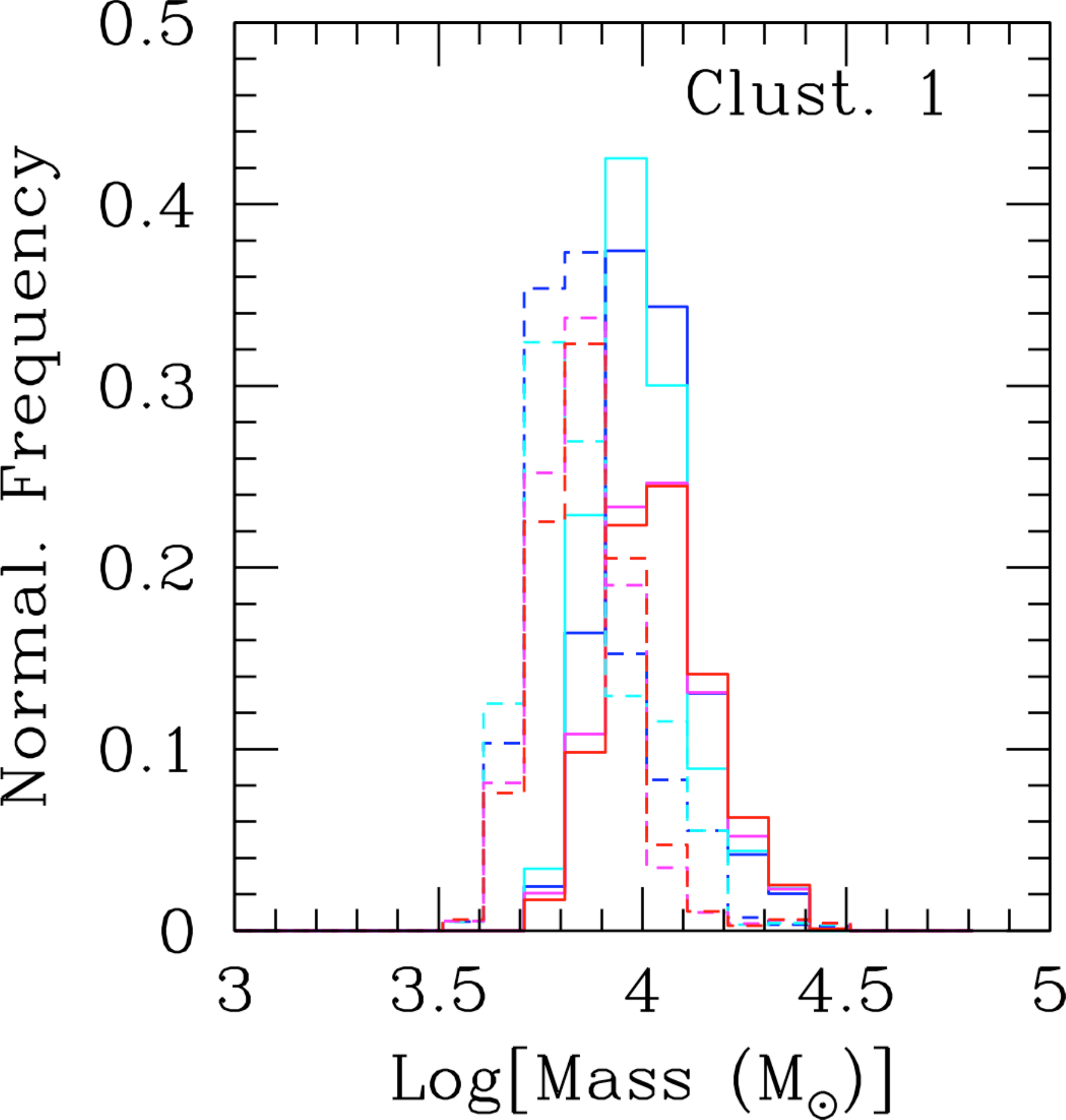}
\caption{Best fit SED (top--left panel) and 99\% confidence histograms for the distribution of ages (top--right panel), color excess (bottom--left panel), and mass (bottom--right panel) for 
Cluster~1. In the best--fit SED plot, the blue points with error bars are the observed photometry,  the magenta points are the synthetic photometry, and the black line 
is the Yggdrasil model, together with dust attenuation and mass normalization, that  provides the best fit (smallest reduced $\chi^2$ value) to the observed photometry. The measured 
photometry in the H--band is not used in the fits, in order to avoid potential contamination by stochastic presence of red supergiant stars; we show, however, the predicted photometric value 
(in red) from the best fit model. In the histograms, the 
continuous lines are for Padova stellar models and the dashed lines for Geneva stellar models. The colors indicate: differential LMC extinction curve (blue),
differential SMC extinction curve (cyan), differential MW extinction curve (magenta), and starburst attenuation curve (red). 
\label{fig4}}
\end{figure}

\begin{figure}
\figurenum{5}
\plottwo{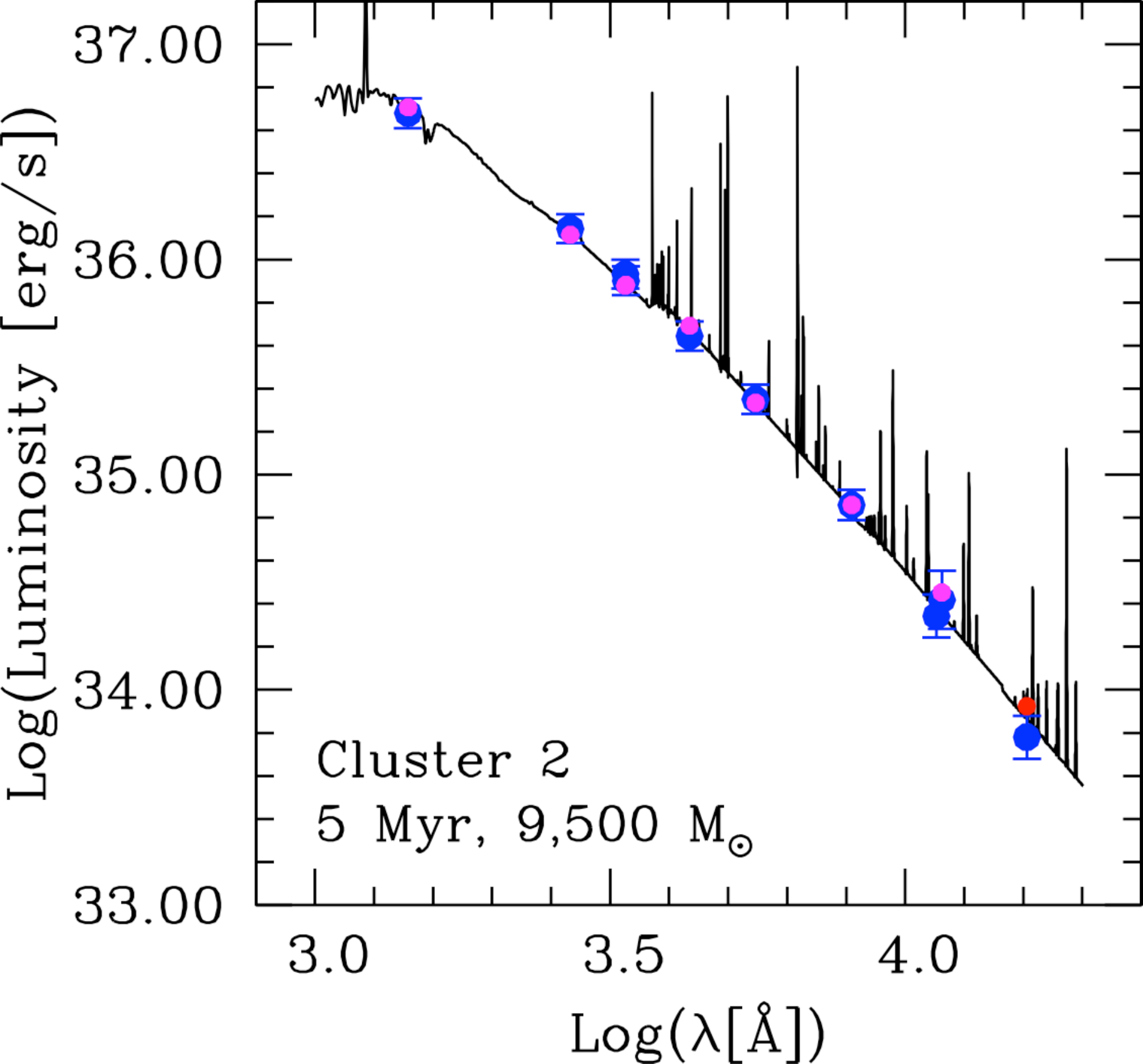}{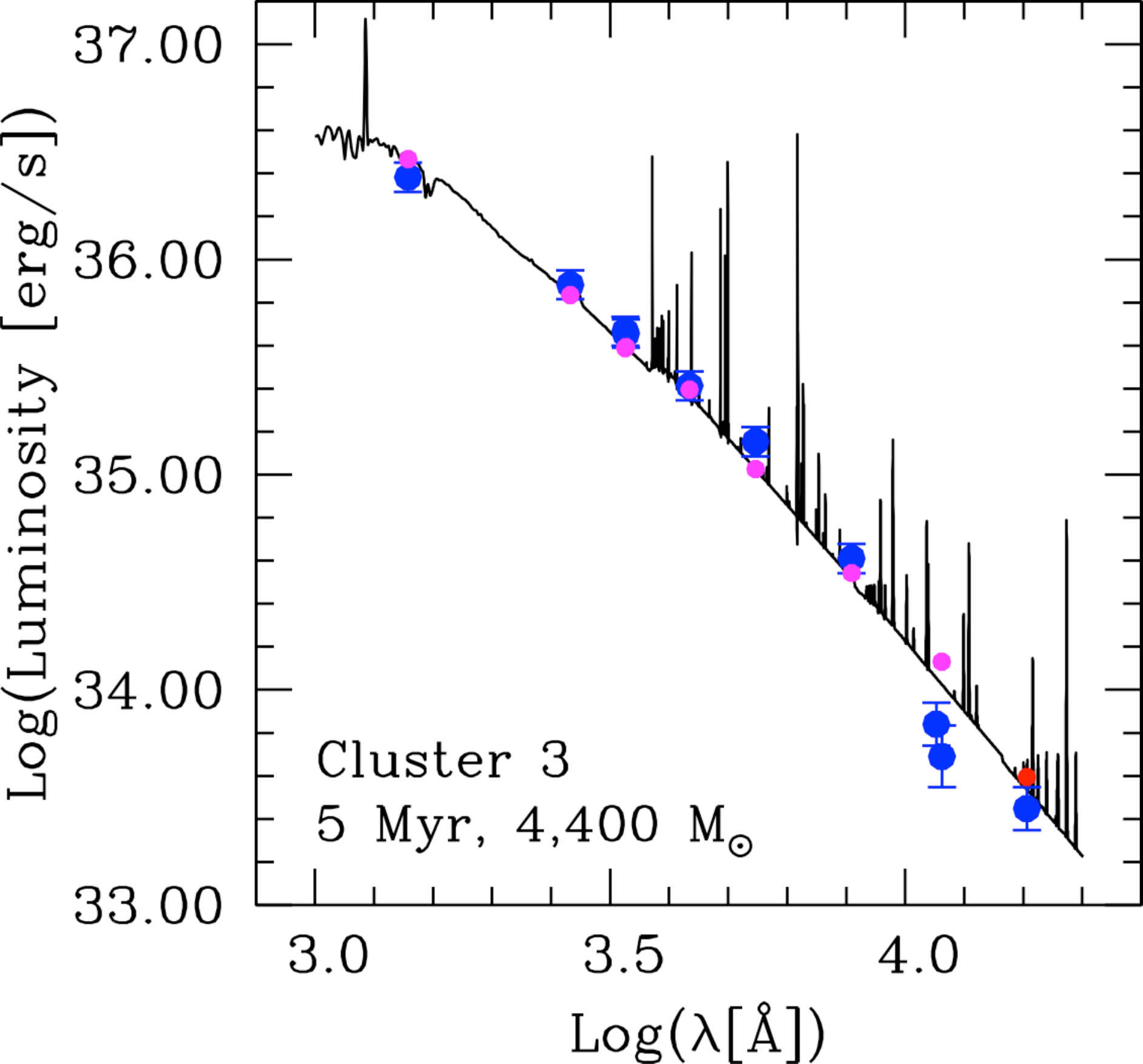}
\plottwo{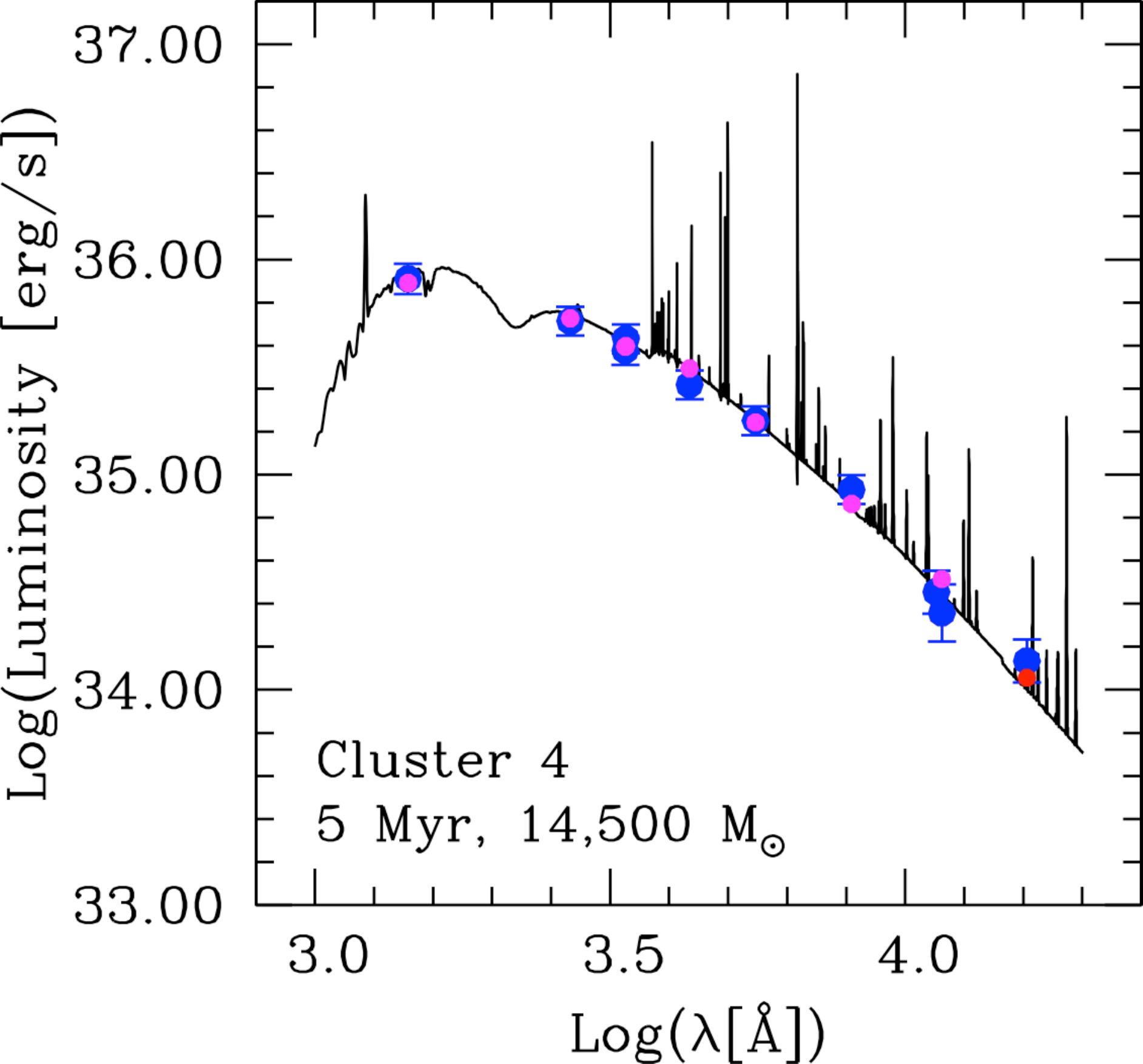}{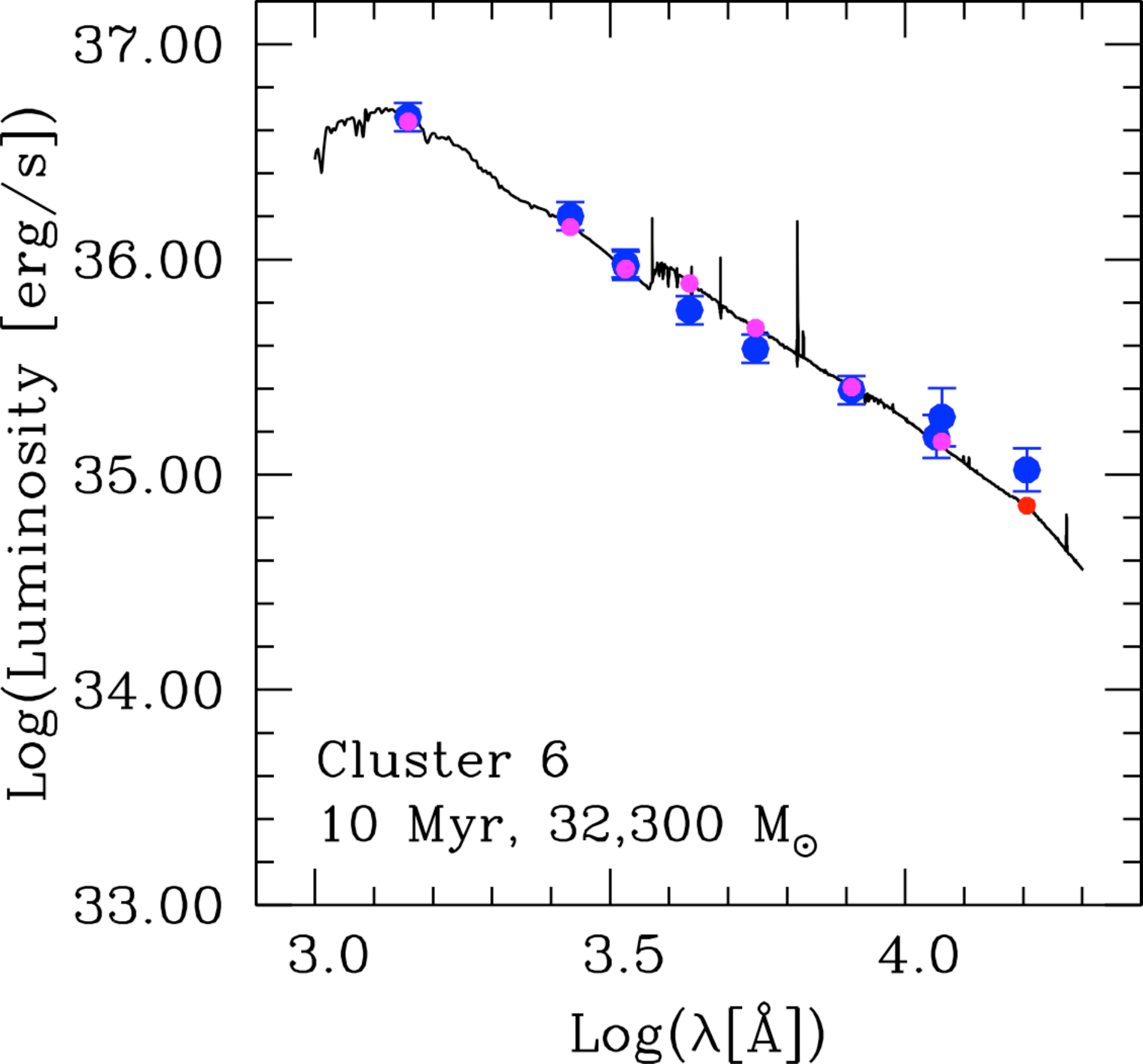}
\caption{Best fit SED plots for (clockwise from top--left) Clusters~2, 3, 4, and 6. The blue points with error bars are the observed photometry,  the magenta points are the synthetic photometry, and the black line 
is the Yggdrasil model, together with dust attenuation and mass normalization, that  provides the best fit (smallest reduced $\chi^2$ value) to the observed photometry. The red point in correspondence 
of the H--band is the predicted synthetic photometry from the best fit model; in all cases, the predicted photometry is within 2~$\sigma$ of the observed photometry, although the H--band data are not used in the fitting. 
\label{fig5}}
\end{figure}

\begin{figure}
\figurenum{6}
\plottwo{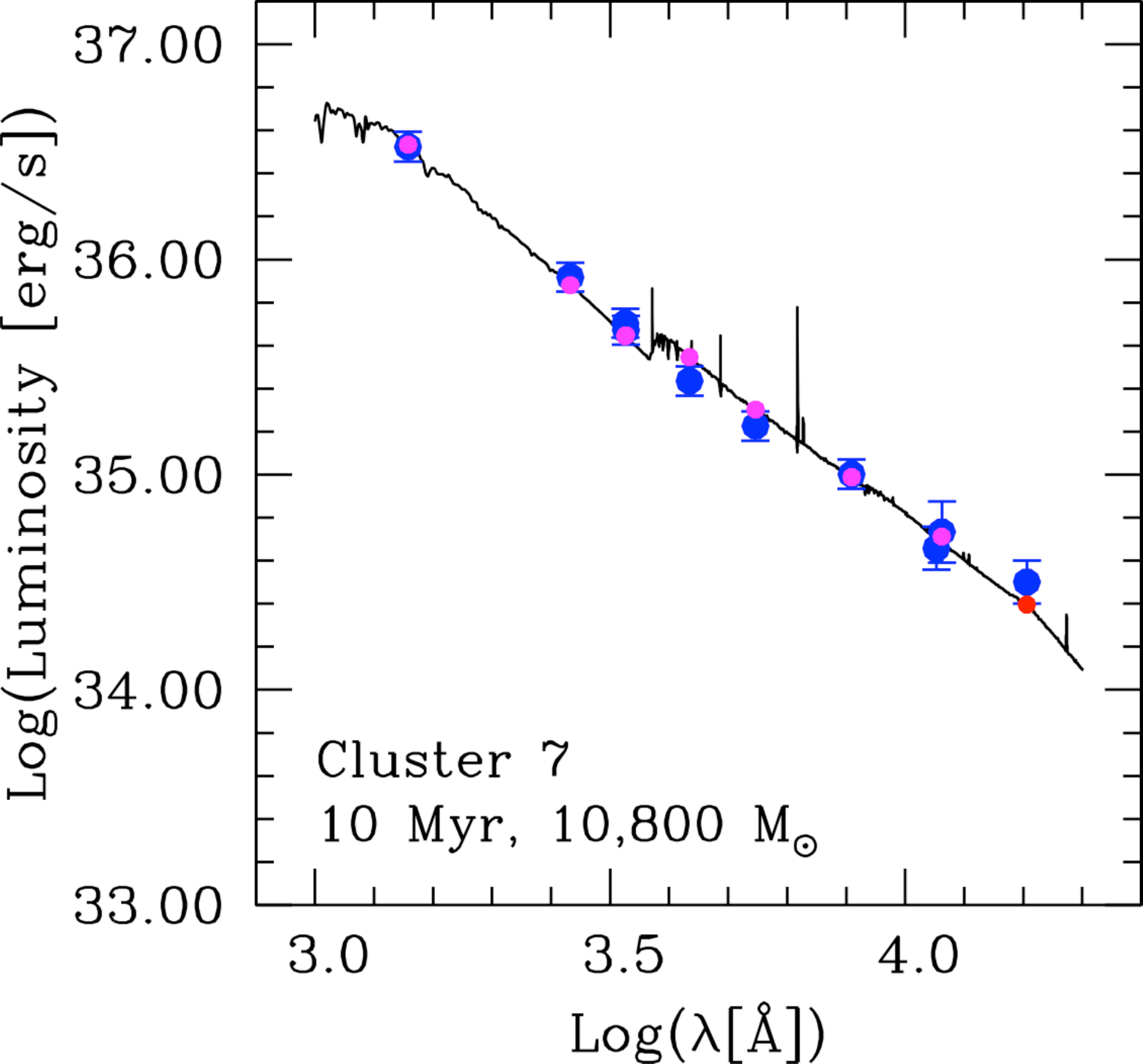}{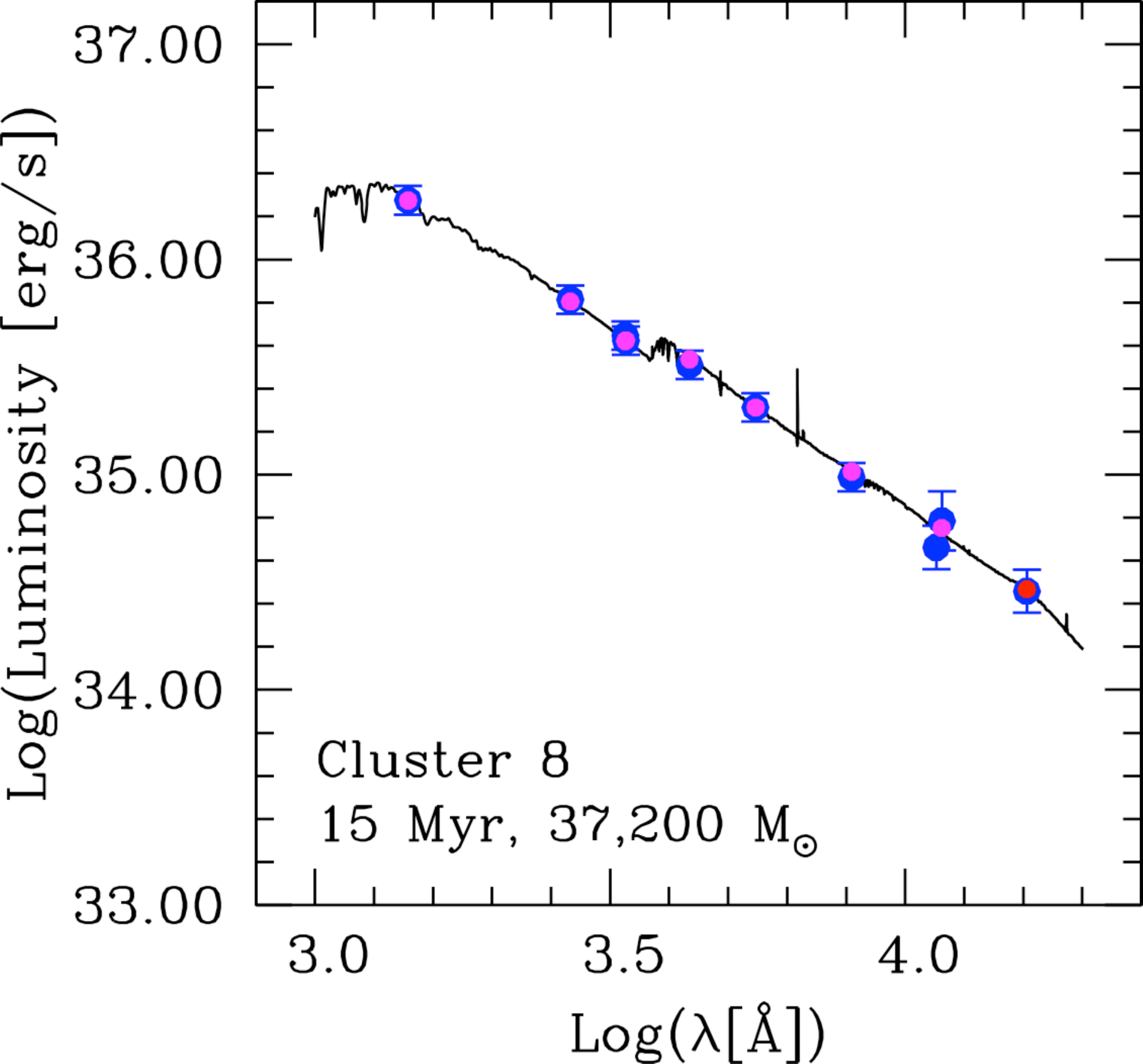}
\plottwo{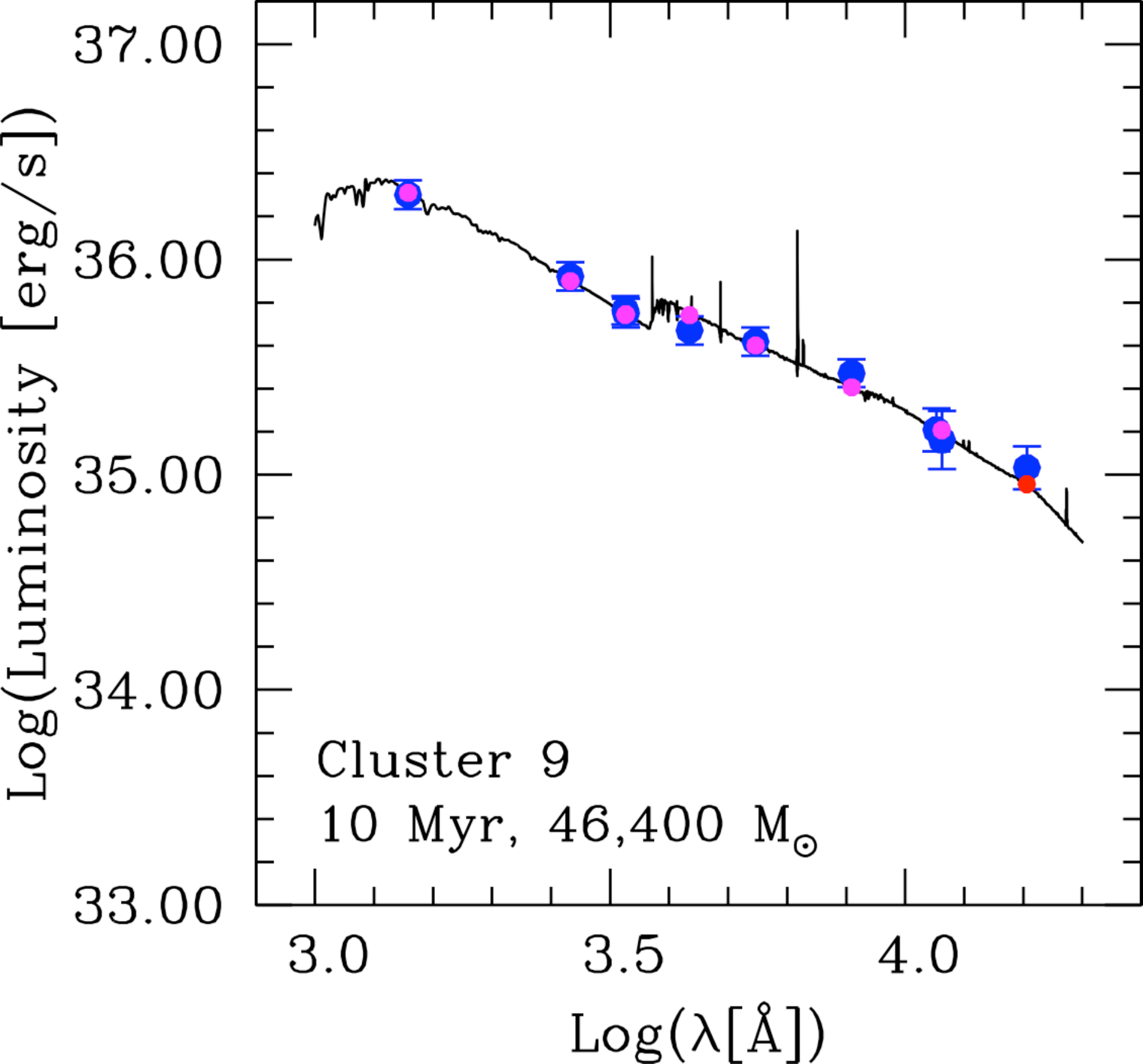}{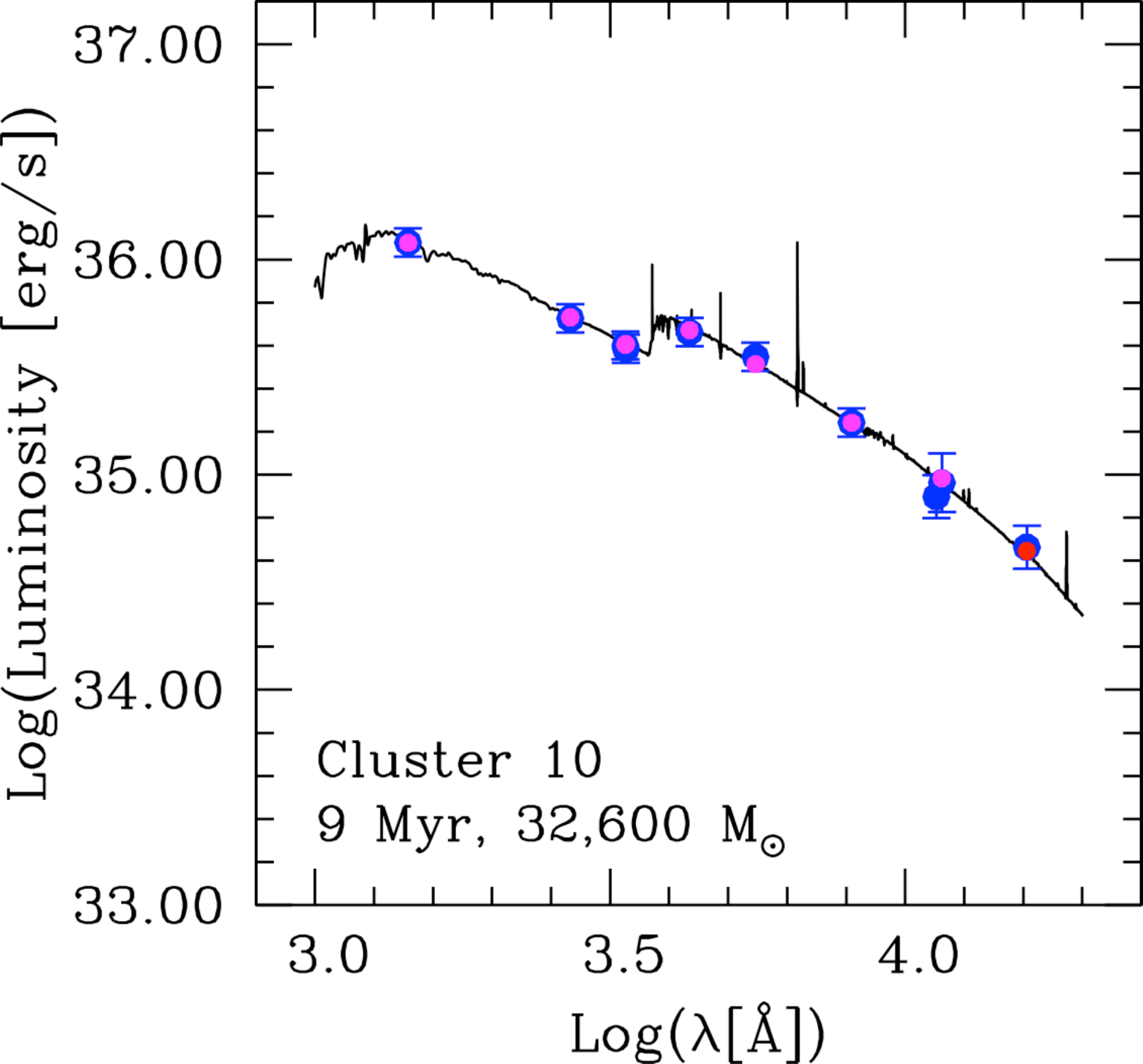}
\caption{As in Figure~\ref{fig5}, but for Clusters~7, 8, 9, and 10.
\label{fig6}}
\end{figure}

The best fit ages, masses, and color excesses, with their 1$\sigma$ uncertainties,  are listed in Table~\ref{tab4} for these clusters\footnote{The masses of all clusters would increase by about 60\% if NGC\,5253 were located at a distance of 4~Mpc, instead of our adopted 3.15~Mpc. Changing the stellar IMF from Kroupa to Salpeter also increases masses by a factor 1.6, for the same 0.1--120~M$_{\odot}$ stellar range.}. For each cluster, we generate separate 
files sorted by  reduced $\chi^2$ values, and listing ages, color excesses, and masses for different combinations of stellar tracks (Geneva, Padova) and extinction/attenuation 
curves (MW, LMC, and SMC, both with and without differential treatment of lines and stellar continuum, and starburst curve). These files are used to determine both the 
best fits and the 99\% confidence histograms, an example of which is given in Figure~\ref{fig4} for cluster~1. The histograms enable us to evaluate the 
uncertainties associated with each parameter, and these are the 1$\sigma$ uncertainties reported in Table~\ref{tab4}, but do not carry information on the best fits (i.e., on which 
of the 14 combinations of stellar tracks and extinction curves provides the best fit to the measured photometry). We infer the best fit values by extracting the model with the 
smallest $\chi^2$ value directly from the files, and the resulting synthetic SEDs and photometry are shown for all nine clusters in Figures~\ref{fig4} (top--left panel), \ref{fig5}, and \ref{fig6}. 

A few common characteristics emerge for all nine clusters from the exercise above. All are better fit by Padova stellar tracks, and, within the limit of validity of 
our foreground dust extinction assumptions, by the differential LMC or by the starburst attenuation curve. In this context, `better' means that the reduced $\chi^2$ is at least 
50\%, and often more than a factor of 2, smaller than for all other solutions. For ages $<$6~Myr, the Padova tracks  cluster around 5~Myr, while the Geneva tracks 
tend to cluster around 3~Myr for the best--fit values. There is also a transition for the best fitting dust extinction/attenuation: younger clusters ($<$6~Myr) 
prefer the differential LMC extinction, while older clusters prefer the starburst attenuation curve, which has the differential 
treatment of lines and stellar continuum `built in'. Thus, differential extinction/attenuation is always required by the best fits solutions, i.e., emission lines are required to 
be more attenuated than the stellar continuum.  In this case, we expect the color excess derived 
from line ratios to be larger than that derived for the stellar continuum from SED fitting. To test this, Table~\ref{tab4} lists side--by--side E(B$-$V) values from SED fitting and 
from emission line ratios (columns~5 and 6, respectively). The two sets of values are generally consistent with each other and, within the large error bars of the line--derived E(B$-$V), we cannot exclude 
that the latter can be larger than the SED--derived E(B$-$V). Indeed, \citet{MonrealIbero2010} finds evidence for differential extinction in NGC\,5253, with the stars being less attenuated than the gas by a factor 0.33. 

The main effect of differential extinction/attenuation between lines and continuum in the SED fits is to reduce the contribution of emission lines to the synthetic photometry in 
the broad/medium band filters, more than what is already accomplished by constructing  models that assume only half of the  ionized gas is in front of the 
clusters. A similar reduction effect can be obtained if the gas covering factor is lower than 0.5; indeed, the aperture correction for the H$\alpha$ line is a factor over 2.5 
larger than  that for the underlying stellar continuum, suggesting a covering factor around 0.4. Furthermore, a decrease in the contribution of 
the metal lines (the major contributors to the broad band filters) can be accomplished by changing the ionization parameter in the CLOUDY models. Thus, 
differential extinction/attenuation should not be considered a unique solution in this case. 

For all clusters, we also show the NICMOS/NIC2/F160W  photometry values predicted by the best--fit SEDs in Figures~\ref{fig4} (top--left panel), \ref{fig5}, and \ref{fig6}. In all nine cases, 
the prediction is within 2~$\sigma$ of the observational value, lending further support to our results. 

As a comparison, we report in Table~\ref{tab4} the best--fit ages and 1$\sigma$ uncertainties as obtained from fitting the photometry from nebular--line--subtracted 
images with Starburst99 population synthesis models. We use a method similar to the one used for the Yggdrasil models to derive ages and uncertainties, with the 
only change that we do not need to apply differential extinction, since the lines are no longer included in the SEDs (the nebular continuum is generally a much smaller 
contribution than the lines). As before, the younger clusters, 1 through 4, are better fit by an LMC extinction curve with Padova$+$AGB tracks, although the Geneva tracks 
give an almost as good best fit in all cases; the Padova and Geneva tracks yield a peak age of 5~Myr and 3~Myr, respectively, which implies differences in the best fit masses 
of roughly a factor of 2 (with the masses from the Geneva tracks being the smaller of the two). Older clusters (6--10) are better fit by Padova stellar tracks with the starburst obscuration 
curve in the nebular--line--subtracted case, as well. The masses and color excesses are also in agreement between the fits performed on the photometry with and without nebular emission 
line: they are well within the 1$\sigma$ uncertainty for E(B$-$V), and are within 70\% of each other for the mass. The main exception is cluster~6, with an estimated mass that is 
a little over two times larger for the nebular--line--subtracted photometry than for the unsubtracted photometry. This cluster shows a more marked tail towards older ages than 
other clusters, which accounts for the discrepancy in the most likely mass value.

The SED--derived ages are listed in Table~\ref{tab4} side--by--side with the ages inferred from the EW(H$\alpha$). The latter age is a short--hand for checking whether 
the ionizing photon rates recovered through the emission line are consistent with those expected from the best--fit SED. For clusters $\sim$10~Myr old and older, i.e., clusters 
6--10, any agreement or discrepancy may be caused by uncertainties, since the H$\alpha$ emission is expected to be low in this age range. It is, however, encouraging 
that, for the most part, there are no major discrepancies between the SED--derived and the EW(H$\alpha$)--derived ages for these clusters. 
For the younger clusters, 1, 2, and 4, the two ages are in good agreement. This result also suggests that ionizing photon leakage outside the HII regions 
and direct dust absorption of ionizing photons are not significant within/around these star clusters. 
For cluster~3, the  SED--derived age is $\sim$5~Myr, but the EW(H$\alpha$) 
suggests values in excess of 7~Myr.  The disagreement between the two age indicators can be ascribed to the discrepancy between the observed and the best--fit photometry in the V and I 
bands (top--left panel of Figure~\ref{fig5}); the stellar continuum used to derive the EW(H$\alpha$) is obtained from the interpolation between these two bands. The observed continuum values are 
higher than the predicted ones, possibly due to untreated uncertainties in the photometry and in the nebular emission, and yield an underestimated EW(H$\alpha$). Within the uncertainties, 
no discrepancy is observed between the predicted  and the measured, attenuation--corrected H$\alpha$ luminosity (see section~5.2).  

Cluster~3 is the one with the smallest mass among the nine analyzed so far, and, at a mean value around 4,500~M$_{\odot}$, may be showing some effects of 
stochastic sampling of the IMF  \citep{Cervino2004,Gazak2013,deGrijs2013}. For instance, stochastic effects may account for the low observed photometry in the nearIR relative to 
expectations from the best fit SED model (Figure~\ref{fig5}): the observed SED is possibly deficient in red supergiants,  which are bright, but rare, stars that become prominent nearIR contributors 
at approximately the age of cluster~3 \citep{Popescu2010,Anders2013}.
The low mass of cluster~3 requires further verification that we are dealing with a bona--fide star cluster, rather than a single, isolated star. From its attenuation--corrected SED, 
we infer that this cluster  contains at least 10 O stars,  thus it is unlikely to be an isolated massive star. 

\begin{figure}
\figurenum{7}
\plottwo{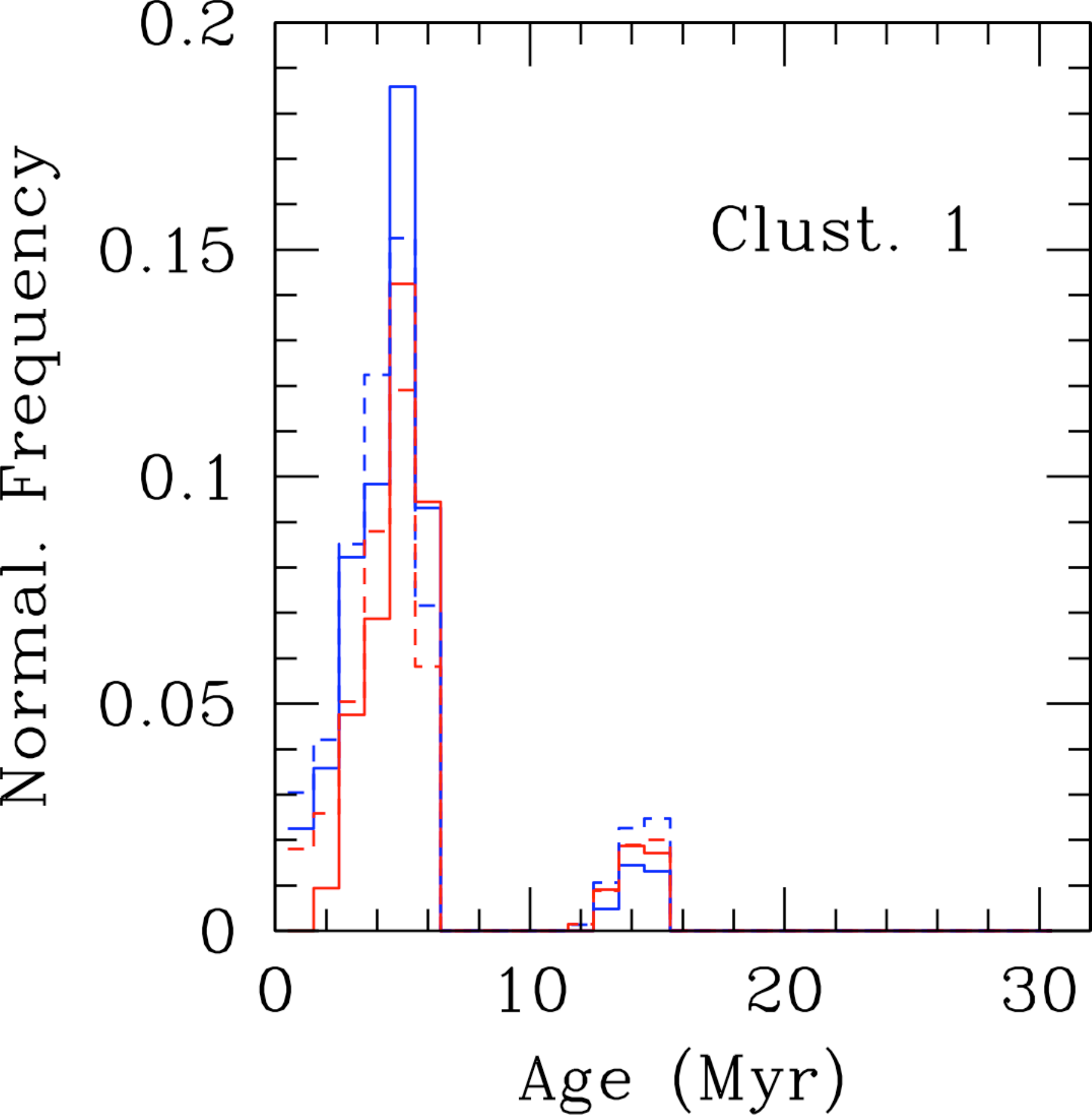}{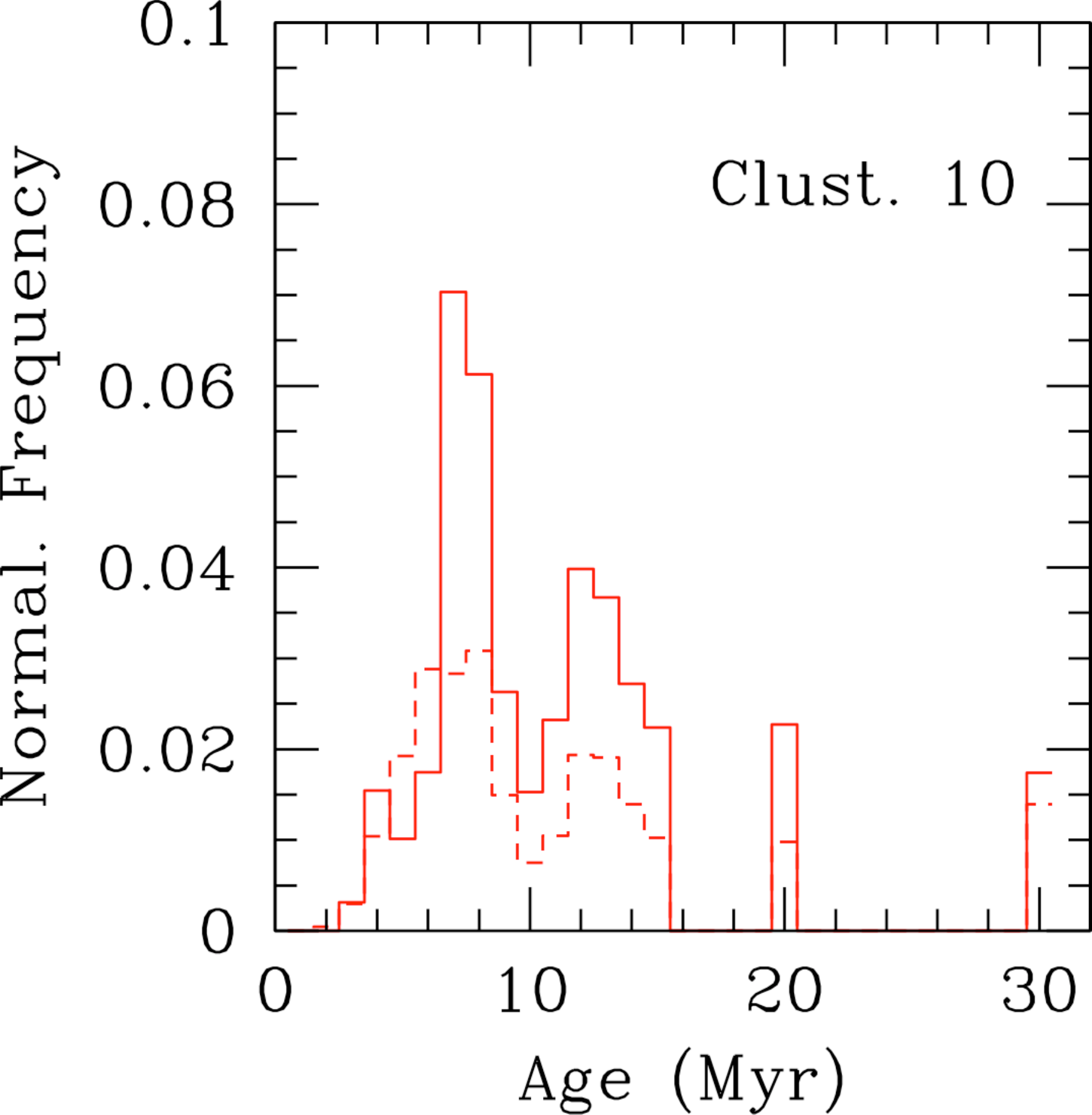}
\caption{Confidence histograms for the distribution of ages for Cluster~1 (left) and Cluster~10 (right) for seven bands SED fits (continuous line) and 5 bands fits (dash line). 
Both a differential LMC extinction curve (blue) and the starburst attenuation curve (red) are used for Cluster~1, while only the starburst curve is used for Cluster~10. For both 
cases, the starburst curve provides consistent results for both seven and five bands fits, also in the case of Cluster~1, which has a better solution with the differential LMC curve. 
\label{fig7}}
\end{figure}

Fitting five bands (NUV, U, B, V, I) recovers physical parameters  that are close to those obtained with the seven bands fits for 7/9 clusters. In all cases, the 
extinction/attenuation model needs to be specified a priori, in order to converge. We run tests using both the starburst curve and the LMC extinction curve with differential 
treatment of lines and continuum for the clusters younger than 6~Myr and the starburst curve only for older star clusters. The goal is to check whether the starburst curve 
can be adopted in all cases. Figure~\ref{fig7} shows the results for Clusters~1 and 10, that are representative of the seven clusters with consistent solutions for seven and 
five bands fits. Cluster~1 displays a preference for young ages independently of the extinction/attenuation model selected (either LMC or starburst), although the LMC 
yields better reduced $\chi^2$ values, by about 50\%. Cluster~10 shows the same double peak, with a range from $\sim$7~Myr to 15~Myr in both cases. 

\begin{figure}
\figurenum{8}
\plottwo{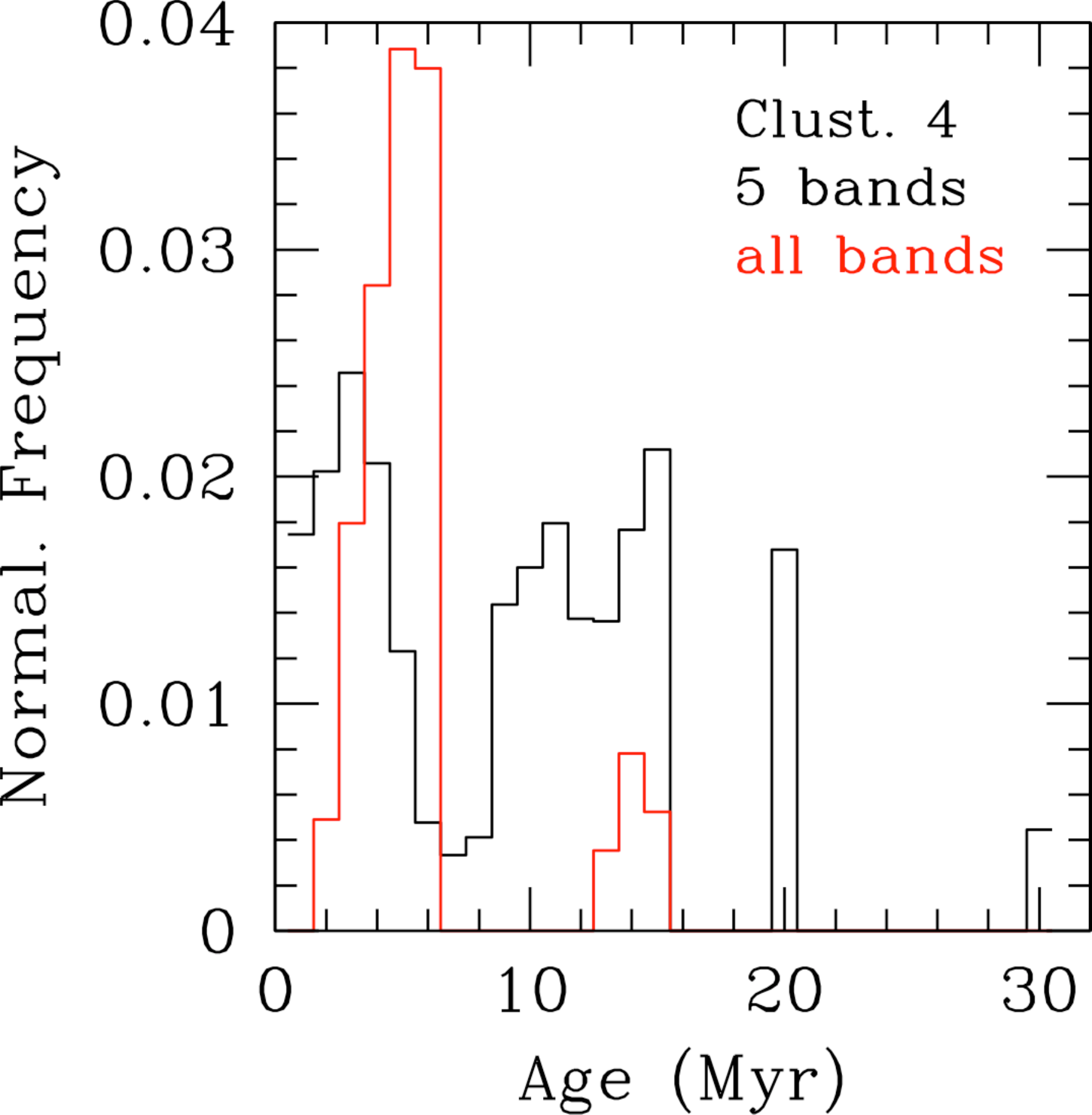}{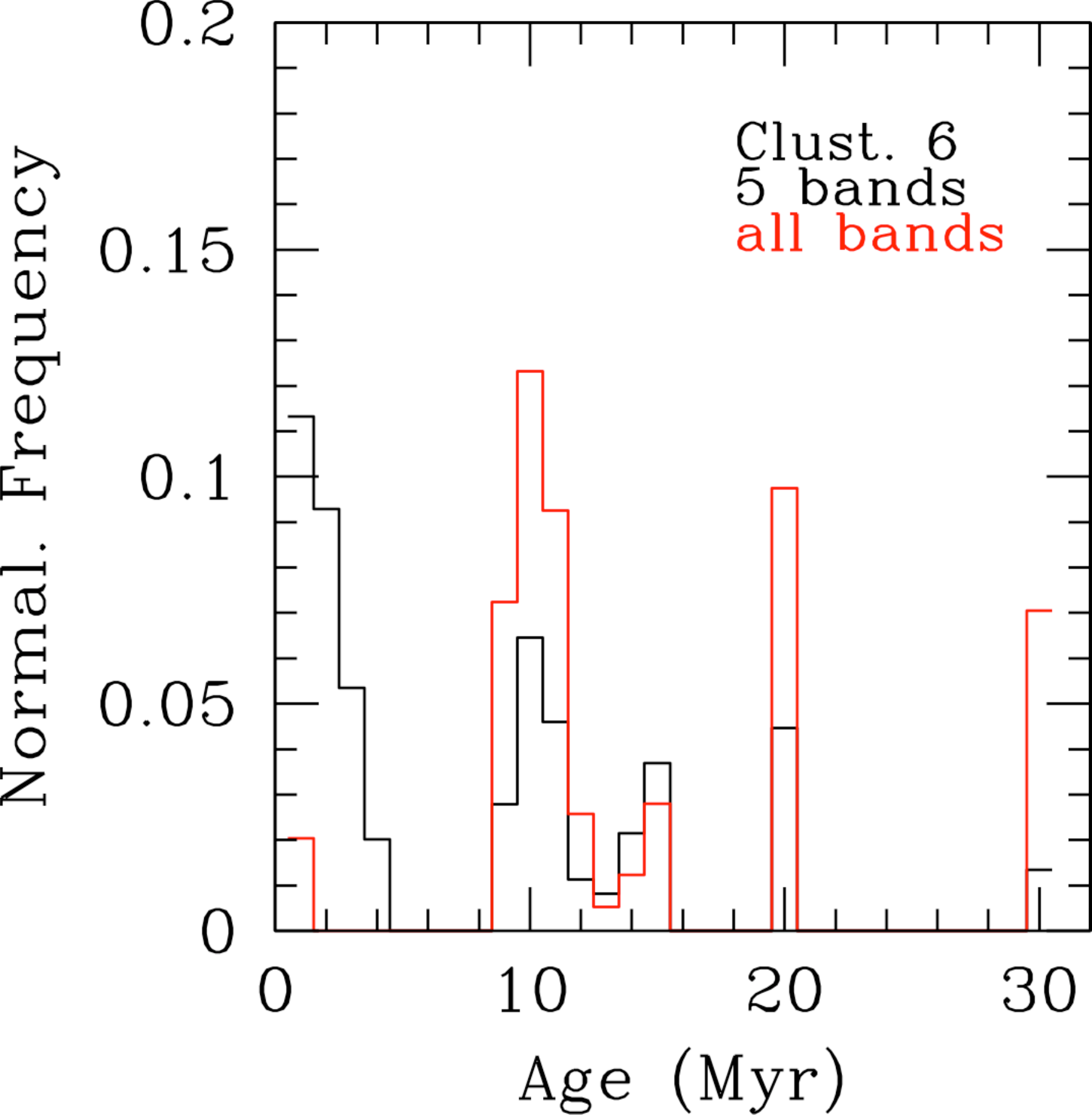}
\plottwo{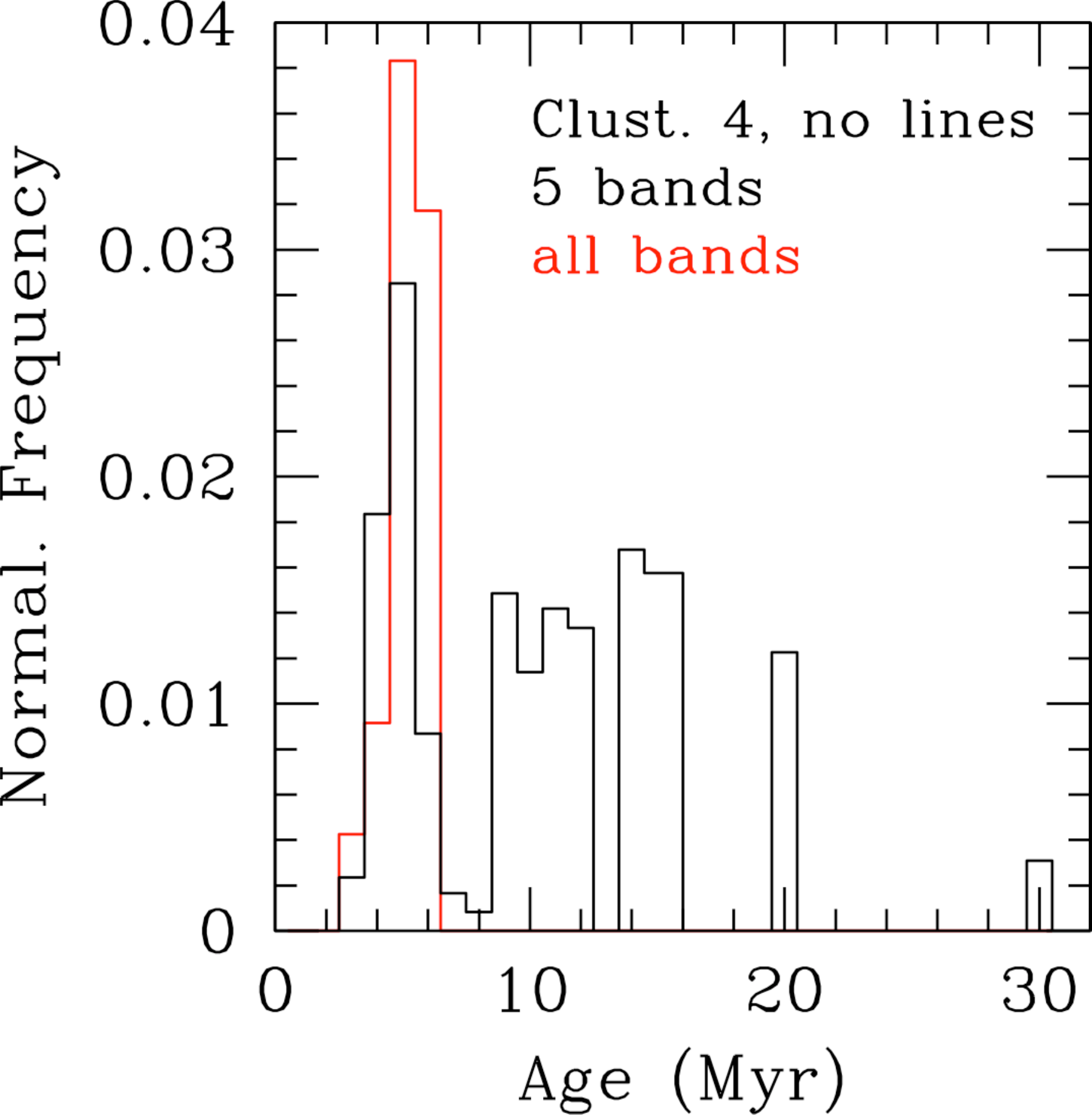}{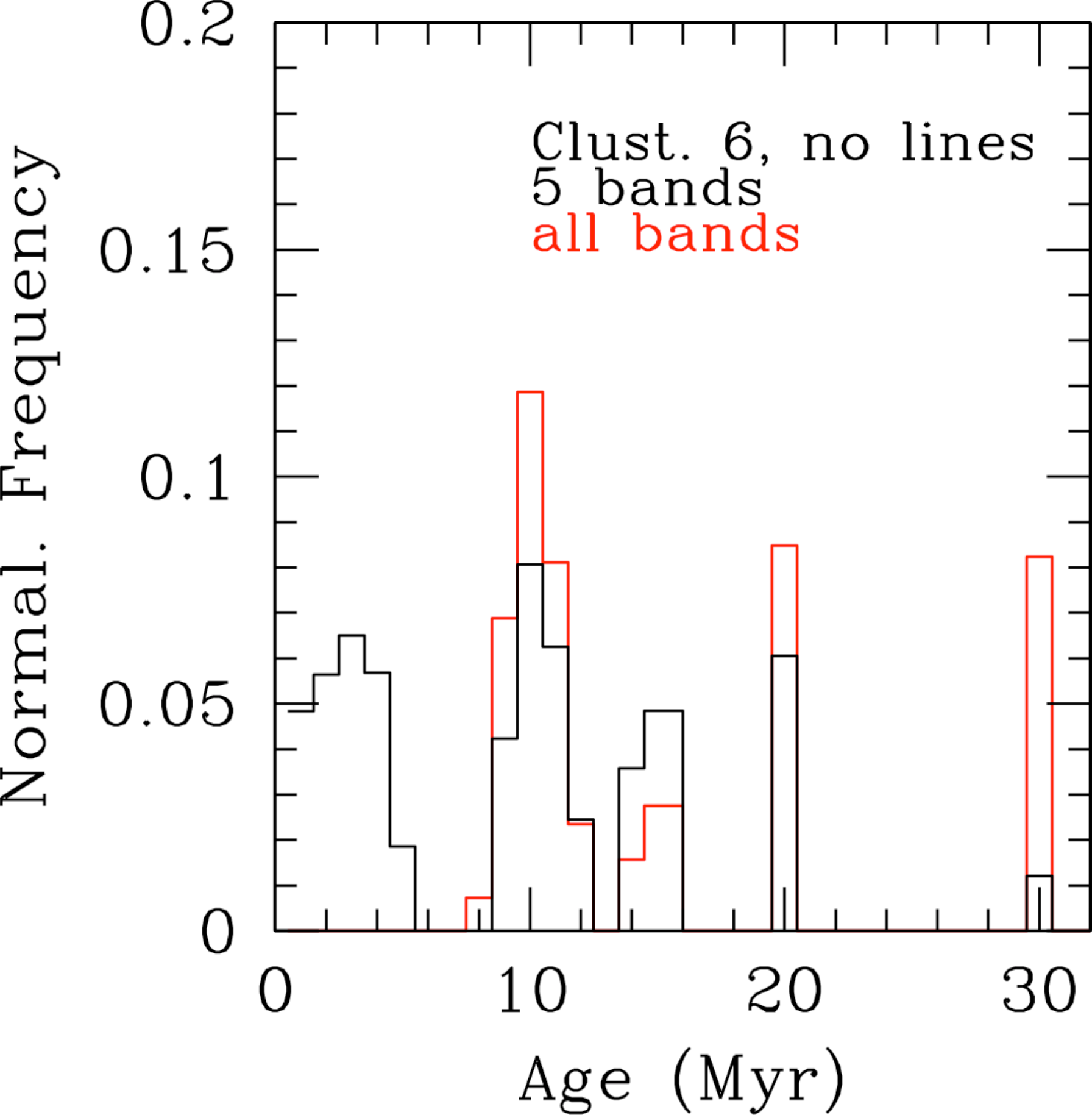}
\caption{Histograms of the 99\% confidence level for the age distributions of Cluster~4 (left--hand--side panels) and Cluster~6 (right--hand--side panels), determined using 
all bands (red histograms) and only the five bands from the NUV to the I (black histograms). The top panels are for photometry that includes emission lines and fits performed 
with Yggdrasil models. The bottom panels are for photometry from which emission lines have been subtracted, and fits are performed with Starburst99 models. The 
agreement between all bands and five bands fits for the most likely age is higher when the emission lines are subtracted from the photometry.
\label{fig8}}
\end{figure}

The remaining two clusters, cluster~4 and cluster~6, yield results that are not as robust as the other seven clusters, when using five band fits. The left panels of Figure~\ref{fig8} 
show the 99\% confidence histograms for cluster~4, for the two cases of seven bands (red) and five bands fits (black). When five bands are used, the young age of cluster~4 is only marginally 
preferred for the Yggdrasil fits (top--left panel of Figure~\ref{fig8}), i.e. when fitting broad/medium band photometry that include emission lines. However, when using Starburst99 to 
fit photometric data from which emission lines have been subtracted, the preferred age for cluster~4 is markedly young, around 4--5~Myr, for both seven band and five band fits 
(bottom--left panel of Figure~\ref{fig8}). Results are more complicated for cluster~6: fitting five bands yields an extremely young age ($\approx$1--2~Myr) when emission 
lines are included in the photometry (top--right panel of Figure~\ref{fig8}), but marginally prefers the older age of $\sim$10~Myr when emission lines are subtracted 
(bottom--right panel of Figure~\ref{fig8}). The older age is preferred by the seven band photometry in both cases. For both cluster~4 and cluster~6, removing emission lines 
from the photometric data yields better agreement in the recovered ages between the restricted and the more extended wavelength ranges. This indicates that 
the method by which emission lines are included in Yggdrasil may not reflect reality in a small fraction of the sources.  Verifying whether the observed effect is more general than 
indicated by our limited analysis will require a larger sample of star clusters.

\subsection{Clusters within the Radio Nebula}

The results from the previous section suggest that the SED fitting yields results that are internally consistent, and can be used to investigate the properties of clusters~5 and 11. 
These clusters are highly attenuated by dust. The hydrogen line ratios reveal significant dust effects, with foreground values A$_V\sim$1.5~mag and 4.7~mag for clusters~5 and 11, respectively. 

\begin{figure}
\figurenum{9}
\plottwo{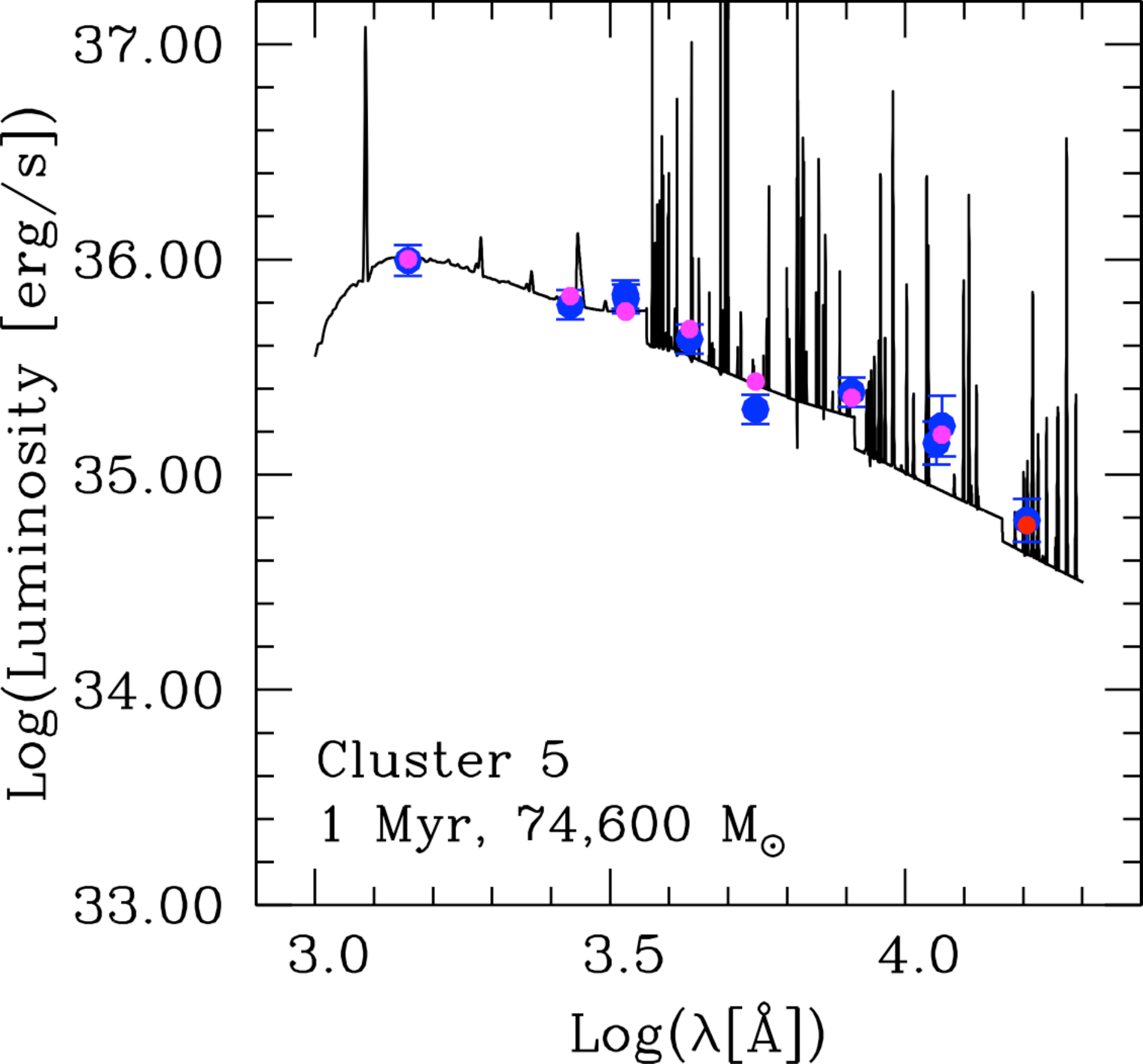}{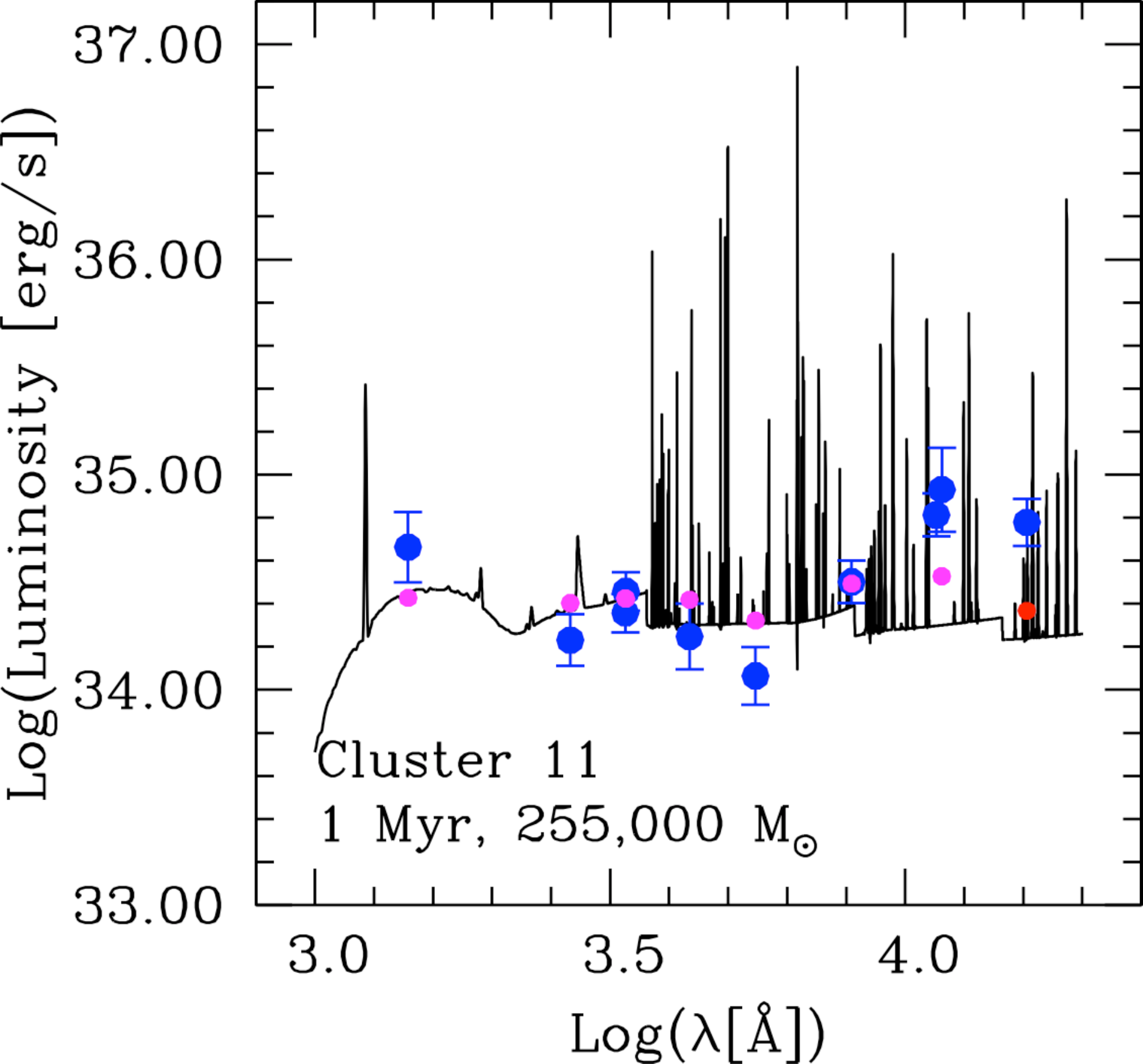}
\caption{Best fit SEDs for Clusters~5 (left panel) and 11 (right panel). As in Figure~\ref{fig2}, the blue points with error bars are the observed photometry,  the magenta points 
are the synthetic photometry, and the black line is the Yggdrasil model, together with dust attenuation and mass normalization, that  provides the best fit (smallest reduced $\chi^2$ value) to 
the observed photometry. For Cluster~5, only foreground dust is required to achieve a reduced $\chi^2\sim$1. For Cluster~11, a combination of both mixed dust, 
with a total dust column of A$_V\sim$48.7~mag, and foreground dust, with A$_V\sim$1.9~mag, are needed to approximate the observed SED. 
\label{fig9}}
\end{figure}

The results of SED fitting for these two clusters are listed in Table~\ref{tab5}, and shown in Figure~\ref{fig9}. Both clusters are extremely young, with a best fit age of 1~Myr, 
which agrees with the age inferred from the EW(H$\alpha$). Both are better represented by Geneva stellar tracks, and by the starburst 
attenuation curve for the foreground dust. As for the other clusters, we have verified that the best fit solution also reproduces the intensity and ratio of the line emission at H$\alpha$, P$\beta$, 
and P$\alpha$. This is shown in Table~\ref{tab6}, where we list, for all clusters, the intrinsic H$\alpha$ luminosity, obtained from the attenuation--corrected measurements of Table~\ref{tab3}, and the 
SED--predicted H$\alpha$ luminosities, obtained from the models and the best--fit ages and masses of Tables~\ref{tab4} and \ref{tab5}. Within the 1~$\sigma$ uncertainties, there is general agreement between the two sets of values.  Major disagreements are discussed below (cluster~5) and in section~6.1 (cluster~9).

For cluster~5 the use of a simple dust geometry, i.e., foreground dust, is sufficient to recover an excellent fit for the SED, with a reduced $\chi^2\sim$1. The color excess derived from 
the SED fitting agrees with the one derived from both line ratios: H$\alpha$/P$\beta$ and H$\alpha$/P$\alpha$. The emission lines are strong enough that, although the formal 
uncertainties are significant (Table~\ref{tab5}), both ratios yield similar extinction values. Despite the goodness of fit, not all observational data points match the model SED: a significant deviation 
($\sim$2$\sigma$) is evident for the V--band (F550M), which may be due to difficulties in measuring the background surrounding the cluster. The observed photometry is below the 
model's value, which explains the unphysically large EW(H$\alpha$) (Table~\ref{tab3}). 

The mass of cluster~5 derived from the SED fit is $\sim$7.5$\times$10$^4$~M$_{\odot}$, confirming that this cluster 
is massive enough to drive the observed ionization. The H$\alpha$ luminosity predicted for cluster~5 is L(H$\alpha$)$\sim$5.7$\times$10$^{39}$~erg~s$^{-1}$, to be compared with the 
attenuation--corrected measured value L(H$\alpha$)=2.8$_{-0.7}^{+1.1}$ $\times$10$^{39}$~erg~s$^{-1}$ (Table~\ref{tab6}) and with the H$\alpha$ luminosity inferred from the free--free measurement of 
\citet{Turner2000}, L(H$\alpha$)$\approx$2--3$\times$10$^{39}$~erg~s$^{-1}$ \citep[the free--free measure likely provides an underestimate of the ionizing photon flux, since 
it is affected by self--absorption, see][]{Meier2002}. The discrepancy between the SED--predicted and the observed, attenuation--corrected H$\alpha$ luminosity can be interpreted as due 
to  either leakage of ionizing photons outside of the HII region or direct 
absorption of ionizing photons by dust. Leakage is supported by the presence of extended ionization around cluster~5, which suggests that ionizing photons 
from this cluster reach further out than what we recover within our photometric apertures. We  
infer that 25\%--50\% of the ionizing photons produced by this star cluster leak out of the region. Direct 
absorption of ionizing photons by dust is also a potential mechanism  in the dense environment surrounding cluster~5. However, we provide arguments both below and in 
section~6.6 against a significant contribution from dust absorption of ionizing photons in this galaxy. 

Cluster~11 requires a more complex approach to dust attenuation, in order to approximate the observed SED. A simple foreground dust model is unable to provide a 
reasonable fit to the observed SED, on account of the SED being mostly flat in L($\lambda$)--versus--$\lambda$. A combination 
of a homogeneous dust--star mixture and a foreground dust screen provide a better, 
albeit not perfect, fit (Figure~\ref{fig9}). Flat or slightly blue SEDs in the presence of significant amounts of dust tend to require models in which the dust is mixed with the stars. 
Foreground or shell models, either homogeneous or clumpy, will generally tend to produce too red SEDs relative to the data  at blue wavelengths and too blue SEDs at red wavelengths \citep{Calzetti2001}. 

The major deviations between the data of cluster~11 and the model are in the V band and in the J and H bands. Like in cluster~5, the V-band observations 
are about 2$\sigma$ below the model, possibly an effect of background placement for the measurements. As in the other cluster, this accounts for the unphysically large 
EW(H$\alpha$) measured for cluster~11 (Table~\ref{tab3}). The other major deviation, in the nearIR bands, is in the opposite direction: the data are more luminous, by slightly more 
than 2$\sigma$, than the model expectation. A discussion of this deviation is deferred to section~6.3. 

Taken at face value, cluster~11 is the most massive among those analyzed here, with a mass  
M$\sim$2.5$\times$10$^5$~M$_{\odot}$. The component of dust that is mixed with the stars in the cluster has a front--to--back optical depth of A$_V\sim$49~mag and follows 
the  Milky Way extinction curve; this curve yields a reduced $\chi^2$ that is at least two times better than any other of the extinction curves tested in this paper. The foreground dust 
component obeys the starburst attenuation curve, with A$_V\sim$1.9~mag. Cluster~11 is therefore associated with about 50~magnitudes of optical depth in dust, although 
the mixed geometry allows for some of the UV--optical radiation to shine through. This optical depth is larger than what  was determined by other authors. \citet{AlonsoHerrero2004} 
derive A$_V\sim$17~mag for this cluster, but they only assume a foreground dust screen, which provides a lower limit to the actual dust optical depth of a region. \citet{MartinHernandez2005} 
derive a value of A$_V\sim$14--17~mag, from the 9.7~$\mu$m silicate absorption feature; however, the relatively large slit they use for their spectroscopic observations 
(1$^{\prime\prime}$.2 $\sim$ 18~pc) is likely to have sampled regions of different optical depth, thus weighting the final result towards less extreme values of A$_V$. \citet{Calzetti1997}
derive a larger range, A$_V\sim$9--35~mag, the upper bound of which is more consistent with the value derived here. 

From the best--fit model SED, the predicted H$\alpha$ luminosity of cluster~11 is L(H$\alpha$)$\sim$2$\times$10$^{40}$~erg~s$^{-1}$, which is comparable to the measured, attenuation--corrected  
 L(H$\alpha$)$\sim$1.8$_{-0.5}^{+0.7}$ $\times$10$^{40}$~erg~s$^{-1}$ (Table~\ref{tab6}). As a reminder,  the full dust model of combined 
mixed and foreground dust is applied to all three emission lines, H$\alpha$, P$\beta$, and P$\alpha$,  to derive internally--consistent values for their luminosities. The free--free emission within 4~pc of the radio peak  corresponds to an intrinsic H$\alpha$ luminosity  L(H$\alpha$)$\sim$1.6--2.0$\times$10$^{40}$~erg~s$^{-1}$ \citep[from the 7~mm measurement of][]{Turner2004}. 
The agreement between the measured, attenuation--corrected H$\alpha$ luminosity and the SED--predicted luminosity provides independent confirmation of the accuracy of our 
results for cluster~11. Within the uncertainty of the 
measurements, the mass and age of cluster~11 are accurate, as are the  dust content and  geometry;  there are no major parts of cluster~11 that are so deeply buried in dust to be unrecoverable
 by our approach. It also suggests that, within the uncertainties,  direct absorption of ionizing photons by dust does not appear to be a dominant mechanism at work in the area where 
 cluster~11 resides, despite the large dust optical depth. 
 
 Because of the red SED of cluster~11, a secondary best--fit solution is provided by a $\sim$100~Myr old star cluster, with a mass M$\sim$3.5$\times$10$^5$~M$_{\odot}$, also mixed with 
dust with optical depth A$_V\sim$42~mag. We reject this solution on account of the strong nebular line emission in coincidence of the location of cluster~11.  However, the uncertainties 
in both line emission and stellar continuum measurements allow the co--existence of two populations at the location of cluster~11: a $\sim$2$\times$10$^5$~M$_{\odot}$, 1~Myr old population 
together with a $\sim$1$\times$10$^5$~M$_{\odot}$, 100~Myr old population. This solution would push against the tolerance of our error bars, but is not formally excluded by the data.

Other sources of uncertainty for the best--fit solutions of cluster~11 include the requirement that emission line intensities and ratios be consistent with the model for the stellar continuum. 
If we allow the stellar continuum to be modeled independently of the emission lines, we obtain a larger range of degeneracies as a consequence of the larger number of degrees of freedom. 
For instance, in the absence of constraints from the emission lines, the dust--star mixed model produces a linear correlation between mass and front--to--back optical depth for 
A$_V>$15~mag: by doubling the total attenuation, one can double the cluster mass recovered with virtually unchanged $\chi^2$ values. 

\subsection{Limitations in the Population Synthesis Models}

There are several properties of massive stars that are not currently included in the population synthesis models used in the present analysis. These are: 
stellar  rotation, binary stars, and Very Massive Stars (VMSs). Most of these properties are still under investigation, and their inclusion in 
models is either in the early stages (e.g., rotation and binaries)  or non existent. We provide here a brief summary of the effects we expect each to have on our results. 

A stellar population containing rotating massive stars will appear younger than a non--rotating counterpart, with an increase in the ionizing flux that can 
be as much as a factor two--five higher \citep{Leitherer2014}. Applied to the star clusters in our sample, such models are expected to yield older 
best--fit ages than what we derive or, alternatively, larger ionizing fluxes at fixed age.

Young massive stars are predominantly found in close binaries, and can interact \citep[e.g.,][]{Sana2012}. The products of such interactions include massive blue stragglers formed 
through mass transfer and mergers, which may be common \citep{deMink2014}, and appear at the upper tail end of the stellar IMF \citep{Schneider2014,Schneider2015}. The 
effects on young star clusters include apparent rejuvenation and age spreads \citep{EldridgeStanway2009}. Our 
data do not include enough information to infer whether either of those effects may be present, but we cannot exclude them either.

The birth mass of stars is limited to $\sim$120--150~M$_{\odot}$ in virtually all population synthesis models, but there is mounting evidence that stars as 
massive as 300~M$_{\odot}$ are present in nearby young star clusters \citep{Crowther2010}. There is still debate on whether these VMSs are 
the result of birth conditions or of mergers, and stellar tracks are being produced in order to further investigate this issue \citep{Yusof2013, Koehler2015}. 
The youngest among our clusters, clusters~5 and 11, are also sufficiently massive that they may contain VMSs.  Presence 
of VMSs increases the ionizing photon flux from a cluster. If the VMSs contributions to clusters~5 and 11 were similar to those found in the LMC cluster 
R136, the ionizing photon flux could be 50\%--100\% larger than the one currently predicted \citep{Crowther2010, Doran2013}. 

One final limitation of our model fitting is that the youngest synthetic population we consider is 1~Myr old, in line with our photometric uncertainties which 
yield a best accuracy of $\sim$1~Myr at those young ages. The formal solutions for  clusters~5 and 11 do not exclude that they could be younger than 1~Myr. 

\section{Discussion}

\subsection{Age Comparisons with the Literature}

The physical parameters of age and mass have been derived before by several authors for most of the star clusters in this study 
\citep{Calzetti1997,Tremonti2001,Harris2004,deGrijs2013}. We compare our derived ages with those previous derivations in Table~\ref{tab7}. We do not compare 
masses, as these are somewhat degenerate with ages and depend on additional assumptions such as the stellar IMF and the adopted distance for NGC\,5253. There 
is general agreement between the ages derived by all authors, with a few exceptions. 

The most notable discrepancies are present for cluster~9, with age estimates that range from 3~Myr \citep{Tremonti2001} to 30--50~Myr \citep{Calzetti1997}. 
The old age derived by \citet{Calzetti1997} is an effect of the limited number of broad bands available to those authors, since they only used 
NUV (centered at $\sim$2250~\AA), V, and I; in particular, the absence of a filter close to the age--sensitive U--band limits the ability to attribute ages to star clusters 
\citep[e.g.,][]{Lee2005}. The younger age derived by \citet{Tremonti2001} is far more puzzling. These authors use UV spectroscopy for deriving cluster ages, leveraging 
the information from the photospheric lines, while all other authors derive their age estimates from colors or SED fitting. The UV spectrum of cluster~9 shows the 
presence of P-Cygni profiles for the lines of [NV]($\lambda$1240~\AA) and [CIV]($\lambda$1550~\AA), a clear sign for the presence of early O--stars and 
ages $\lesssim$5~Myr. However, the long slit of the HST/STIS only `grazes' the outskirts of cluster~9, with only 1/30th of the light from this cluster captured by the UV 
spectrum \citep{Tremonti2001}. Possibly, the UV spectrum is targeting a smaller cluster in the periphery of cluster~9. The presence of one or more interlopers is 
supported by the anomalously large (for its age and location) color excess from emission lines for cluster~9 (Table~\ref{tab3}); these large values for E(B$-$V) are 
usually found in correspondence of much younger star clusters. Despite the large attenuation value, the emission lines in correspondence of the cluster are still 
weak for its mass (Table~\ref{tab6}), and there is no evidence for additional ionized gas in the form of shells or arcs in the region; this further supports the older age, $\approx$10~Myr, for cluster~9. 

For cluster~10, the main discrepancy is given by the age reported by \citet{Calzetti1997}, 50--60~Myr, while our determination and those from other authors 
suggest an age around 8--15~Myr. As in cluster~9, the discrepancy is due to the limited number of photometric bands available to \citet{Calzetti1997}. An early 
suggestion that clusters~9 and 10 could be younger than the 30--60~Myr age inferred by \citet{Calzetti1997} came from \citet{Strickland1999}, based on the measured 
X--ray sizes and luminosities of the super bubbles in NGC\,5253. 

The young ages of the dusty clusters~5 and ~11  have been known for quite some time \citep[e.g,][]{Meurer1995,Schaerer1997, Calzetti1997}, but the realization that the region hosts  
two separate clusters instead of a single one is more recent \citep[e.g.][]{AlonsoHerrero2004}. \citet{AlonsoHerrero2004}, specifically, favors the 
presence of two star clusters over other interpretations, such as that cluster~5 may be a reflection nebula generated by cluster~11 \citep{Turner2004}. Our analysis also favors the interpretation of 
two separate star clusters. The morphology of cluster~5 is consistent with that of a compact star cluster: a 
slightly--resolved ($\sim$0.1$^{\prime\prime}$ corresponding to a deconvolved size of $\sim$1.2~pc), symmetric, and centrally concentrated source, similar to other clusters 
in the area, including those in the same $\sim$1$^{\prime\prime}$ region as cluster~5.  At least an additional 6--7 fainter 
star cluster candidates are visible in the $\sim$0.6$^{\prime\prime}$ area surrounding cluster~5, suggesting that this one is the most prominent one in an association of stars 
and/or star clusters. 

Both clusters have been characterized as having ages 
$\sim$3--3.4~Myr by \citet{AlonsoHerrero2004}, based on HST nearIR imaging and ground--based spectroscopy; the  inferred masses are $\sim$6.6$\times$10$^4$~M$_{\odot}$
for cluster~5  and $\sim$3.9$\times$10$^5$--1.3$\times$10$^6$~M$_{\odot}$ for cluster~11, when reported to our adopted distance and stellar IMF. For cluster~5, the difference between  
our mass and the mass derived by \citet{AlonsoHerrero2004} can be entirely attributed to differences in the derived ages and extinctions. For cluster~11, a major component of 
the discrepancy is the fraction of hot dust contributing to the K--band: \citet{AlonsoHerrero2004} adopt a fraction ranging from 0\% to 70\%, but we derive a fraction closer to 
90\% from our best fit (see section~6.3). The differences in age and dust column density and geometry account for the remaining portion of the discrepancy. Even with these 
discrepancies, the masses we derive for both clusters and those of \citet{AlonsoHerrero2004} agree to better than 60\%, when using their more 
conservative estimate for cluster~11. 

The area surrounding clusters~5 and 11 hosts a half--dozen Wolf--Rayet stars, mainly of the younger WN type. This 
would suggest presence of stars/clusters in the narrow age 
range $\sim$2.5--3.5~Myr \citep{Schaerer1997, MonrealIbero2010, Westmoquette2013}. This is not necessarily in contradiction with the $\sim$1~Myr age we derive 
for clusters~5 and 11. The relatively low spatial resolution of ground--based spectroscopy, $\sim$1$^{\prime\prime}$.5--2$^{\prime\prime}$ \citep[20--30~pc, or about the 
size of the panels in Figure~\ref{fig2}; see a summary description in ][]{Westmoquette2013}, has not enabled accurate location of the stars or clusters responsible for 
the Wolf--Rayet emission features. This suggests two possible scenarios. In one scenario, the WN stars co--exist in the region with clusters~5 and 11, and these clusters 
are the most recent `products' of ongoing star formation over the past few Myr. In the second scenario, the WN features may  originate from  
VMSs \citep{Crowther2010}, hosted in the extremely young clusters~5 and 11.  

\citet{Whitmore2011} suggest a method based on the morphology of the H$\alpha$ emission in order to classify star clusters according to their ages: the gas morphology 
is compact and coincident with the star cluster for ages $<$a few Myr; has a small ring--like structure in clusters up to $\sim$5~Myr of age; has a large, well--formed ring surrounding the cluster for ages in the range 5--10~Myr; and is virtually absent in clusters older than 10~Myr. \citet{Whitmore2014a} adds earlier stages to this classification by including proto--cluster phases based on CO appearance. When evaluated according to the criteria of these papers, our clusters form a well defined sequence, with a close agreement between our SED--fitting ages and the  morphological ages, and with clearly identifiable stages from 2 for cluster~11 (embedded cluster) to 5 for clusters 6--to--10 (intermediate/old cluster). 

\subsection{Global Properties of the Clusters}

The total H$\alpha$  luminosity of NGC5253, corrected for foreground Milky Way extinction and [NII] contribution, but {\em uncorrected for internal dust attenuation}, 
is L(H$\alpha$)$_{total}=$2.2$\times$10$^{40}$~erg~s$^{-1}$ 
\citep{Kennicutt2008}. About 15\% of the H$\alpha$ is associated with shock--ionization \citep{Hong2013}, implying that the photo ionized H$\alpha$ luminosity is 
L(H$\alpha$)$_{phot}$=1.9$\times$10$^{40}$~erg~s$^{-1}$. The sum of the observed H$\alpha$ luminosity from the star clusters analyzed in this work, also uncorrected 
for internal dust attenuation, is 
L(H$\alpha$)$_{clusters}=$1.43$\times$10$^{39}$~erg~s$^{-1}$ (from Table~\ref{tab3}), or about 7.5\% of the total H$\alpha$. Thus, the brightest 10 clusters contribute 
almost 10\% of the total (observed) H$\alpha$ luminosity in this galaxy; indeed, although there are almost 150 young star clusters in the central region of NGC\,5253, the vast 
majority tend to have low mass, i.e., $\lesssim$10$^4$~M$_{\odot}$ \citep{deGrijs2013}. 

A similar argument can be made for the FUV luminosity. The total luminosity density contained within the ACS/SBC frame at $\sim$1,500~\AA, corrected for foreground Milky Way extinction, 
but uncorrected for internal dust attenuation, 
is L(1500\AA)$\sim$5.8$\times$10$^{38}$~erg~s$^{-1}$~\AA$^{-1}$. The GALEX FUV luminosity density of NGC\,5253 
is L(1500\AA)$_{total}\sim$6.6$\times$10$^{38}$~erg~s$^{-1}$~\AA$^{-1}$, i.e., only about 15\% larger than the amount contained in the HST image. We assume this 15\% discrepancy to be 
the upper limit to the amount of FUV light outside of the ACS/SBC frame, since both the SBC/F125LP filter and the GALEX/FUV filter have pivot wavelengths at $\sim$1,500\AA. 
The total contribution from the 11 clusters investigated here is L(1500\AA)$_{clusters}\sim$2.7$\times$10$^{37}$~erg~s$^{-1}$~\AA$^{-1}$, or about 4\%--5\% of the total. Star clusters 
contribute about 20\% of the total observed UV light in nearby star--forming and starburst galaxies \citep{Meurer1995,Maoz1996}. Thus, the 11 bright clusters  represents about 20\%--25\% of this contribution. 

The patchy extinction in the center of NGC\,5253 makes it difficult to convert the above numbers to intrinsic luminosities. We adopt a hybrid approach, by combining the ionizing photon 
flux measured from the free--free emission within the radio nebula \citep{Meier2002,Turner2004} with 
the virtually extinction--free H$\alpha$ emission outside the radio nebula \citep{Calzetti2004}. The two combined yield a total, attenuation--corrected H$\alpha$ luminosity L(H$\alpha$)$_{phot,corrected}\sim$7.1--8.2$\times$10$^{40}$~erg~s$^{-1}$ for the photo--ionized component, about 20\%--23\% of which is from outside the radio nebula. The attenuation--corrected H$\alpha$ luminosity is, thus, $\approx$3.5--4 times larger than the observed one for NGC\,5253. The SED--predicted H$\alpha$ luminosity for all the 11 clusters is L(H$\alpha$)$_{clusters,corrected}\sim$2.6$\times$10$^{40}$~erg~s$^{-1}$ (Table~\ref{tab6}), 
 to which clusters~5 and ~11 contribute 22\% and 76\%, respectively, and the remaining 9 clusters only contribute a total of 2\%. 
These 9 clusters provide $\sim$3\% of the ionizing photon flux {\em outside} the radio nebula; the remaining flux is provided by the smaller clusters and UV--bright diffuse stars that populate the region \citep{Hong2013}. Within the radio nebula, clusters~5 and 11 supply about 40\%--50\% of the detected ionizing photon flux, i.e., they come short of providing the full ionizing flux by a 
factor $\sim$2. We further discuss this discrepancy in section~6.4.

If unimpeded by dust, cluster~11 would have an absolute magnitude M$_{V}$(Vega)$\sim -$12.8, which is  about 2--3$\sigma$ above 
the mean of the  M$_{brightest}$--SFR  relation for star clusters,  for the SFR value of NGC\,5253 \citep{Whitmore2014b}; this is in the bright envelope, but not outside the range of 
statistically possible values. 
Indeed, the dust--free absolute V magnitude of cluster~11 is consistent with the results of \citet{Billett2002}, who find that dwarf galaxies tend to host massive 
clusters more frequently than do their more massive spiral counterparts. \citet{Billett2002} normalize 
all star clusters to a fiducial age of 10~Myr; if cluster~11 is `aged' to 10 Myr, its dust--free absolute magnitude would be M$_V\sim -$12.6.

\subsection{The Near Infrared Excess of Cluster~11}

The heavily dust--attenuated cluster~11 is anomalously bright in the J and H bands, about a factor 2--2.5 than what predicted from the best fit stellar population models. There 
are four possible causes for this, which we will analyze in turn: (1) excess nebular emission relative to our default model assumption; (2) red supergiants; (3) Pre--Main--Sequence 
stars; and (4) hot dust emission. 
Option~4 has been already considered by other authors for this region, and previously seen in other embedded clusters \citep{Johnson2004}, but will be re--analyzed here in light of our model 
for the stellar and dust mixture of this cluster. 

For option~1, we re--run our best--fit programs using models that implement nebular continuum and line emission with 100\% covering factor, to mimic a tightly confined HII 
region around a star cluster. In all cases, we find that the best fit solution has a reduced $\chi^2$ that is at least twice as worse as the cases with 50\% covering factor. The reason 
is because the nebular continuum and line emission contribute to the optical bands as well as the infrared ones, thus requiring a higher degree of dust attenuation to produce 
the observed SED shape. The higher attenuation pushes the UV model further away from the data, while not compensating enough for the high value of the nearIR data, and 
ultimately resulting in a poorer fit. Thus, a 50\% covering factor is in better agreement with the data even in this more extreme case, and is in agreement with our measurement 
procedure for the stellar continuum and emission lines (in which we have applied larger values of the aperture corrections to the lines than to the continuum). 

The presence of red supergiants (option~2) is attractive because these contribute significantly in the J and H bands. However, cluster~11 is too young to include red 
supergiants, which would need to be originating from another cluster/location, along the line--of--sight of cluster~11. These red supergiants would also need to be 
associated with a star cluster that is otherwise heavily obscured at wavelengths shorter than J and does not emit ionizing photons, since the entire P$\beta$ and P$\alpha$ 
luminosities are fully accounted for by cluster~11. The 100~Myr old population in the two--populations solution discussed in section~5.2 is  too old to still contain red 
supergiant stars. In summary,  we disfavor this scenario, although we cannot exclude it completely. 

Pre--Main--Sequence stars (option~3) are likely to be present in a young cluster, and to provide a significant contribution to the infrared emission. However, 
the effect corresponds to an increase of $\sim$30\% in the K--to--V luminosity ratio relative to a population without Pre--Main--Sequence stars \citep{Zackrisson2001}; this 
is much smaller than that the order--of--magnitude increase observed (see below). 

Option~4, the presence of hot dust surrounding the P$\alpha$ peak, has been studied already by previous authors, including \citet{VanziSauvage2004} and \citet{AlonsoHerrero2004}. 
\citet{VanziSauvage2004} used Adaptive--Optics near infrared observations at Ks (2.16~$\mu$m) and L' (3.78~$\mu$m) to determine and measure  a peak of emission in these two 
bands in correspondence of the location of Clusters~5 and 11. The observations have FWHM=0.4$^{\prime\prime}$, which is comparable to the separation between the two 
clusters. \citet{VanziSauvage2004}  modeled the  SED of cluster~5 at optical wavelengths combined with their and other measurements at longer wavelengths, employing a 
physically--based model for the dust emission. \citet{VanziSauvage2004}, however, did not have measurements at wavelengths between I and Ks, and were not aware of the presence of 
Cluster~11, so the accuracy of their  results is difficult to assess. Since  cluster~5  does not show any evidence for near infrared excess based on our SED modeling, we assign 
those authors'  Ks and L'  emission to cluster~11.

\citet{AlonsoHerrero2004} measured the emission from the entire central region, targeting an area of about 23--24~pc (1.5$^{\prime\prime}\times$1.6$^{\prime\prime}$, larger 
at wavelengths longer than 5~$\mu$m), including both Clusters~5 and 11. They also used a physically--based model to account for both the stellar and dust emission of this 
region, but their model had difficulties in accounting for the emission below the H band. They ascribed this difficulty to the complexity of the stellar and dust geometry in the region. 

Since we are interested mainly in explaining the excess emission at J and H for cluster~11, we use the measurements in the Ks and L' bands from 
\citet[their Table~3]{VanziSauvage2004}, which are obtained with a smaller aperture size than those of \citet{AlonsoHerrero2004}. We attempt a simple modeling 
of the emission of the dust surrounding this cluster. For the same reason, we do not include measurements at longer wavelengths, because the lower resolution 
requires larger apertures, and more contamination from the region surrounding cluster~11. 
Using our distance for NGC\,5253, we derive L(2.16~$\mu$m)=2.0$\times$10$^{35}$~erg~s$^{-1}$~\AA$^{-1}$ and 
L(3.78~$\mu$m)=1.9$\times$10$^{36}$~erg~s$^{-1}$~\AA$^{-1}$. An extrapolation of our best--fit stellar population (plus dust attenuation) model shows that the 
stellar contribution to the 2.16~$\mu$m emission is about 10\%, and is insignificant in the longer wavelength band. 

\begin{figure}
\figurenum{10}
\plotone{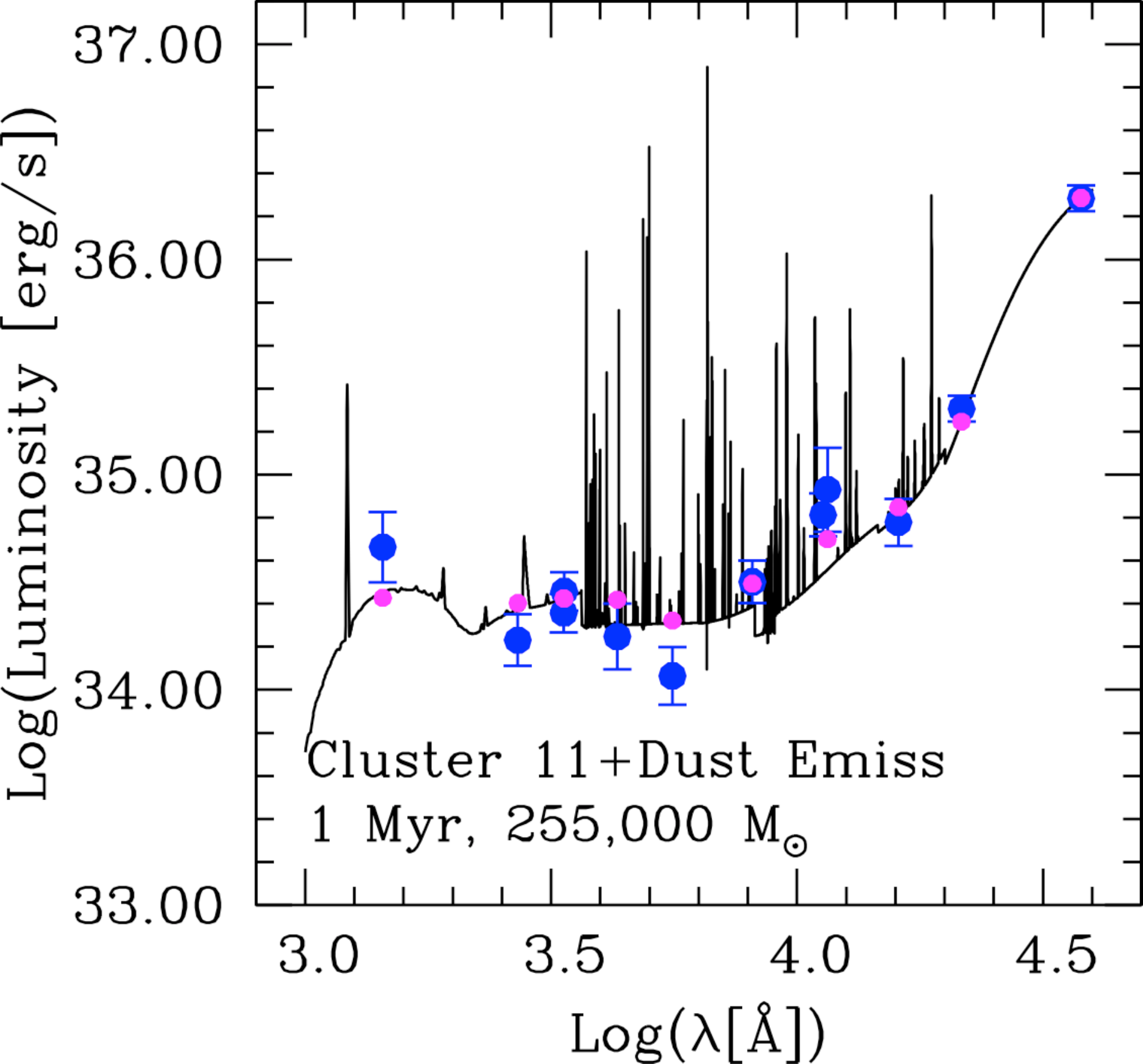}
\caption{Best fit SEDs for Cluster~11, including  stellar, nebular, and dust emission, plus dust attenuation for the stellar and nebular components. The best fit for the stellar, nebular and dust attenuation components are the same as in Figure~\ref{fig9}, while the dust emission is the combination of two modified black--bodies, with temperatures T$_{d,1}\sim$1,100~K and T$_{d,2}\sim$440~k, and emissivity $\epsilon$=1.8. The two data points longward of 1.6~$\mu$m are from \citet{VanziSauvage2004}. Virtually all of the excess 
in the J and H bands can be accounted for with this simple dust emission model.
\label{fig10}}
\end{figure}

Adding a two--component dust emission model  to the stellar$+$dust attenuation model reproduces  the data from $\sim$1,500~\AA \ 
to $\sim$4~$\mu$m reasonably well, as shown in Figure~\ref{fig10}. The two--component dust model consists of 
two modified black--bodies with temperatures T$_{d,1}\sim$1,100~K 
and T$_{d,2}\sim$440~k, and emissivity $\epsilon$=1.8.  The dust emission model parameters are not the result of a rigorous fitting 
procedure, thus the derived temperatures are approximate. Furthermore, an emissivity with power law of 1.8 is a gross approximation below 20~$\mu$m \citep{Draine2003}, 
but changing the dust emissivity has a small effect on the derived temperatures, for our restricted wavelength range. Our two values for the dust temperature, 1,100~K and 
440~K,  bracket the value of T=570~K derived by  \citet{VanziSauvage2004} for the inner shell of their cocoon dust model. Additionally, we have measurements in both the 
J and H bands, which  further enable us to place constraints on the dust emission in this range and derive a higher temperature of 1,100~K. 

There is still some unaccounted for excess in the J--band, although we cannot assess its magnitude, given the size of 
our uncertainties. Thus, emission from hot dust surrounding cluster~11 appears to be capable of  explaining most of the observed  excess in the J--band and beyond
for the SED of this star cluster.  A more physically--based model for the dust emission would be required to infer additional properties for the dust, but this is 
beyond the scope of the present work.

We infer the dust mass associated with these two temperature components, as a sanity check to the reasonability of our dust SED fit. This mass will be a lower limit to 
the actual dust mass associated with those components, since we are in the optically thick regime:  the dust optical depth along the direction of cluster~11 is 
$\tau_d$(4~$\mu$m)$\gtrsim$1.2 , which we derive using the gas column density discussed in the next section and the dust emissivity value at $\sim$4~$\mu$m 
from \citet{Draine2004}; we use the LMC dust emissivity curve from this paper, on account of the low metallicity in NGC\,5253. 
We derive  M$_{dust}$(4~$\mu$m)$\gtrsim$10~M$_{\odot}$, more than 99\% of which is contributed by the cooler of the two dust components, T$_{d,2}\sim$440~K. This mass is a tiny fraction 
of the total dust mass in this region, $\sim$10$^5$~M$_{\odot}$, as derived by probing the cooler dust SED with a  peak around 30--40~$\mu$m 
\citep{VanziSauvage2004}.  It  is, however, consistent with the the expectation that only the dust located in close proximity of the massive stars will be heated to high temperatures. 
For reference, the amount of hot dust surrounding Ultracompact HII regions in the Milky Way is calculated to be around a few to a few tens of M$_{\odot}$ \citep{Walsh1999}, 
consistent with our lower limit for the mass in hot dust associated with cluster~11. 

The presence of dust as hot as T$_d\approx$1,100~K in proximity of cluster~11 supports our SED fitting approach of using a mixed geometry for the stars 
and dust. In this type of geometry, the dust is likely located close to  the UV--emitting stars needed to heat it to high temperatures. Albeit large, the value of the dust temperature 
is still below the dust sublimation temperature of $\approx$2,000~K \citep{Kobayashi2011}, and 
is not dissimilar from that found for Ultracompact HII regions in the Milky Way, where single massive stars can heat a shell of dust in the natal cloud up to $\sim$1,500--1,800~K 
\citep{Walsh1999}.

\subsection{Energy Balance Within the Radio Nebula}

Clusters~5 and ~11 are the most prominent ones in the the radio nebula. These clusters have strong hydrogen 
emission line intensity and concentrated free--free emission \citep{AlonsoHerrero2004,Turner2004}, although several other fainter cluster or massive star candidates are visible in the 
immediate area ($\le$0.6$^{\prime\prime}\sim$9~pc from cluster~5).  

The two clusters share a very young age, $\sim$1~Myr, and a common shell of foreground dust, with optical depth A$_V\sim$1.9~mag. However, they 
also show significant differences. Cluster~5 is about 3.5 times less massive than cluster~11, and is not as deeply buried in dust. Its location in a less dust--enshrouded  
environment than cluster~11 accounts for its prominence as the H$\alpha$ peak emitter in the galaxy, and for the possibility that 25\%--50\% of its ionizing 
photons may be leaking out of the surrounding HII region  (section~5.2).  

Cluster~11, with a best--fit mass of 2.5$\times$10$^5$~M$_{\odot}$, is the behemoth in 
this dwarf galaxy, although not as extreme as inferred  in earlier estimates, where values as large as $\sim$10$^6$~M$_{\odot}$ have been suggested. 
While more massive than cluster~5, cluster~11 is significantly fainter because it is immersed in a dust cloud with front--to--back optical depth A$_V\sim$49~mag. 

Clusters~5 and 11 contribute about 50\% of the ionization in the radio nebula and about 35\% of the total in the galaxy, with most of it coming from cluster~11.  
While they provide a significant fraction of the ionizing photons in both the radio nebula and in NGC\,5253, they do not provide the totality.  This poses a potential issue of 
energy balance, which we discuss in the context of the radio nebula. Within this $\sim$20~pc region (Figure~\ref{fig2}), clusters~5 and 11 
account for 50\% of the free--free emission, for almost the totality of the {\em observed}  H$\alpha$ emission, but only for 
23\% of the {\em observed} P$\alpha$ emission. The rest of the P$\alpha$ emission is spread throughout the region of the radio nebula and has a diffuse morphology, as 
already observed by \citet{AlonsoHerrero2004}. 

If we double the mass of cluster~11 in order to compensate for the factor $\sim$2 discrepancy in the free-free emission, the observed stellar continuum photometry 
is reproduced with a front--to--back optical depth A$_V\sim$100~mag for the dust cloud in which the star cluster is immersed. 
The H$\alpha$ emission requires the same optical depth, but  the P$\alpha$ emission only needs A$_V\sim$20~mag, in order to recover the measured luminosity. 
The discrepancy in the dust optical depth of the two emission lines suggests that this is not a viable scenario. Self--consistent solutions for the available data using 
a $\sim$5$\times$10$^5$~M$_{\odot}$ star cluster can only be obtained by modeling independently the stellar continuum and the ionized emission, as discussed in section~5.2. 
We disfavor such solutions, on the basis that the lines and continuum of cluster~11 appear spatially correlated. 

Our original solution, with a cluster mass of 2.5$\times$10$^5$~M$_{\odot}$ and front--to--back A$_V\sim$50~mag, can close the gap between predicted and measured 
ionizing photon flux in the radio nebula, if young stellar populations 
produce  more ionizing photons than what accounted for by the models we use.  Models that include VMSs  
\citep{Crowther2010, Doran2013} have the potential to increase the ionizing photon flux by up to a factor of 2, with smaller impact on the stellar 
continuum luminosities. The amount of diffuse P$\alpha$ in the radio nebula suggests that about half of the ionizing photons leak out of the clusters into the region. 
This is similar to the fraction of  ionizing photons leakage found by \citet{Johnson2009} for the compact, dusty clusters in the starburst galaxy SBS~0335--052. The observed 
H$\alpha$/P$\alpha$/free--free intensity ratios outside the two clusters can be fully explained by foreground dust with color excess E(B$-$V)$\sim$2~mag, a factor almost three lower than the 
total extinction in cluster~11. The  gas outflow present in the area, as inferred from the broad component in the 
H$\alpha$ emission \citep{MonrealIbero2010,Westmoquette2013}, may create favorable conditions for leakage of ionizing photons, by causing the ISM to become porous. 

The diffuse nature of the P$\alpha$ emission outside of clusters~5 and 11 further suggests that this emission may  originate from other stars and/or clusters in the region,  
some of which may be themselves buried in dust and undetected except at wavelengths longer than a few micron.

\subsection{The Environment of the Radio Nebula}

As already indicated by a number of previous studies (see Introduction), the radio nebula is  the youngest and most active area of 
star formation in NGC\,5253. It is also the region of intersection between the central starburst and the dust lane, which accounts for its significant dust content. 

The dust cloud enshrouding  cluster~11 may appear exceptional, with its A$_V$=49~mag, but it is comparable to clouds in other galaxies, including those in the Milky Way \citep[see, e.g., the 
Ultracompact HII region W3(OH),][]{Turner1984}.    
Adopting the total hydrogen column density--to--color excess relation of \citet{Bohlin1978}, rescaled to the lower metallicity of NGC\,5253,  the optical depth of the cloud 
corresponds to a hydrogen column density N(H)$\sim$2$\times$10$^{23}$~cm$^{-2}$. Although large, this value is comparable to that observed in some 
of the massive dense clouds towards the center of the Milky Way \citep[e.g.][]{Kauffmann2013}. If distributed uniformly throughout a region of about 1~pc size, the observed gas column 
density would correspond to a dust density $\rho_d\sim$7$\times$10$^{-22}$~g~cm$^{-3}$, only a factor of a few lower than the dust densities typical of Ultracompact HII regions in 
the Milky Way \citep{Walsh1999}. As an independent check, the H$_2$ density calculated by \citet{Turner2015} corresponds to  $\rho_{gas}\sim$1.56$\times$10$^{-19}$~g~cm$^{-3}$, or a dust/gas ratio$\sim$0.0045  for the cloud of cluster~11, which is only slightly lower than the values measured for nearby galaxies \citep{Draine2007}. This  ratio could be lower still, since the H$_2$ density is derived from the CO(3--2) transition, which yields a lower limit to the actual molecular gas density. High--resolution HI maps of the region only show HI absorption, which sets a lower limit of 5$\times$10$^{20}$~cm$^{-2}$ to the HI column density \citep{Kobulnicky1995}.  

The virial mass contained in a region about 43~pc$\times$23~pc, roughly the size of the region displayed in either panel of Figure~\ref{fig2}, is about 1.2--1.3$\times$10$^6$~M$_{\odot}$ \citep[][scaled to our distance]{Turner2015}. If clusters~5 and 11 provide most of the stellar mass in the region, the star formation efficiency is 
SFE$\sim$0.25--0.30, about a factor 2 lower than the estimate\footnote{The star formation efficiency decreases to SFE$\sim$0.15, if we adopt a top--heavy stellar IMF with a lower cut--off of 1~M$_{\odot}$; this cut--off is lower than the 3~M$_{\odot}$ proposed by \citet{Turner2015}.} of \citet{Turner2015}. The SFE we derive is comparable to the SFE of clusters in our own Milky Way, but larger than the SFE$\sim$5\%--10\% measured when including the entire molecular cloud  \citep{Lada2003}. Whether or not clusters~5 and 11 will remain bound depends on both the value of the local (cluster--size) SFE 
and the ratio of the gas removal to the crossing timescales \citep{Parmentier2009}.  

If they remain bound and do not lose significant mass, clusters~5 and 11  are massive enough 
that they could be progenitors of globular clusters. Globular clusters, like those in the halo of our own Milky Way and other galaxies, have masses in 
the range $\approx$10$^4$--10$^6$~M$_{\odot}$ \citep{Fall2009}. The mass loss of an evolving cluster depends on a number of factors, including the nature/origin 
of the second population \citep{Schaerer2011}, but under most scenarios, both clusters~5 and 11 would retain sufficient mass to remain within the range of globular 
clusters. Clusters~5 and 11 are two of the youngest among the very massive clusters (super--star--clusters) detected in nearby galaxies, such as those in 
He2--10 \citep{Kobulnicky1999}, NGC1569 \citep{Hunter2000}, M82 \citep{Smith2006}, SBS0335--052 \citep{Johnson2009},  the Antennae  \citep{Whitmore2010},  and NGC1705 \citep{Martins2012}, 
to name a few. As such, they provide important case studies to test theories of multiple populations in globular clusters 
\citep[e.g.][]{deMink2009, Bastian2015}.

\subsection{The Starburst in NGC\,5253}

The youth of the two clusters in the radio nebula  
has an interesting implication for the estimate of the SFR in this galaxy from its TIR luminosity, L$_{TIR}$=3.7$\times$10$^{42}$~erg~s$^{-1}$. 
The majority of the TIR emission in NGC\,5253 originates from the same region, as suggested by the point--like emission 
in the Spitzer images \citep{Dale2009} and by mid--IR imaging \citep{Gorjian2001}. If we use a scaling factor between L(TIR) and SFR appropriate for a population as 
young as 2~Myr \citep{Calzetti2013}, we obtain SFR(TIR)=0.31~M$_{\odot}$~yr$^{-1}$, a factor almost 3 larger than what would be inferred using the standard calibration. The higher SFR(TIR) is consistent with the range of SFRs derived from the free--free emission, SFR$\sim$0.3--0.36~M$_{\odot}$~yr$^{-1}$, suggesting that, within the uncertainties, we see no obvious indication for significant direct absorption of ionizing photons by dust. Adding the UV, which originates in separate regions from the IR, we get a total SFR for NGC\,5253 SFR=0.37~M$_{\odot}$~yr$^{-1}$, when using a timescale of 10~Myr for the SFR(UV), which is appropriate for the youth of the UV--emitting area.

\begin{figure}
\figurenum{11}
\plotone{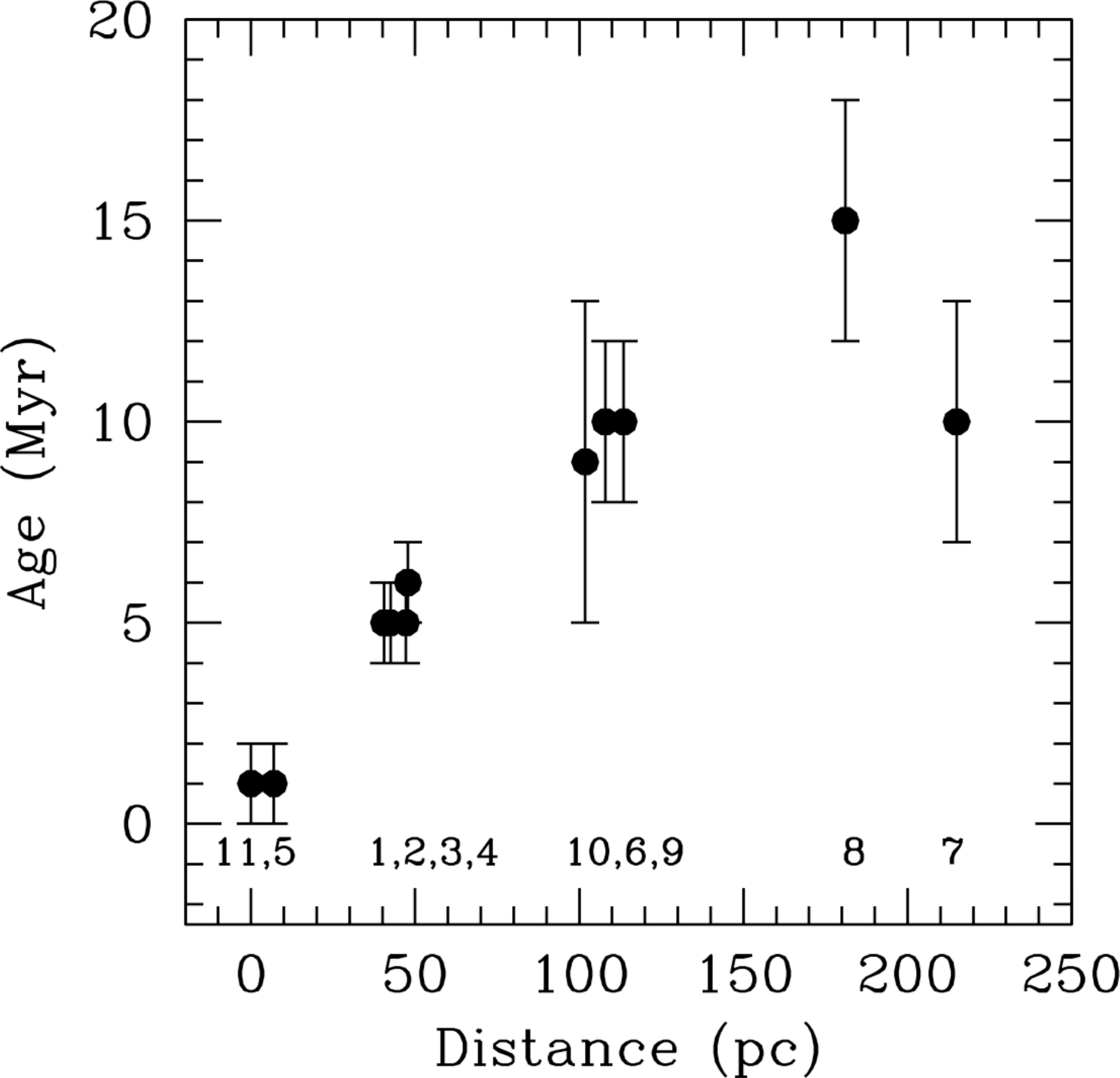}
\caption{The ages, with their 1~$\sigma$ uncertainty,  of the 11 star clusters as a function of their distance from cluster~11.  For each data point, the corresponding cluster 
number is indicated at the bottom of the figure. A systematic trend is observed, with the older clusters 
located further away from the location of cluster~11, i.e., from the region of most active star formation in the galaxy. 
\label{fig11}}
\end{figure}

Moving away from the central radio nebula, the clusters increase in age for increasing distance from clusters~5 and 11 (Figures~\ref{fig1b} and \ref{fig11}). 
Clusters 1--4 have best--fit ages around 5--6~Myr, while clusters 
7--10 are consistent with ages $\sim$9--15~Myr.  The `association' of clusters 1--4, all with comparable ages and spread over a small area of $\sim$170~pc$^2$, agrees with the micro--level 
hierarchy picture discussed by \citet{Efremov1998};  these authors find  that in the Magellanic Clouds small regions form stars over a short period of time.  
UV--bright star clusters to the North of the radio nebula have ages in the range 4--8~Myr \citep{Tremonti2001}. \citet{deGrijs2013} also recovers 
typically young ages for the star clusters in the central region of NGC\,5253, between a few 10$^6$~yr and a few 10$^7$~yr. This suggests a progressive concentration 
of the most recent event of star formation from a larger, $\sim$300~pc, region to the smaller area, $\sim$20~pc,  of the radio nebula. Covering 300~pc in $\sim$15~Myr requires a propagation velocity of the perturbation 
of about 20~km~s$^{-1}$. This is larger than the typical sound speed of a few km/s of the ISM, but is consistent with the velocities of a few tens of km/s expected for the propagation of turbulence
in a multi--phase ISM \citep[e.g.][]{Bournaud2011}. External triggers, such as the recent--past interaction with M\,83 and/or gas infall, are required for such large--scale propagating disturbances, although internal triggers \citep[e.g., shocks from earlier events that produce sequential star formation, see, ][]{Whitmore2010} may also play a role. The region within the central $\approx$300~pc of 
NGC\,5253 is likely to have been the stage of one continuous and connected event of star formation over at least the past $\sim$15~Myr. This is in agreement with other estimates 
that indicate relatively constant levels of SFR over the past few hundred Myr \citep{McQuinn2010,Harbeck2012}, although these studies cannot discriminate between  continuous or sporadic star formation 
over this timescale.

The energetics that drive the ISM kinematics in NGC\,5253 are also consistent with a continuous star formation event over at least this timescale. 
The measured H$\alpha$ luminosity of the shocked gas, 
L(H$\alpha$)$_{shock}\sim$3$\times$10$^{39}$~erg~s$^{-1}$ \citep[][, rescaled to our distance]{Hong2013}, corresponds to a mechanical luminosity of 
L$_{mech}\sim$7$\times$10$^{40}$~erg~s$^{-1}$ \citep{Rich2010}, for radiative shocks and the shock velocity measured by \citet{Marlowe1995}. For constant star formation and our adopted Kroupa IMF, the 
mechanical energy requirements can be satisfied with SFR$\sim$0.2M$_{\odot}$~yr$^{-1}$ over the past 10--15~Myr.  The five star clusters in this age range, 6--to--10, correspond to a cumulative mass of $\sim$1.6$\times$10$^5$~M$_{\odot}$ (about 1/2 of the sum of the masses of 5 and 11), and may have contributed, in the past, about 1/3 of the mechanical energy needed to support the shocks. The rest of the energy requirement has come from the remaining young star clusters and diffuse stars in the region.  When they will have aged enough to start 
producing supernova explosions, clusters~5 and 11 will be able to supply about 2/3 of the mechanical energy requirement of the present--day shocks. 

\section{Summary and Conclusions}

The combination of new and archival HST imaging data of  the central 300~pc in the starburst galaxy NGC\,5253 has enabled us to derive,  with 
unprecedented accuracy,  ages, masses, and extinctions for the brightest among the star clusters in the starburst. The HST data  cover the full wavelength range from $\sim$1500~\AA\ to 1.9~$\mu$m in 10 continuum bands and 3 narrow bands, which include the emission lines of H$\alpha$, P$\beta$, and P$\alpha$. The 11 star clusters analyzed here include the two young clusters 
located within the central dusty radio nebula, in addition to nine clusters distributed across the UV--bright starburst. 

The multi--wavelength SED of each cluster 
has been fit with models that include stellar populations, gas emission, and dust attenuation (and, for cluster~11, dust emission). 
Models that include only foreground dust are sufficient to produce excellent fits for 
the 10 bright clusters, while the heavily dust extincted cluster~11 requires a geometry that includes a combination of foreground and mixed dust. 

Because of its location within a region of high optical depth, this is the first time the SED of cluster~11 has been fit shortward of the J--band. This is also the first time extensive stellar continuum photometry, from the UV to the nearIR, is available for a natal cluster. Albeit counter--intuitive, cluster~11 is detectable in all optical and UV bands, mainly because dust mixed with 
stars dims the light more readily than reddening it \citep{Calzetti2001}. Like for 
all other clusters analyzed in this paper, the SED of cluster~11 can be fit with a standard \citet{Kroupa2001} IMF in the mass range 0.1--120~M$_{\odot}$, and does not 
require truncations or other modifications. The resulting best--fit age, mass, and total dust attenuation of cluster~11 yield a self--consistent picture of this source, also 
in agreement with independent constraints. 

All clusters studied here are younger than $\sim$15~Myr, in agreement with previous results, with a systematic trend for the age to become 
younger  when moving from the outskirts of the starburst towards its center, where the radio nebula is located. The picture that emerges is for the current burst of 
star formation to have been a continuous event over the past $\sim$15~Myr, and to have concentrated inward, forming a hierarchical `age' structure similar to 
those discussed by  \citet{Efremov1998} for the LMC.  Dwarf starburst galaxies, including NGC\, 5253, do tend, in fact,  to display a single dominant region of 
connected star formation hierarchical in structure \citep{Elmegreen2014}. SFR indicators calibrated for the youth of the starburst yield 
SFR$\sim$0.4$_{\odot}$~yr$^{-1}$ for the galaxy, or about three times higher than using standard SFR calibrations. 

The nine star clusters located outside the radio nebula have masses in the range 0.5--5$\times$10$^4$~M$_{\odot}$, and dust attenuations A$_V\lesssim$1~mag. The 
relatively low A$_V$ value is consistent with the location of the clusters in the UV--bright region of the starburst. Other indicators show dust attenuation to be low in this region in general. The nine clusters provide today a minimal contribution, less than 2\%, to the total ionizing photons in the galaxy. 

The two most massive clusters in our sample, clusters~5 and 11, are both located inside the dusty radio nebula. Their large masses,  
7.5$\times$10$^4$~M$_{\odot}$ and 2.5$\times$10$^5$~M$_{\odot}$, respectively, are consistent with having been born in a high--density environment. Both clusters are extremely young, with best fit ages 1$\pm$1~Myr. Neither has reached a stage when supernovae are starting to affect the local environment, including blowing out the natal dust cloud, which accounts for their significant dust attenuation. Energy balance between the ionizing photon rate recovered from the free--free emission and the total far--infrared luminosity suggests that, albeit buried in a significant dust cloud, the clusters in the radio nebula do not suffer from significant direct absorption of ionizing photons by dust. 

Both clusters are behind a dust layer of A$_V<$2~mag, which is likely the outer dust shell of the radio nebula. Cluster~5 is behind only this dust layer, which is still sufficiently transparent to enable this cluster to be the H$\alpha$ peak in the galaxy. Cluster~11 is attenuated by an additional dust cloud with front--to--back optical depth A$_V\sim$49~mag, mixed with its stars. 
This dust cocoon has many characteristics typical of those observed in Milky Way Ultracompact HII regions, including a maximum temperature of $\sim$1,100~K for the dust closest to the massive stars. It is also responsible 
for about 60\%, 65\%, and 90\% of the emission observed in the J, H, and K bands, respectively.

The mass of cluster~11 places it a factor $\approx$2--4 below the $\sim$0.6--1.2$\times$10$^6$~M$_{\odot}$ value inferred in previous studies. Except for its `dusty' condition, owed presumably to its youth, this cluster's other characteristics are relatively normal. For instance, it is within 2-3$\sigma$ of the relation between the most massive 
clusters and SFR in galaxies, and is similar to massive clusters  recovered in other dwarf galaxies. Its mass is in--between those of the scaled OB association NGC604 in M\,33 \citep{Maiz2001} and the massive cluster NGC346 in the SMC \citep{Portegies2010}. For the level of dust attenuation and for its youth, cluster~11 is reminiscent of the super--star--cluster WS~80 in the Antennae galaxy \citep{Whitmore2002}, although it is also about 16 times less massive. Based on the self--consistency of several indicators, we do not believe to be missing a large fraction of the mass of cluster~11.

Clusters~5 and 11 provide about half of the ionizing photons in the radio nebula, and about one third of the ionizing flux in the entire galaxy. The remaining 50\% of the nebula's ionizing photons are diffuse, suggesting leakage of ionizing photons from the  immediate surroundings of the two clusters into the nebula's region.  Once they 
start producing supernovae, clusters~5 and 11 will supply about 2/3 of its mechanical luminosity requirements to support the current level of shocks. Thus, by themselves, the two clusters can 
sustain a significant fraction of the energy requirements of NGC\,5253. The remaining energy needs to be supplied by other sources, which may include: other stars and star 
clusters within the area of the radio nebula, Very Massive Stars \citep{Crowther2010,Doran2013}, and/or rotating massive stars \citep{Leitherer2014}.   
The potential presence of VMSs is consistent with the detection of  Wolf--Rayet features of WN type in the region of the radio nebula. Follow--up high--spatial--resolution 
spectroscopy of these clusters has the potential to reveal  signatures of rare star formation products and address this open question.

\acknowledgments

Based on observations made with the NASA/ESA Hubble Space Telescope, obtained  at the Space Telescope Science Institute, which is operated by the 
Association of Universities for Research in Astronomy, Inc., under NASA contract NAS 5--26555. These observations are associated with program \# 13364. 
Support for program \# 13364 was provided by NASA through a grant from the Space Telescope Science Institute.

Based also on observations made with the NASA/ESA Hubble Space Telescope, and obtained from the Hubble Legacy Archive, 
which is a collaboration between the Space Telescope Science Institute (STScI/NASA), the Space Telescope European Coordinating 
Facility (ST-ECF/ESA) and the Canadian Astronomy Data Centre (CADC/NRC/CSA).

This research has made use of the NASA/IPAC Extragalactic Database (NED) which is operated by the Jet
Propulsion Laboratory, California Institute of Technology, under contract with the National Aeronautics and Space
Administration.

Part of this work was conducted while D.C. was a Raymond and Beverley Sackler Distinguished Visitor at the Institute of 
Astronomy, University of Cambridge (UK), and an Overseas Fellow at the Churchill College (Cambridge, UK).
D.C. acknowledges the kind hospitality of both the Institute and the College.  
A.S.E. was supported by the Taiwan, R.O.C. Ministry of Science and Technology grant MoST 102-2119-M-001-MY3. 
M.F. acknowledges support by the Science and Technology Facilities Council [grant number ST/L00075X/1].  
D.A.G. kindly acknowledges financial support by the German Research Foundation (DFG) through
grant GO\,1659/3-2. 
E.Z. acknowledges research funding from the Swedish Research Council (project 2011-5349).

\clearpage

\begin{deluxetable}{lrrllrrll}
\tablecolumns{9}
\rotate
\tabletypesize{\footnotesize}
\tablecaption{Characteristics of the HST Images of NGC\,5253.\label{tab1}}
\tablewidth{0pt}
\tablehead{
\colhead{Instrument/Camera\tablenotemark{a}} & \colhead{Pixel Size\tablenotemark{a}} & \colhead{Ang. Res.\tablenotemark{a}}  & \colhead{Field of View\tablenotemark{a}} 
& \colhead{Filter\tablenotemark{b}}  & \colhead{Pivot Wavelength\tablenotemark{b}} & \colhead{Exposure Time\tablenotemark{c}} 
& \colhead{Date Obs.\tablenotemark{d}} & \colhead{Program\tablenotemark{d}}
\\
\colhead{} & \colhead{($^{\prime\prime}$)}  & \colhead{($^{\prime\prime}$)} 
& \colhead{($^{\prime\prime}\times^{\prime\prime}$)} & \colhead{} & \colhead{(\AA)} 
& \colhead{(s)} & \colhead{}  & \colhead{}   
\\
\colhead{(1)} & \colhead{(2)} & \colhead{(3)} & \colhead{(4)} & \colhead{(5)} & 
\colhead{(6)} & \colhead{(7)} & \colhead{(8)} & \colhead{(9)} 
\\
}
\startdata
\hline
ACS/SBC     & 0.030 & 0.097 & 35$\times$31      & F125LP & 1438.2 & 2660.0 (665.$\times$4) &  2009--03--07 & GO--11579 \\
WFC3/UVIS & 0.040 & 0.070 & 162$\times$162  & F275W & 2710.1 & 2448.0 (816.$\times$3) & 2013--08--28 & GO-13364\\
                      &             &            &                                & F336W & 3354.8 & 2346.0 (782.$\times$3) & 2013--08--28 & GO-13364\\
ACS/HRC    & 0.025 & 0.060  & 29$\times$26       & F330W & 3362.7 & 1796.0 (449.$\times$4) &  2006--02--20 & GO--10609 \\
                      &             &            &                                 & F435W & 4311.0 & 600.0 (150.$\times$4) &  2006--02--20 & GO--10609 \\
                      &             &            &                                 & F550M & 5579.8 & 800.0 (200.$\times$4) &  2006--02--20 & GO--10609 \\
                      &             &            &                                 & F658N(H$\alpha+$[NII]) & 6583.7 & 240.0 (60.$\times$4) &  2006--02--20 & GO--10609 \\
                      &             &            &                                 & F814W & 8115.4 & 368.0 (92.$\times$4) &  2006--02--20 & GO--10609 \\
WFC3/IR     & 0.130  & 0.22    & 123$\times$136  & F110W & 11534.0 & 597.6 (199.2$\times$3) & 2011--07--26 & GO--12206\\
                      &             &            &                                & F128N (P$\beta$)& 12832.0 & 1497.6 (499.2$\times$3) & 2011--07--26 & GO--12206\\
NICMOS/NIC2 & 0.075 & 0.13 & 19$\times$19  & F110W & 11292.0 & 96.0 (24.$\times$4)  & 1998--01--04 & GO--7219\\
                      &             &            &                                 & F160W & 16071.0 &96.0 (24.$\times$4)  & 1998--01--04 & GO--7219\\
                      &             &            &                                 & F187N (P$\alpha$)& 18747.8 &256.0 (64.$\times$4)  & 1998--01--04 & GO--7219\\
                      &             &            &                                 & F190N & 18986.0 &256.0 (64.$\times$4)  & 1998--01--04 & GO--7219\\
\hline
\enddata

\tablenotetext{a}{Instrument/Detector combination, together with its native pixel scale, the angular resolution expressed as the Point Spread Function (PSF) FWHM, and 
Field-of-View (FoV). Note that in some cases, the HLA serves image products at slightly different pixel scale than the native one. The PSF FWHM 
is directly measured on our images.}
\tablenotetext{b}{For each filter, the pivot wavelength is also listed. The narrow band filters F658N, F128N, and F187N target the emission lines of 
H$\alpha$(6563\AA)$+$[NII](6548,6584\AA), P$\beta$(12818\AA), and P$\alpha$(18756\AA), respectively. The F190N narrow--band image is used 
for the stellar continuum subtraction of the F187N image. }
\tablenotetext{c}{The total exposure time, shown in units of seconds, is usually the result of 3 or 4 combined (dithered) exposures, with individual times as indicated.}
\tablenotetext{d}{Date in which the observation took place, and name of the HST observing program that obtained the data.}
 \end{deluxetable}

\clearpage

\begin{deluxetable}{lrr}
\tablecolumns{3}
\tabletypesize{\normalsize}
\tablecaption{Foreground Extinction Correction Values\label{tab2}}
\tablewidth{0pt}
\tablehead{
\colhead{Filter} 
& \colhead{Band} 
& \colhead{k($\lambda$)\tablenotemark{a}} 
\\
}
\startdata
F125LP & FUV & 8.54\\
F275W  & NUV & 6.29\\
F336W &  U       & 5.07\\
F330W &  U       & 5.06\\
F435W  &  B      &  4.21\\
F550M  &  V      & 3.05\\
F658N & H$\alpha$ & 2.54\\
F814W & I         & 1.80\\
F110W(NIC2) & J & 1.03\\
F110W(WFC3) & J & 1.00\\
F128N & P$\beta$ & 0.84\\
F160W & H       & 0.58\\
F187N & P$\alpha$ & 0.46\\
\enddata

\tablenotetext{a}{The extinction curve for the Milky Way, k($\lambda$), expressed as: F$_{obs}$($\lambda$)= F$_{int}$($\lambda$) 10$^{-0.4 E(B-V) k(\lambda)}$. The values 
of the extinction curve are from the parametrization of \citet{Fitzpatrick1999}, with total--to--selective extinction value R$_V$=3.1. }
\end{deluxetable}

\clearpage

\begin{deluxetable}{lrrrrrrrrrrr}
\tablecolumns{12}
\rotate
\tabletypesize{\tiny}
\tablecaption{Cluster Location and Photometry\label{tab3}}
\tablewidth{0pt}
\tablehead{
\colhead{Field} & \colhead{\# 1}  & \colhead{\# 2} & \colhead{\# 3} & \colhead{\# 4} & \colhead{\# 5} 
& \colhead{\# 6} & \colhead{\# 7} & \colhead{\# 8} & \colhead{\# 9} & \colhead{\# 10} & \colhead{\# 11} 
\\
\colhead{(1)} & \colhead{(2)} & \colhead{(3)} & \colhead{(4)} & \colhead{(5)}  
& \colhead{(6)} & \colhead{(7)} & \colhead{(8)} & \colhead{(9)} & \colhead{(10)} & \colhead{(11)} & \colhead{(12)} 
\\
}
\startdata
\hline
RA(J2000)\tablenotemark{a} & 13:39:55.919 & 13:39:55.901  & 13:39:55.960  & 13:39:55.858  & 13:39:55.986  & 13:39:55.965  &  13:39:55.833 & 13:39:55.351  &  13:39:55.512 &  13:39:55.563 &  13:39:55.951\\
DEC(J2000)\tablenotemark{a} & $-$31:38:27.09 & $-$31:38:27.18 & $-$31:38:27.57 & $-$31:38:27.41 & $-$31:38:24.54 & $-$31:38:31.51 & $-$31:38:38.49 & $-$31:38:33.54 & $-$31:38:29.35  & $-$31:38:28.92 & $-$31:38:24.45 \\
Cross--IDs\tablenotemark{b}                                      & 4, 4, 95               & 4, 4, 90?               & ..., 8, 106    & ..., ..., ... & 5, 1, 87              & 1, 2, 129      & ..., 27, 156 & 6, 9, 36         & 3, 3, 38               & 2, 5, 45         & ..., ..., ... \\
L(F125LP)\tablenotemark{c}                                      & 36.56(0.07) & 36.68(0.07) & 36.38(0.07) & 35.91(0.07)         &36.00(0.07) &36.66(0.07) &36.52(0.07)  &36.28(0.07)  &36.30(0.07) &36.08(0.07) &34.66(0.16)\\
L(F275W)\tablenotemark{c}                                       & 36.07(0.07)  & 36.14(0.07) & 35.88(0.07) & 35.71(0.07)        &35.79(0.07)  &36.20(0.07) &35.92(0.07) &35.81(0.07)  &35.92(0.07)  &35.73(0.07) &34.23(0.12)\\
L(F336W)\tablenotemark{c}                                        & 35.87(0.07) & 35.93(0.07) & 35.67(0.07) & 35.58(0.07)        &35.84(0.07)  &35.98(0.07) &35.70(0.07) &35.65(0.07)  &35.77(0.07)  &35.60(0.07) &34.36(0.09)\\
L(F330W)\tablenotemark{c}                                        & 35.87(0.07) & 35.90(0.07) & 35.66(0.07) & 35.63(0.07)        &35.82(0.07)  &35.97(0.07) &35.67(0.07) &35.62(0.07)  &35.75(0.07)  &35.59(0.07) &34.45(0.09)\\
L(F435W)\tablenotemark{c}                                        & 35.62(0.07) & 35.64(0.07) & 35.41(0.07) &35.42(0.07)         &35.63(0.07)  &35.77(0.07) &35.44(0.07) &35.51(0.07)  &35.67(0.07)  &35.66(0.07) &34.25(0.15)\\
L(F550M)\tablenotemark{c}                                        & 35.36(0.07) & 35.35(0.07) & 35.15(0.07) & 35.25(0.07)        &35.30(0.07) &35.59(0.07)  &35.23(0.07) &35.31(0.07)  &35.62(0.07)  &35.55(0.07) &34.06(0.13)\\
L(F814W)\tablenotemark{c}                                        & 34.88(0.07) & 34.86(0.07) & 34.61(0.07) & 34.93(0.07)        &35.38(0.07) &35.39(0.07) &35.00(0.07) &34.99(0.07)  &35.47(0.07)   &35.24(0.07) &34.50(0.10)\\
L(N/F110W)\tablenotemark{c}                                    & 34.35(0.10) & 34.34(0.10) & 33.84(0.10) & 34.45(0.10)        &35.15(0.10) &35.18(0.10) &34.66(0.10) &34.66(0.10)   &35.21(0.10)  &34.90(0.10) &34.81(0.11)\\
L(W/F110W)\tablenotemark{c}                                   & 34.44(0.14) & 34.42(0.14)  & 33.69(0.14) & 34.36(0.13)       &35.23(0.14)  &35.27(0.13) &34.73(0.14) &34.78(0.14)  &35.16(0.13)  &34.96(0.14) &34.93(0.19)\\
L(F160W)\tablenotemark{c}                                        & 33.77(0.10) & 33.78(0.10) & 33.45(0.10) & 34.13(0.10)       &34.79(0.10)  &35.02(0.10) &34.50(0.10) &34.46(0.10)  &35.03(0.10)  &34.66(0.10) &34.78(0.11)\\
L(H$\alpha$)\tablenotemark{d}                                  & 38.26(0.11) & 38.06(0.11) & 37.04(0.11) & 37.63(0.11)       &38.96(0.11)  & $<$35.36   &36.77(0.11) &$\lesssim$35.36  &37.00(0.11) &36.17(0.13) &38.17(0.11) \\
L(P$\beta$)\tablenotemark{d}                                    & 37.17(0.19) & 37.01(0.19)  & 36.08(0.21) & 36.62(0.19)       &38.05(0.19)  & $<$35.54   &36.01(0.22) &$<$35.54       &37.22(0.19) &$<$35.54    &37.93(0.19)  \\
L(P$\alpha$)\tablenotemark{d}                                  & 37.25(0.13) & 37.08(0.13)  & 35.90(0.17) & 36.78(0.13)       &38.42(0.13)  &35.92(0.17) &35.86(0.18) &36.01(0.16)  &36.97(0.13) &35.94(0.17) &38.52(0.13)   \\
EW(H$\alpha$)\tablenotemark{e}                               &450--1140   &290--730       & 40--110        &200--300            &$>$3200       & 0                    &30--40        &$\approx$1    &10--30          &$\sim$5        &$>$3200     \\
E(B$-$V)$_{H\alpha/P\beta}$\tablenotemark{f}   & 0.23(0.32)     & 0.29(0.32)    & 0.43(0.35)    & 0.35(0.32)         & 0.49(0.32)   & ...                    &0.71(0.36) & ...                    &2.16(0.32)    & ...                  &1.48(0.32)   \\
E(B$-$V)$_{H\alpha/P\alpha}$\tablenotemark{f} & 0.00(0.20)    & 0.00(0.20)    & 0.00(0.24)     & 0.10(0.20)         & 0.47(0.20)   & ...                    &0.03(0.25) &1.92(1.22)     &1.10(0.20)    &0.86(0.26)    &1.54(0.20)   \\
\enddata

\tablenotetext{a}{Astrometry obtained from the LEGUS WFC3/UVIS/F336W image.}
\tablenotetext{b}{The identification of the clusters in the papers of \citet{Calzetti1997} (first number),  \citet{Harris2004} (second number), and \citet{deGrijs2013} (third number). Clusters 1 and 2 have been identified 
as a single star cluster in the lower--resolution data of the first two papers. The cross--identification of cluster~3 with cluster 90 of \citet{deGrijs2013} is uncertain. 
Clusters~5 and 11 are located within the radio--bright nebula of \citet{Turner2000} and \citet{Turner2004}, see Figure~\ref{fig2}.}
\tablenotetext{c}{Broad/medium band luminosity densities, in units of erg~s$^{-1}$~cm$^{-2}$~\AA$^{-1}$, with the 1$\sigma$ uncertainties in parenthesis. The two measurements in the F110W filter from the NICMOS and WFC3 images are  indicated as N/F110W and W/F110W, respectively. The listed photometry is from  5--pixel (0.125$^{\prime\prime}$) radius aperture
measurements, corrected to infinite aperture and for foreground Milky Way extinction (E(B$-$V)=0.049, see text). The luminosity densities are calculated adopting a distance of 3.15~Mpc for 
NGC\,5253.}
\tablenotetext{d}{Emission line luminosity, in units of erg~s$^{-1}$~cm$^{-2}$, with the 1$\sigma$ uncertainties in parenthesis. The luminosities are derived as described 
in section~3, also corrected to infinite aperture and for foreground Milky Way extinction (E(B$-$V)=0.049, see text), and calculated adopting a distance of 3.15~Mpc.}
\tablenotetext{e}{The equivalent width (EW) of H$\alpha$, in \AA, calculated from the ratio of the emission line flux  to the stellar continuum flux density, derived as described in section~3. The range of EWs indicates the ratio obtained both before (smaller value) and after (larger value) aperture correction.}
\tablenotetext{f}{The color excess derived from the indicated emission lines, under the simple assumption of a foreground dust screen, using the selective extinction values from Table~\ref{tab2}. See text for more details.}
\end{deluxetable}

\clearpage

\begin{deluxetable}{lrrrrrr}
\tablecolumns{7}
\rotate
\tabletypesize{\small}
\tablecaption{Physical Parameters of the Clusters Outside the Radio Nebula \label{tab4}}
\tablewidth{0pt}
\tablehead{
\colhead{Cluster} & \colhead{Age$_{SED}$\tablenotemark{a} } & \colhead{Age$_{SED-lines}$\tablenotemark{a} } & \colhead{Age$_{EW(H\alpha)}$\tablenotemark{b}} & \colhead{Mass$_{SED}$\tablenotemark{c} } 
& \colhead{E(B$-$V)$_{SED}$\tablenotemark{d} } & \colhead{E(B$-$V)$_{lines}$\tablenotemark{e} } 
\\
\colhead{(\#)} & \colhead{(Myr)} & \colhead{(Myr)}  & \colhead{(Myr)} & \colhead{(10$^4$~M$_{\odot}$) } 
& \colhead{(mag)} & \colhead{(mag) } 
\\
\colhead{(1)} & \colhead{(2)} & \colhead{(3)} & \colhead{(4)} & \colhead{(5)}  & \colhead{(6)} & \colhead{(7)} 
\\
}
\startdata
\hline
\# 1 & 5$^{+1}_{-2}$ & 5$^{+1}_{-1}$ & 4.6--5.6 & 1.05$^{+0.28}_{-0.22}$ & 0.12$^{+0.03}_{-0.03}$ & 0.12$\pm$0.19 \\
\# 2 & 5$^{+1}_{-2}$ & 5$^{+1}_{-2}$ & 5.1--6.1 & 0.91$^{+0.31}_{-0.22}$ & 0.08$^{+0.03}_{-0.03}$ & 0.15$\pm$0.19\\
\# 3 & 5$^{+1}_{-0}$ & 5$^{+1}_{-2}$ &7.5--10.1 & 0.46$^{+0.11}_{-0.10}$ & 0.04$^{+0.02}_{-0.02}$ &0.22$\pm$0.21 \\
\# 4 & 6$^{+0}_{-2}$ & 5$^{+1}_{-1}$ & 6.0--6.5 & 1.62$^{+0.52}_{-0.48}$ & 0.32$^{+0.04}_{-0.04}$ & 0.22$\pm$0.19\\
\# 6 & 10$^{+2}_{-1}$ & 10$^{+8}_{-2}$ & $>$30 & 3.24$^{+1.33}_{-0.94}$ & 0.12$^{+0.04}_{-0.02}$ & ...\\
\# 7 & 10$^{+3}_{-1}$ & 10$^{+4}_{-2}$& 10--11 & 1.15$^{+0.30}_{-0.56}$ & 0.05$^{+0.04}_{-0.03}$ & 0.37$\pm$0.22 \\
\# 8 & 15$^{+3}_{-3}$ & 16$^{+4}_{-4}$ & $>$30 & 2.88$^{+1.64}_{-0.84}$ & 0.16$^{+0.04}_{-0.04}$ & 1.92$\pm$1.22 \\
\# 9 & 10$^{+2}_{-1}$ & 10$^{+4}_{-2}$ & 11--16 & 5.13$^{+2.12}_{-1.50}$ & 0.26$^{+0.06}_{-0.04}$ & 1.63$\pm$0.19\\
\# 10 & 9$^{+5}_{-2}$ & 12$^{+5}_{-4}$ & $>$16 & 3.63$^{+3.22}_{-1.34}$ & 0.26$^{+0.04}_{-0.04}$ & 0.86$\pm$0.26 \\
\enddata

\tablenotetext{a}{The age, and 1~$\sigma$ uncertainty, from the SED fitting of the photometry with the Yggdrasil models for column~2, and from the SED fitting of the nebular--line--subtracted photometry 
with the Starburst99 models for column~3.}
\tablenotetext{b}{The age inferred from the EW(H$\alpha$) listed in Table~\ref{tab3}.}
\tablenotetext{c}{The cluster mass as derived from the SED fitting with the Yggdrasil models.}
\tablenotetext{d}{The color excess, with its 1~$\sigma$ uncertainty,  derived from SED fitting with the Yggdrasil models.}
\tablenotetext{e}{The color excess from the emission lines, reported as the average between the two values listed in Table~\ref{tab3}. When only one value is present, that value is the one reported here.}
\end{deluxetable}

\clearpage

\begin{deluxetable}{lrrrrrr}
\tablecolumns{7}
\rotate
\tabletypesize{\small}
\tablecaption{Physical Parameters of the Clusters Within the Radio Nebula \label{tab5}}
\tablewidth{0pt}
\tablehead{
\colhead{Cluster} & \colhead{Age$_{SED}$\tablenotemark{a} } & \colhead{Age$_{EW(H\alpha)}$\tablenotemark{b}} & \colhead{Mass$_{SED}$\tablenotemark{c} } 
& \colhead{E(B$-$V)$_{SED,mix}$\tablenotemark{d} } & \colhead{E(B$-$V)$_{SED,fore}$\tablenotemark{d} } & \colhead{E(B$-$V)$_{lines}$\tablenotemark{e} } 
\\
\colhead{(\#)} & \colhead{(Myr)} & \colhead{(Myr)}  & \colhead{(10$^4$~M$_{\odot}$) } 
& \colhead{(mag)} & \colhead{(mag)} & \colhead{(mag)}
\\
\colhead{(1)} & \colhead{(2)} & \colhead{(3)} & \colhead{(4)} & \colhead{(5)}  & \colhead{(6)} & \colhead{(7)} 
\\
}
\startdata
\hline
\# 5 & 1$^{+1}_{-1}$ & $<$1 & 7.46$^{+0.20}_{-0.27}$ & \nodata & 0.46$^{+0.04}_{-0.04}$ & 0.48$\pm$0.19 \\
\# 11 & 1$^{+1}_{-1}$ & $<$1 & 25.5$^{+6.7}_{-4.2}$ & 15.7$^{+3.5}_{-2.0}$ & 0.46$^{+0.06}_{-0.02}$ & 1.51$\pm$0.26 \\
\enddata

\tablenotetext{a}{The age, and 1~$\sigma$ uncertainty, from the SED fitting of the photometry.}
\tablenotetext{b}{The age inferred from the EW(H$\alpha$) listed in Table~\ref{tab3}.}
\tablenotetext{c}{The cluster mass as derived from the SED fitting.}
\tablenotetext{d}{The color excess, with its 1~$\sigma$ uncertainty,  derived from SED fitting. The two values of E(B$-$V) are for the mixed dust--stars and foreground dust cases. Cluster~11 requires both dust 
geometries to be present in order to account for the shape and luminosity of its observed SED.}
\tablenotetext{e}{The color excess from the emission lines, reported as the average between the two values listed in Table~\ref{tab3}. }
\end{deluxetable}

\clearpage

\begin{deluxetable}{lrr}
\tablecolumns{3}
\tabletypesize{\small}
\tablecaption{Intrinsic and Predicted H$\alpha$ Luminosities \label{tab6}}
\tablewidth{0pt}
\tablehead{
\colhead{Cluster} & \colhead{Log [L(H$\alpha$)$_{intrinsic}$]\tablenotemark{a} } & \colhead{Log [L(H$\alpha$)$_{predicted}$]\tablenotemark{b} }  
\\
\colhead{(\#)} & \colhead{(erg~s$^{-1}$)} & \colhead{(erg~s$^{-1}$)} 
\\
\colhead{(1)} & \colhead{(2)} & \colhead{(3)}  
\\
}
\startdata
\hline
\# 1 & 38.38$^{+0.18}_{-0.12}$ & 38.13$^{+0.39}_{-0.31}$\ \\
\# 2 & 38.21$^{+0.19}_{-0.15}$ & 38.07$^{+0.72}_{-0.31}$\\
\# 3 & 37.62$^{+0.21}_{-0.21}$ & 37.77$^{+0.72}_{-0.30}$\\
\# 4 & 37.85$^{+0.19}_{-0.19}$ & 37.99$^{+0.73}_{-0.19}$\\
\# 5 & 39.45$^{+0.15}_{-0.12}$ & 39.76$^{+0.04}_{-0.04}$\\
\# 6 & $<$35.36                            & 37.26$^{+0.45}_{-1.16}$\\
\# 7 & 37.15$^{+0.22}_{-0.22}$ & 36.81$^{+0.45}_{-0.68}$\\
\# 8 & $<$37.31                            & 36.38$^{+0.41}_{-0.31}$\\
\# 9\tablenotemark{c}   & 38.62$^{+0.19}_{-0.19}$ & 37.46$^{+0.46}_{-0.92}$\\
\# 10&36.81$^{+0.26}_{-0.26}$ & 37.56$^{+0.49}_{-1.06}$\\
\# 11&40.25$^{+0.15}_{-0.13}$ & 40.29$^{+0.10}_{-0.08}$\\
\enddata

\tablenotetext{a}{The  intrinsic H$\alpha$ luminosity of each cluster, in log scale, including 1~$\sigma$ uncertainties, derived from the measurements listed in Table~\ref{tab3}, corrected for dust attenuation using eq.~(1) and the color excess values listed in Tables~\ref{tab4} and \ref{tab5}. For cluster~11, both eq.~(1) and (2) are used to remove the effects of dust attenuation. }
\tablenotetext{b}{The predicted H$\alpha$ luminosity, in log scale, from the best-fit ages and masses of each cluster, including uncertainties (Tables~\ref{tab4} and \ref{tab5}). Except for clusters~5 
and 11, the main contributor to the uncertainty in the predicted L(H$\alpha$) is the uncertainty in the best-fit age. For clusters~5 and 11, the main contributor to the overall uncertainty in L(H$\alpha$) 
is the uncertainty in the best--fit mass, since the ionizing photon rate is fairly constant for ages $\lesssim$2.5~Myr. The H$\alpha$ luminosities are related to the ionizing photon rate Q(H$^o$) via: L(H$\alpha$) [erg~s$^{-1}$] =  1.37$\times$10$^{-12}$ Q(H$^o$) [s$^{-1}$] \citep{Leitherer1999, Calzetti2013}. }
\tablenotetext{c}{The intrinsic H$\alpha$ luminosity of cluster~9 is likely over predicted, by about an order of magnitude, due to the presence of a highly dust--attenuated, younger interloper (see discussion in section~6.1). If using the color excess derived from the SED fits, rather than the line ratios (Table~\ref{tab4}), the intrinsic H$\alpha$ luminosity of cluster~9 decreases to Log[L(H$\alpha$)$_{intrinsic}$]=37.23$_{-0.12}^{+0.13}$, in better agreement with the predicted H$\alpha$ luminosity.}
\end{deluxetable}

\clearpage

\begin{deluxetable}{lrrrrrr}
\tablecolumns{6}
\tabletypesize{\small}
\tablecaption{Comparison with the Literature \label{tab7}}
\tablewidth{0pt}
\tablehead{
\colhead{Cluster} & \colhead{Age\tablenotemark{a} } & \colhead{Age$_{C97}$\tablenotemark{b} } & \colhead{Age$_{T01}$\tablenotemark{c}} & \colhead{Age$_{H04}$\tablenotemark{d}} & \colhead{Age$_{dG13}$\tablenotemark{e}} 
\\
\colhead{(\#)} & \colhead{(Myr)} & \colhead{(Myr)}  & \colhead{(Myr)} & \colhead{(Myr)} & \colhead{(Myr)} 
\\
\colhead{(1)} & \colhead{(2)} & \colhead{(3)} & \colhead{(4)} & \colhead{(5)}  & \colhead{(6)}  
\\
}
\startdata
\hline
\# 1 & 5$^{+1}_{-2}$ & 2.5--4.4 & ...& 1--5 & ...\\
\# 2 & 5$^{+1}_{-2}$ & 2.5--4.4 & ...& 1--5 & ...\\
\# 3 & 5$^{+4}_{-2}$ & ... & ... & 3--5 & ... \\
\# 4 & 6$^{+1}_{-2}$ & ... & ... & ... & ... \\
\# 5 & 1$^{+1}_{-1}$  & $<$3 & 2$^{+0.7}_{-0.8}$& 1--3 & ... \\
\# 6 & 10$^{+8}_{-2}$ & 8--12 & ... & 10--11 & 12--16 \\
\# 7 & 10$^{+4}_{-2}$ & ... & ... & 6 & ... \\
\# 8 & 15$^{+4}_{-3}$ & 10--17 & ... & 9--14 & 12--16 \\
\# 9 & 10$^{+5}_{-2}$ & 30--50 & 3$^{+0.9}_{-0.9}$& 11--14 & 6--8 \\
\# 10 & 9$^{+7}_{-2}$ & 50--60 & 8$^{+2.5}_{-0.9}$& 8--15 & ... \\
\# 11 & 1$^{+1}_{-1}$ & ... & ... & ... & ... \\
\enddata

\tablenotetext{a}{The  ages, and 1~$\sigma$ uncertainty, of the clusters in this work from the SED fitting and the EW(H$\alpha$) constraints combined together.}
\tablenotetext{b}{The ages derived by \citet{Calzetti1994}.}
\tablenotetext{c}{The ages derived by \citet{Tremonti2001}, using UV spectroscopy from the HST/STIS instrument.}
\tablenotetext{d}{The ages derived by \citet{Harris2004}.}
\tablenotetext{e}{The ages derived by \citet{deGrijs2013}.}
\end{deluxetable}

\clearpage

\clearpage

\clearpage

\clearpage

\clearpage 

\clearpage 

\clearpage 

\clearpage 

\clearpage 

\clearpage 

\clearpage 

\clearpage 


\begin{thebibliography}{}
\bibitem[Alonso--Herrero et al.(2004)]{AlonsoHerrero2004} Alonso--Herrero, A.,  Takagi, T., Baker, A.J., Rieke, G.H., Rieke, M.J., Imanishi, M., \& Scoville, N.Z. 2004, \apj, 612, 222
\bibitem[Anders et al.(2013)]{Anders2013} Anders, P., Kotulla, R., de Grijs, R., \& Wicker, J. 2013, \apj, 778, 138
\bibitem[Asplund et al.(2009)]{Asplund2009} Asplund, M., Grevesse, N., Sauval, A.J., \& Scott, P. 2009, \araa, 47, 481
\bibitem[Bastian, Cabrera--Ziri, \& Salaris(2015)]{Bastian2015} Bastian, N., Cabrera--Ziri, I., \& Salaris, M. 2015, \mnras, 449, 3333
\bibitem[Beck et al.(1996)]{Beck1996}  Beck, S.C., Turner, J.L., Ho, P.T.P., Lacy, J.H., \& Kelly, D.M. 1996, \apj, 457, 610
\bibitem[Billett, Hunter \& Elmegreen(2002)]{Billett2002} Billett, O.H., Hunter, D.A., \& Elmegreen, B.G. 2002, \aj, 123, 1454
\bibitem[Bohlin, Savage \& Drake(1978)]{Bohlin1978} Bohlin R.C., Savage B.D., Drake J.F. 1978, \apj, 224, 132
\bibitem[Bournaud et al.(2011)]{Bournaud2011} Bournaud, F., Chapon, D., Teyssier, R., Powell, L.C., Elmegreen, B.G., Elmegreen, D.M.., Duc, P.A., Contini, T., Epinat, B., \& Shapiro, K.L. 2011, \apj, 730, 4
\bibitem[Bresolin(2011)]{Bresolin2011} Bresolin, F. 2011, \apj, 729, 56
\bibitem[Caldwell \& Phillips(1989)]{Caldwell1989} Caldwell, N., \& Phillips, M.M. 1989, \apj, 338, 789
\bibitem[Calzetti(2001)]{Calzetti2001} Calzetti, D. 2001, \pasp, 113, 1449
\bibitem[Calzetti(2013)]{Calzetti2013} Calzetti, D. 2013, in `Secular Evolution of Galaxies', J.Falc\'on-Barroso \& J.H. Knapen eds. (Cambridge, UK: Cambridge University Press), 419
\bibitem[Calzetti, Kinney \& Storchi-Bergmann(1994)]{Calzetti1994} Calzetti, D., Kinney, A.L., \& Storchi-Bergmann, T. 1994, \apj, 429, 582
\bibitem[Calzetti et al.(1997)]{Calzetti1997} Calzetti, D., Meurer, G.R., Bohlin, R.C., Garnett, D.R.,  Kinney, A.L., Leitherer, C., \& Storchi--Bergmann, T. 1997, \aj, 114, 1834 
\bibitem[Calzetti et al.(1999)]{Calzetti1999} Calzetti, D., Conselice, C.J., Gallagher, J.S. III, \&  Kinney, A.L.  1999, \aj, 118, 797 
\bibitem[Calzetti et al.(2000)]{Calzetti2000} Calzetti, D., Armus, L., Bohlin, R.C., Kinney, A.L., Koornneef, J., \& Storchi-Bergmann, T. 2000, \apj, 533, 682
\bibitem[Calzetti et al.(2004)]{Calzetti2004} Calzetti, D., Harris, J., Gallagher, J.S., Smith, D., Conselice, C.J., Homeier, N., \& Kewley, L. 2004, \aj, 127, 1405
\bibitem[Calzetti et al.(2015)]{Calzetti2015} Calzetti, D., Lee, J.C., Sabbi, E., Adamo, A., Smith, L.J., Andrews, J.E., Ubeda, L., Bright, S.N., Thilker, D., Aloisi, A., et al. 
2015, \aj, 149, 51 
\bibitem[Campbell, Terlevich \& Melnick(1986)]{Campbell1986} Campbell A., Terlevich R., \& Melnick J., 1986, \mnras, 223, 811
\bibitem[Chandar et al.(2005)]{Chandar2005} Chandar, R., Leitherer, C., Tremonti, C.A., Calzetti, D., Aloisi, A., Meurer, G.R., \&  de Mello, D. 2005, \apj, 628, 210
\bibitem[Cresci, Vanzi \& Sauvage(2005)]{Cresci2005} Cresci G., Vanzi L., \& Sauvage M. 2005, \aap, 433, 447
\bibitem[Cervi\~no \& Luridiana(2004)]{Cervino2004}  Cervi\~no, M., \& Luridiana, V. 2004, \aap, 413, 145
\bibitem[Cook et al.(2014)]{Cook2014} Cook, D.O., Dale, D.A., Johnson, B.D., van Zee, L., Lee, J.C., Kennicutt, R.C., Calzetti, D., Staudaher, S.M., \& Engelbracht, C.W. 2014, \mnras, 445, 899
\bibitem[Crowther et al.(2010)]{Crowther2010} Crowther, P.A., Schnurr, O., Hirschi, R., Yusof, N., Parker, R.J., Goodwin, S.P., \& Kassim, H.A. 2010, \mnras, 408, 731
\bibitem[Dale et al.(2009)]{Dale2009} Dale, D.A., Cohen, S.A., Johnson, L.C., Schuster, M.D., Calzetti, D., Engelbracht, C.W., Gil de Paz, A., Kennicutt, R.C., Lee, J.C.,  Begum, A., et al. 2009, \apj, 703, 517
\bibitem[Davidge(2007)]{Davidge2007} Davidge, T.J. 2007, \aj, 134, 1799
\bibitem[de Grijs et al.(2013)]{deGrijs2013} de Grijs, R., Anders, P., Zackrisson, E., \& \"Ostlin, G. 2013, \mnras, 431, 2917
\bibitem[de Mink et al.(2009)]{deMink2009} de Mink, S.E., Pols, O.R., Langer, N., \& Izzard, R.G. 2009, \aap, 507, L1 
\bibitem[Doran et al. (2013)]{Doran2013} Doran, E.I., Crowther, P.A., de Koter, A., Evans, C.J., McEvoy, C., Walborn, N.R., Bastian, N., Bestenlehner, J.M., Graefener, G., 
Herrero, A., et al. 2013, \aap, 558, A134
\bibitem[Draine(2003)]{Draine2003} Draine, B.T. 2003, \araa, 41, 241
\bibitem[Draine(2004)]{Draine2004} Draine, B.T. 2004, in `The Cold Universe', eds. A.W. Blain, F. Combes, B.T. Draine, D. Pfenniger \& Y. Revaz, 
Saas-Fee Advanced Course 32 (Springer-Verlag), 213 
\bibitem[Draine et al.(2007)]{Draine2007} Draine, B.T., Dale, D.A., Bendo, G., Gordon, K.D., Smith, J.D.T., Armus, L., Engelbracht, C.W., Helou, G., Kennicutt, R.C., Li, A., et 
al. 2007, \apj, 663, 866
\bibitem[de Mink(2014)]{deMink2014} de Mink, S.E., Sana, H., Lange, N., Izzard, R.G., \& Schneider, F.R.N. 2014, \apj, 782, 7
\bibitem[Eldridge \& Stanway(2009)]{EldridgeStanway2009} Eldridge, J.J., \& Stanway, E.R. 2009, \mnras, 400, 1019
\bibitem[Efremov \& Elmegreen(1998)]{Efremov1998} Efremov, Y.N., \& Elmegreen, B.G. 1998, \mnras, 299, 588
\bibitem[Elmegreen et al.(2014)]{Elmegreen2014} Elmegreen, D.M., Elmegreen, B.G., Adamo, A., Aloisi. A. Andrews, J., Annibali, F., Bright, S.N.,, Calzetti, D., Cignoni, M.,  Evans, A.S., et al. 2014, \apj, 787, L15
\bibitem[Fall, Chandar \& Whitmore(2009)]{Fall2009} Fall, S.M., Chandar, R., \& Whitmore, B.C. 2009, \apj, 704, 453
\bibitem[Ferland et al.(2013)]{Ferland2013} Ferland, G.J., Porter, R.L., van Hoof, P.A.M. , Williams, R.J.R. , Abel, N.P. , Lykins, M.L. , Shaw, G., Henney, W.J., \& 
Stancil, P.C. 2013, RMexAA, 49, 1 (The 2013 Release of Cloudy)
\bibitem[Fitzpatrick(1999)]{Fitzpatrick1999} Fitzpatrick, E.L. 1999, \pasp, 111, 63
\bibitem[Freedman et al.(2001)]{Freedman2001} Freedman, W.L., Madore, B.F., Gibson, B.K., Ferrarese, L., Kelson, D.D.,  et al. 2001, ApJ, 553, 47
\bibitem[Gazak et al.(2013)]{Gazak2013} Gazak, J.Z., Bastian, N., Kudritzki, R.-P., Adamo, A., Davies, B., Plez, B., \& Urbaneja, M.A., 2013, \mnras, 430, L35
\bibitem[Girardi et al.(2000)]{Girardi2000} Girardi, L., Bressan, A., Bertelli, G., \& Chiosi, C. 2000, \aap Supp., 141, 371
\bibitem[Gorjian, Turner \& Beck(2001)]{Gorjian2001} Gorjian, V., Turner, J.L., \& Beck, S.C. 2001, \apj, 554, L29
\bibitem[Grimes et al.(2009)]{Grimes2009} Grimes, J.P., Heckman, T., Aloisi, A., Calzetti, D., Leitherer, C., Martin, C.L., Meurer, G., Sembach, K., \& Strickland, D. 2009, \apjs, 
181, 272
\bibitem[Harbeck, Gallagher \& Crnojevic(2012)]{Harbeck2012} Harbeck, D., Gallagher, J.S. III,  \& Crnojevic, D. 2012, \mnras, 422, 629
\bibitem[Harris et al.(2004)]{Harris2004} Harris, J., Calzetti, D., Gallagher, J.S., Smith, D.A., \& Conselice, C.J. 2004, \apj, 603, 503
\bibitem[Hong et al.(2013)]{Hong2013} Hong, S., Calzetti, D., Gallagher, J.S., III; Martin, C.L., Conselice, C.J., \& Pellerin, A. 2013, \apj, 777, 63	
\bibitem[Hunter et al.(2000)]{Hunter2000} Hunter, D.A., O'Connell, R.W., Gallagher, J.S., \& Smecker--Hane, T.A. 2000, \aj, 120, 2383
\bibitem[Johnson et al.(1999)]{Johnson1999} Johnson, K.E., Vacca, W.D., Leitherer, C., Conti, P.S., \& Lipscy, S.J. 1999, \aj, 117, 1708
\bibitem[Johnson et al.(2004)]{Johnson2004} Johnson, K.E., Indebetouw, R., Watson, C., \& Kobulnicky, H.A. 2004, \aj, 128, 610
\bibitem[Johnson, Hunt \& Reines(2009)]{Johnson2009} Johson, K.E., Hunt, L.K., \& Reines, A.E. 2009, \aj, 137, 3788 
\bibitem[Karachentsev et al.(2007)]{Karachentsev2007} Karachentsev I. D. et al., 2007, \aj, 133, 504
\bibitem[Kauffmann et al.(2013)]{Kauffmann2013} Kauffmann, J., Pillai, T., \& Zhang, Q. 2013, \aap, 765, L35
\bibitem[Kennicutt \& Evans(2012)]{KennicuttEvans2012} Kennicutt, R.C., \& Evans, N.J. 2012, \araa,  50, 531
\bibitem[Kennicutt et al.(2008)]{Kennicutt2008} Kennicutt, R.C., Lee, J.C., Funes, S.J. , Sakai, S., \& Akiyama, S. 2008, \apjs, 178, 247
\bibitem[Kobayashi et al.(2011)]{Kobayashi2011} Kobayashi, H., Kimura, H., Watanabe, S., Yamamoto, T., \& M\"uller, S. 2011, {\em Earth, Planets, and Space}, 63, 1067
\bibitem[Kobulnicky \& Skillman(1995)]{Kobulnicky1995} Kobulnicky H.A., \& Skillman, E.D., 1995, \apj, 454, L121
\bibitem[Kobulnicky et al.(1997)]{Kobulnicky1997} Kobulnicky H.A., Skillman E.D., Roy J.-R.,Walsh J.R., \& Rosa M.R., 1997,
\apj, 277, 679
\bibitem[Kobulnicky \& Johnson(1999)]{Kobulnicky1999} Kobulnicky H.A., \& Johnson, K.E., 1999, \apj, 527, 154
\bibitem[Koehler et al.(2015)]{Koehler2015} Koehler, K., Langer, N., de Koter, A,; de Mink, S.E., Crowther, P.A., Evans, C.J., Grafener, G., Sana, H., Sanyal, D., Schneider, F.R.N., \&  Vink, J.S. 2015, \aap, 573, A71
\bibitem[Koekemoer et al.(2006)]{Koekemoer2006} Koekemoer, A.M., McLean, B., McMaster, M., \& Jenkner, H. 2006, in `The 2005 HST calibration workshop: Hubble after the transition to two-gyro mode', Proceedings of a workshop held at the Space Telescope Science Institute, Baltimore, Maryland; A. M. Koekemoer, P. Goudfrooij, and L. L. Dressel eds., p. 384
\bibitem[Kreckel et al.(2013)]{Kreckel2013} Kreckel, K.,Groves, B., Schinnerer, E., Johnson, B.D., Aniano, G., Calzetti, D., Croxall, K.V., Draine, B.T., Gordon, K.D., Crocker, A.F., 
et al. 2013, \apj, 771, 62
\bibitem[Kroupa(2001)]{Kroupa2001} Kroupa, P. 2001, \mnras, 322, 231
\bibitem[Lada \& Lada(2003)]{Lada2003} Lada, C.J., \& Lada, E.A. 2003, \araa, 41, 67
\bibitem[Lee, Chandar \& Whitmore(2005)]{Lee2005} Lee, M.G., Chandar, R., \& Whitmore, B.C. 2005, \aj, 130, 2128
\bibitem[Leitet et al.(2013)]{Leitet2013} Leitet, E., Bergvall, N., Hayes, M., Linn\'e, S., \& Zackrisson, E. 2013, \aap, 553, A106
\bibitem[Leitherer et al.(1999)]{Leitherer1999} Leitherer, C., Schaerer, D., Goldader, J.D., Gonzalez Delgado, R.M, Robert, C., Kune, D.F., de Mello, D.F.,  Devost, D., 
\&  Heckman, T.M. 1999, \apjs, 123, 3
\bibitem[Leitherer et al.(2014)]{Leitherer2014} Leitherer, C., Eckstr\"om, S., Meynet, G., Schaerer, D., Agienko, K.B., \& Levesque, E.M., 2014, \apjs, 212, 14
\bibitem[Lopez-Sanchez et al.(2007)]{LopezSanchez2007} Lopez--Sanchez, A.R., Esteban C.,  Garcia-Rojas, J., Peimbert, M., \& Rodriguez, M., 2007, \apj 656, 168
\bibitem[Lopez-Sanchez et al.(2012)]{LopezSanchez2012} Lopez--Sanchez, A.R., Koribalski, B.S., van Eymeren J., Esteban C., Kirby
E., Jerjen H., \& Lonsdale N., 2012, \mnras, 419, 1051
\bibitem[Maiz-Apellaniz(2001)]{Maiz2001} Maiz-Apellaniz, J. 2001, \apj, 563, 151
\bibitem[Maoz et al.(1996)]{Maoz1996} Maoz, D., Barth, A.J., Sternberg, A., Filippenko, A.V., Ho, L.C., Macchetto, F.D,; Rix, H.-W., \& Schneider, D.P. 1996, \aj, 111, 2248
\bibitem[Marlowe et al.(1995)]{Marlowe1995} Marlowe A.T., Heckman T.M., Wyse R.F.G., \& Schommer R., 1995, \apj, 438, 563
\bibitem[Martin(1998)]{Martin1998} Martin, C.L., 1998, \apj, 506, 222
\bibitem[Martins et al.(2012)]{Martins2012} Martins, F., Foerster Schreiber, N.M., Eisenhauer, F., \& Lutz, D. 2012, \aap, 547, A17
\bibitem[Martin--Hernandez, Schaerer \& Sauvage(2005)]{MartinHernandez2005} Martin--Hernandez, N.L., Schaerer, D., \& Sauvage, M. 2005, \aap, 449, 2005
\bibitem[McQuinn et al.(2010)]{McQuinn2010} McQuinn, K.B.W. et al., 2010, \apj, 724, 49
\bibitem[Meier, Turner \& Beck(2002)]{Meier2002} Meier, D.S., Turner, J.L., \& Beck, S.C. 2002, \aj, 124, 877
\bibitem[Meurer et al.(1995)]{Meurer1995} Meurer, G.R., Heckman, T.M., Leitherer, C., Kinney, A., Robert, C., \& Garnett, D.R., 1995, \aj, 110, 2665
\bibitem[Meynet et al.(1994)]{Meynet1994} Meynet, G., Schaller, G., Schaerer, D., \& Charbonnel, C., 1994, \aap Supp., 103, 97
\bibitem[Monreal--Ibero et al.(2010)]{MonrealIbero2010} Monreal--Ibero A., Võlchez J.M., Walsh J.R., \& Mu\~noz-Tu\~non C., 2010, \aap,
517, 27
\bibitem[Monreal--Ibero, Walsh \& Vilchez(2012)]{MonrealIbero2012} Monreal--Ibero, A., Walsh, J. R., \& V'lchez, J. M. 2012, \aap, 544, A60
\bibitem[Monreal--Ibero et al.(2013)]{MonrealIbero2013} Monreal--Ibero, A., Walsh, J.R., Westmoquette, M.S., \& V'lchez, J.M. 2013, \aap, 553, A57
\bibitem[Moorwood \& Glass(1982)]{Moorwood1982} Moorwood, A.F.M., \& Glass, I.M. 1982, \aap, 115, 84
\bibitem[Moustakas \& Kennicutt(2006)]{Moustakas2006} Moustakas, J.,\&  Kennicutt, R.C.,  2006,  \apjs, 164, 81
\bibitem[Parmentier et al.(2009)]{Parmentier2009} Parmentier, G., Goodwin, S.P., Kroupa, P., \& Baumgardt, H. 2009, \apj, 678, 347
\bibitem[Pellerin \& Robert(2007)]{Pellerin2007} Pellerin, A., \& Robert, C. 2007, \mnras, 381, 288
\bibitem[Popescu \& Hanson(2010)]{Popescu2010} Popescu, B., \& Hanson, M.M. 2010, \apj, 713, L21
\bibitem[Portegies Zwart et al.(2010)]{Portegies2010} Portegies Zwart, S.F., McMillan, S.L.W., \& Gieles, M.\ 2010, \araa, 48, 431 
\bibitem[Reines et al.(2010)]{Reines2010} Reines, A.E., Nidever, D.L., Whelan, D.G., \& Johnson, K.E. 2010, \apj, 708, 26
\bibitem[Rich et al.(2010)]{Rich2010} Rich, J.A., Dopita, M.A., Kewley, L.J., \& Rupke, D.S.N. 2010, \apj, 721, 505 
\bibitem[Rieke, Lebofsky, \& Walker(1988)]{Rieke1988} Rieke, G.H., Lebofsky, M.J., \& Walker, C.E. 1988, \apj, 325, 679
\bibitem[Sakai et al.(2004)]{Sakai2004} Sakai, S., Ferrarese, L., Kennicutt, R.C., \& Saha, A. 2004, \apj, 608, 42
\bibitem[Sana et al.(2012)]{Sana2012} Sana, H., de Mink, S.E., de Koter, A., Langer, N., Evans, C.J., Gieles, M., Gosset, E., Izzard, R.G., Le Bouquin, J.-B., \& Schneider, F.R.N., 2012, {\bf Science}, 337,  444
\bibitem[Schaerer et al.(1997)]{Schaerer1997} Schaerer ,D., Contini, T., Kunth ,D., \& Meynet, G., 1997, \apj, 481, L75 
\bibitem[Schaerer \& Charbonnel(2011)]{Schaerer2011} Schaerer, D., \& Charbonnel, C. 2011, \mnras, 413, 2297
\bibitem[Schlafly \& Finkbeiner(2011)]{SchlaflyFinkbeiner2011} Schlafly, E.F., \& Finkbeiner, D.P., 2011, \apj, 737, 103
\bibitem[Schneider et al.(2014)]{Schneider2014} Schneider, F.R.N., Izzard, R.G., de Mink, S.E., Langer, N., Stolte, A., de Koter, A., Gvaramadze, V.V., Hussmann, B., Liermann, A., \&  Sana, H. 2014, \apj, 780 117
\bibitem[Schneider et al.(2015)]{Schneider2015} Schneider, F.R.N., Izzard, R.G., Langer, N., \& de Mink, S.E. 2015, \apj, 805, 20
\bibitem[Smith et al.(2006)]{Smith2006} Smith, L.J., Westmoquette, M.S., Gallagher, J.S., O'Connell, R.W., Rosario, D.J., \&  de Grijs, R. 2006, \mnras, 370, 513
\bibitem[Storchi-Bergmann, Kinney \& Challis(1995)]{StorchiBergmann1995} Storchi-Bergmann, T., Kinney, A. L., \& Challis, P. 1995, \apjs, 98, 103
\bibitem[Strickland \& Stevens(1999)]{Strickland1999} Strickland, D.K., \& Stevens, I.R. 1999, \mnras, 306, 43
\bibitem[Summers et al.(2004)]{Summers2004} Summers, L.K., Stevens, I.R., Strickland, D.K., \& Heckman, T.M. 2004, \mnras, 351, 1
\bibitem[Thim et al.(2003)]{Thim2003} Thim F., Tammann G.A., Saha A., Dolphin A., Sandage A., Tolstoy E., \& Labhardt L., 2003, \apj, 590, 256
\bibitem[Turner \& Welch(1984)]{Turner1984} Turner, J.L., \& Welch, W.J. 1984, \apj, 287, L81
\bibitem[Turner, Beck \& Hurt(1997)]{Turner1997} Turner, J.L., Beck, S.C., \& Hurt, L.R. 1997, \apj, 474, L11
\bibitem[Turner, Beck \& Ho(2000)]{Turner2000} Turner, J.L., Beck, S.C., \& Ho, P.T.P. 2000, \apj, 532, L109
\bibitem[Turner \& Beck(2004)]{Turner2004} Turner, J.L., \& Beck, S.C. 2004, \apj, 602, L85
\bibitem[Turner et al.(2015)]{Turner2015} Turner, J.L., Beck, S.C., Bendford, D.J., Consiglio, S.M., Ho, P.T.P., Kovacs, A., Meier, D.S., \& Zhao, J.-H. 2015, Nature, 519, 331
\bibitem[Tremonti et al.(2001)]{Tremonti2001} Tremonti, C.A., Calzetti, D., Leitherer, C., \& Heckman, T.M. 2001, \apj, 555, 322
\bibitem[van den Bergh(1980)]{vdBergh1980} van den Bergh, S. 1980, \pasp, 92, 122
\bibitem[Vanzi \& Sauvage(2004)]{VanziSauvage2004} Vanzi, L., \& Sauvage, M. 2004, \aap, 415, 509
\bibitem[Vazquez \& Leitherer(2005)]{Vazquez2005} Vazquez, G.A., \& Leitherer, C. 2005, \apj, 621, 695 
\bibitem[Walsh \& Roy(1989)]{WalshRoy1989} Walsh, J. R., \& Roy, J.-R. 1989, \mnras, 239, 297
\bibitem[Walsh et al.(1999)]{Walsh1999} Walsh, A.J., Burton, M.G., Hyland, A.R., \& Robinson, G. 1999, \mnras, 309, 905
\bibitem[Westmoquette et al.(2013)]{Westmoquette2013}  Westmoquette, M.S., James, B., Monreal--Ibero, A., \& Walsh, J.R. 2013, \aap, 550, A88
\bibitem[Whitmore \& Zhang(2002)]{Whitmore2002} Whitmore, B.C., \& Zhang, Q. 2002, \aj, 124, 1418
\bibitem[Whitmore et al.(2010)]{Whitmore2010} Whitmore, B.C., Chandar, R., Schweizer, F., Rothberg, B., Leitherer, C.,Rieke, M., Rieke, G., Blair, W.P., Mengel, S., \& 
Alonso--Herrero, A., 2010, \aj, 140, 75
\bibitem[Whitmore et al.(2011)]{Whitmore2011} Whitmore, B.C., Chandar, R., Kim, H., Kaleida, C., Mutchler, M., Stankiewicz, M., Calzetti, D., Saha, A., O'Connell, R., Balick, B., et al. 2011, \apj, 729, 78
\bibitem[Whitmore et al.(2014a)]{Whitmore2014a} Whitmore, B.C., Brogan, C., Chandar, R., Evans, A., Hibbard, J., Johnson, K., Leroy, A., Privon, G., Remijan, A., \& Sheth, K., 2014,a \apj, 795, 156
\bibitem[Whitmore et al.(2014b)]{Whitmore2014b} Whitmore, B.C., Chandar, R., Bowers, A.S., Larsen, S., Lindsay, K., Ansari, A., \& Evans, J. 2014b, \aj, 147, 78
\bibitem[Yusof et al.(2013)]{Yusof2013} Yusof, N., Hirschi, R., Meynet, G., Crowther, P.A., Ekstr{\"o}m, S., Frischknecht, U., Georgy, C., Abu Kassim, H., \& Schnurr, O. 2013, \mnras, 433, 1114
\bibitem[Zackrisson et al.(2001)]{Zackrisson2001} Zackrisson, E., Bergvall, N., Oloffson, K., \& Siebert A. 2001, \aap, 375, 814
\bibitem[Zackrisson et al.(2011)]{Zackrisson2011} Zackrisson, E., Rydberg, C.-E., Schaerer, D., \"Ostlin, G., \& Tuli, M. 2011, \apj, 740, 13 
\bibitem[Zastrow et al.(2011)]{Zastrow2011} Zastrow, J., Oey, M.S., Veilleux, S., McDonald, M., \& Martin, C.L. 2011, \apj, 741, L17
\bibitem[Zastrow et al.(2013)]{Zastrow2013} Zastrow, J., Oey, M.S., Veilleux, S., \& McDonald, M. 2013, \apj, 779, 76
\end{thebibliography}
\end{document}